\documentclass{scrartcl}
\setkomafont{author}{\sffamily}
\setkomafont{date}{\sffamily}

\usepackage[utf8]{inputenc}
\usepackage[T1]{fontenc}
\usepackage[margin=24mm]{geometry}
\usepackage[svgnames]{xcolor}

\usepackage{multicol,microtype,lmodern,enumitem,subcaption,graphicx}
\setlist{leftmargin=*}

\usepackage{mathtools,amssymb}

\usepackage[colorlinks=true,allcolors=DarkBlue]{hyperref}
\usepackage{cleveref}
\crefname{section}{sec.}{secs.}
\crefname{figure}{fig.}{figs.}
\crefname{appendix}{app.}{apps.}
\crefname{opt}{option}{options}

\usepackage{mmacells}
\mmaSet{morelst={breaklines=true},leftmargin=3.5em}

\usepackage{tikz,pgfplots}
\pgfplotsset{compat=1.8}
\usetikzlibrary{patterns,decorations.markings,arrows.meta, decorations.pathmorphing,backgrounds,positioning,intersections}
\tikzset{
    cross/.style={fill=white,path picture={\draw[black] (path picture bounding box.south east) -- (path picture bounding box.north west) (path picture bounding box.south west) -- (path picture bounding box.north east);}},
    dressed/.style={fill=white,postaction={pattern=north east lines}},
    momentum/.style 2 args={->,semithick,yshift=5pt,shorten >=5pt,shorten <=5pt},
    loop/.style 2 args={thick,decoration={markings,mark=at position {#1} with {\arrow{>},\node[anchor=\pgfdecoratedangle-90,font=\footnotesize,] {$p_{#2}$};}},postaction={decorate}}
}

\usepackage[backend=biber,sorting=none,citestyle=numeric-comp,giveninits=true,doi=false,date=year,url=false,backref=true,maxbibnames=5]{biblatex}
\renewbibmacro{in:}{}

\DeclareFieldFormat{eprint:arxiv}{\href{http://arxiv.org/abs/#1}{#1}}
\DeclareFieldFormat{pages}{\mkfirstpage{#1}}

\BeforeStartingTOC[toc]{\begin{multicols}{2}}
\AfterStartingTOC[toc]{\end{multicols}}

\setcapindent{0ex}

\newcommand{\dif}{\mathrm{d}}
\newcommand{\reals}{\mathbb{R}}
\newcommand{\integers}{\mathbb{Z}}
\renewcommand{\vec}[1]{\boldsymbol{#1}}
\newcommand{\crefrel}[2]{\overset{\mathclap{\scalebox{0.6}{\labelcref{#1}}}}{#2}}
\newcommand{\csqrt}[1]{\sqrt{\smash[b]{#1}}}

\newenvironment{mleqn}
	{\begin{equation}\begin{aligned}}
	{\end{aligned}\end{equation}\ignorespacesafterend}

\let\Re\relax
\DeclareMathOperator{\Re}{Re}
\let\Im\relax
\DeclareMathOperator{\Im}{Im}
\DeclareMathOperator{\Tr}{Tr}
\DeclareMathOperator{\sign}{sign}
\DeclareMathOperator*{\Res}{Res}
\DeclareMathOperator{\disc}{disc}
\DeclareMathOperator{\vol}{vol}
\DeclareMathOperator*{\sumint}{%
\mathchoice%
	{\ooalign{$\displaystyle\sum$\cr\hidewidth$\displaystyle\int$\hidewidth\cr}}
	{\ooalign{\raisebox{.14\height}{\scalebox{.7}{$\textstyle\sum$}}\cr\hidewidth$\textstyle\int$\hidewidth\cr}}
	{\ooalign{\raisebox{.2\height}{\scalebox{.6}{$\scriptstyle\sum$}}\cr$\scriptstyle\int$\cr}}
	{\ooalign{\raisebox{.2\height}{\scalebox{.6}{$\scriptstyle\sum$}}\cr$\scriptstyle\int$\cr}}
}

\newcommand{\tfie}{\bar{I}}
\newcommand{\tfim}{I}
\newcommand{\tfimdl}{\tilde I}
\newcommand{\tfjm}{J}
\newcommand\pmint\oint
\newcommand\mmint\sumint

\begin{document}

\title{Functional Renormalization\\Analytically Continued}
\author{\href{mailto:janosh.riebesell@gmail.com}{Janosh Riebesell}}
\date{December 14, 2017}
\maketitle

\begin{center}
	\noindent Master thesis carried out at the \href{https://www.thphys.uni-heidelberg.de}{ITP Heidelberg}. Supervised by\\[1ex]
	\href{stefan.floerchinger@thphys.uni-heidelberg.de}{Stefan Flörchinger}, \href{m.scherer@thphys.uni-heidelberg.de}{Michael Scherer} and \href{c.wetterich@thphys.uni-heidelberg.de}{Christof Wetterich}.\\[1ex]
	Submitted to Heidelberg University's department of physics and astronomy.
\end{center}

\begin{abstract}
	\noindent We discuss a method to analytically continue functional renormalization group equations from imaginary Matsubara frequencies to the real frequency axis as developed in \cite{floerchinger2012analytic}. In this formalism, we discuss the analytic structure of the flowing action and the propagator for a theory of scalar fields with $O(N)$ symmetry. We go on to show how it is possible to derive and solve flow equations for real-time properties such as particle decay widths. The treatment is fully Lorentz-invariant and enables an improved, self-consistent derivative expansion in Minkowski space.
\end{abstract}
\vfil
{\hypersetup{hidelinks}\tableofcontents}

\newpage
\section{Introduction}

Field-theoretic infinities first arose in Lorentz's work \cite{lorentz1909theory} on classical electrodynamics of point particles in the early 20th century. Over the ensuing decades, divergencies proved so persistent and prevalent all across quantum field theory (QFT) that physicists were forced to develop elaborate machinery to extract sensical predictions out of a minefield of singularities. This line of research resulted in the methods of renormalization and regularization. For years the community saw these as objectionable means by which to work around an inability to develop a more well-behaved description of reality. By mid-century, it had reached a point where many physicists thought QFT had to be discarded outright since so many of its predictions aborted in infinities.

Over the course of the 1970s, this view underwent a dramatic shift. In 1971 Wilson published a seminal paper \cite{wilson1971renormalization} on what is now known as the Wilsonian interpretation of renormalization. According to Wilson, infinities in field theory are merely the result of feeding a fundamentally flawed assumption -- the pretense of knowing the fundamental laws of physics down to arbitrarily small distance scales -- into the otherwise functioning machinery of QFT. Instead, Wilson argued, we are ignorant of the correct microscopic degrees of freedom and the laws governing their dynamics. Hence we should view our models of reality as no more than effective descriptions of nature that remain approximately valid down to some cutoff scale at which new physics emerges.

The ensuing change of perspective was so far-reaching that nowadays QFT is heralded as the most successful achievement of theoretical physics to date. Renormalization and regularization are generally accepted as essential tools that allow us to predict nature on experimentally accessible scales even though our models breakdown at smaller distances.

And yet despite its profound impact on our understanding of modern physics, the Wilsonian renormalization group largely failed to manifest itself in actual applications. Over the first 20 years, it remained a mostly formal construct until the unceasing strive for a capable and versatile approach to non-perturbative problems in QFT culminated in Wetterich's 1993 functional formulation of renormalization \cite{wetterich1993average}. This has since proven a workable handle for applying Wilson's renormalization group to practical computations and specific models.

These days applications range from cold gases \cite{diehl2010functional,floerchinger2010functional,boettcher2012ultracold,strack2008renormalization,floerchinger2009superfluid,floerchinger2010modified,floerchinger2011efimov} and critical phenomena \cite{wetterich2001effective,berges2002non,honerkamp2001magnetic,zinn1996quantum,amit2005field,wilson1983renormalization,pelissetto2000critical,fisher1974renormalization,berges1996critical,berges1996critical,bagnuls2001exact,gersdorff2001nonperturbative} to quantum chromodynamics \cite{schaefer2008renormalization,pawlowski2006functional,gross1981qcd,braun2006chiral,braun2011phase} and quantum gravity \cite{reuter1998nonperturbative,reuter2007functional,litim1999gauge,gies2012introduction,weinberg1979general,niedermaier2007asymptotic,oriti2009approaches,litim2008fixed,christiansen2014fixed,percacci2011short,reuter2012quantum,reuter2013asymptotic,nagy2014lectures,percacci2009asymptotic}. The functional renormalization group (FRG) has proven especially successful in scenarios that are difficult to treat with other methods such as massless degrees of freedom induced by spontaneous symmetry breaking (Higgs mechanism) or the question of asymptotic safety in quantum gravity. It now stands among the most powerful tools to solve non-perturbative problems in modern physics.

However, certain shortcomings remain. So far, the formalism has been explored mostly in Euclidean space where it describes either static, classical statistical field theories or quantum fields in the imaginary-time formalism. Besides avoiding the path-integral's notorious real-time sign problem, Euclidean space offers the important practical advantage that most propagators and higher correlation functions exhibit but a single isolated singularity at vanishing momentum \cite{floerchinger2012analytic}.

This state of affairs leaves something to be desired, however. Real-time physics takes place in Minkowski, not Euclidean space. We therefore expect the FRG's spectrum of applications to benefit immensely from an extension to this new domain. After all, real-time correlation functions hold the key to dynamical observables such as the spectral function which contains information about resonances, the mass spectrum and transport coefficients of a theory \cite{pawlowski2015real}.

Initial attempts at gathering real-time information from the FRG kept the formalism confined to Euclidean space, used its machinery to compute imaginary-time observables at the macroscopic scale $k = 0$ and resorted to numerical techniques such as Padé approximants or the maximum entropy method to perform the analytic continuation to Minkowski space based on limited numerical Euclidean data. Not only does this constitute an ill-defined problem, these approaches also entail a bias about the continuation and require highly accurate Euclidean data. As a result, this approach suffers from a systematic as well as a numerical constraint \cite{pawlowski2015real}. The reconstruction problem can only be overcome by a real-time formulation of the theory.

There is no fundamental obstruction that prevents us from extending the FRG to Minkowski space \cite{floerchinger2012analytic}. However, in practice a number of problems arise. Because the Lorentz-invariant four-momentum square $p^2 = -p_0^2 + \vec{p}^2$ is no longer positive semi-definite, the question arises which modes actually correspond to the infrared and which to the ultraviolet. This renders the problem of how to construct an appropriate regulator function non-trivial, particularly if Lorentz invariance is to be preserved. Also, the Euclidean space flowing action $\Gamma_k$ can be shown to approach the microscopic action $S$ for large cutoff scales which serves as a convenient initial condition for the Wetterich equation. This is not necessarily the case in Minkowski space.	

In this work, we give a detailed introduction to an approach developed in \cite{floerchinger2012analytic} with which to overcome these difficulties and calculate dynamical properties from real-time functional renormalization. The formalism uses a linear response framework where the analytic continuation from imaginary Matsubara frequencies to real frequencies is carried out on the level of the flow equations rather than on the final result at $k = 0$, since the former are available in analytic form while the latter can only be attained numerically. This procedure is then applied to the scalar $O(N)$ model.	

We proceed as follows. \Cref{sec:theoretical foundations} recounts the basics of Wetterich's functional formulation of renormalization. \Cref{sec:fr in ms} focuses on its peculiarities in Minkowski space. The analytic continuation of flow equations is carried out in \cref{sec:analytic continuation of flow equations}. \Cref{sec:matsubara summation,sec:momentum integration} respectively present the Matsubara summation and momentum integration performed in terms of conveniently defined threshold functions. Finally, \cref{sec:numerical results} holds numerical results and \cref{sec:conclusions} states our conclusions.

\section{Theoretical Foundations}
\label{sec:theoretical foundations}

In 1973, Wegner and Houghton \cite{wegner1973renormalization} were the first to combine Wilson's intuitive understanding of renormalization with the functional methods of quantum field theory. Over the following two decades, this initially rather formal marriage was developed further \cite{polchinski1984renormalization} and rendered viable for practical applications in 1993 by Wetterich's discovery of an exact evolution equation \cite{wetterich1993exact} for the so-called flowing action $\Gamma_k$. It serves as the central object in FRG to determine the properties of a theory, including its excitation spectrum, symmetries, dynamics and conserved quantities. Wetterich's surprisingly intuitive functional differential equation for $\Gamma_k$ demonstrated that the scale dependence of the flowing action is generated solely by one-loop fluctuations of the regularized propagator.	

$\Gamma_k$ is constructed from the microscopic action $S$ by adding to it an infrared regulator $R_k$ with associated renormalization scale $k$. This is where scale dependence enters the formalism. The modification bestows upon $\Gamma_k$ the remarkable property of continuously interpolating between the microscopic action $S = \lim_{k \to \Lambda} \Gamma_k$ at high energies ($\Lambda$ is some ultraviolet cutoff that regularizes the theory) and the quantum effective action $\Gamma = \lim_{k \to 0} \Gamma_k$ at macroscopic scales -- a process known as transition to complexity. True to the spirit of Wilson's original formulation of renormalization, $R_k$ implements during this procedure a smooth decoupling of high-momentum modes while also acting as an infrared regulator in theories with massless particles -- a major advantage when dealing with spontaneous symmetry breaking in the $O(N)$ model.

This section briefly introduces the most important aspects of Euclidean space functional renormalization. Setting out from the Feynman path integral, \cref{sec:wetterich equation} derives Wetterich's equation for a system of scalar fields $\phi_a$, $a \in \{1,\dots,N\}$. \Cref{sec:average action,sec:regulator} list important properties of the flowing action $\Gamma_k$ and the regulator $R_k$. \Cref{sec:truncations} presents different truncation schemes that make practical applications of this formalism possible. Finally, in \cref{sec:potential flow} we construct an exact flow equation for the effective potential $U_k$ -- the most important quantity when it comes to equilibrium physics such as the ground state and the low-lying mass spectrum.

\subsection{Wetterich Equation}
\label{sec:wetterich equation}

We discuss the derivation of Wetterich's functional renormalization group equation following \cite{wetterich1993exact,wetterich2001effective}. The partition function for a theory of $N$ scalar fields $\phi_a(x)$, $a \in \{1,\dots,N\}$ in $d$ Euclidean dimensions with microscopic action $S[\phi]$ in the presence of the source $\tfjm_a(x)$ reads
\begin{equation}
	Z[J]
	= \int \mathcal{D} \phi \, e^{-S[\phi] + J \cdot \phi},
\end{equation}
We specify the action together with some ultraviolet reference scale $\Lambda$ much larger than all other physical scales \cite{tetradis1994critical}. ($\Lambda$ will be the scale at which we initialize our flow equations.) The scalar product sums over field components and integrates over all space
\begin{equation}
	J \cdot \phi
	= \int_x \, \tfjm_a(x) \, \phi_a(x)
	= \int_p \, \tilde{J}_a(p) \, \tilde\phi_a(-p),
\end{equation}
where
\begin{equation}
	\int_x
	= \int_{\reals^d} \dif^d x,
	\qquad
	\int_p
	= \int_{\reals^d} \frac{\dif^d p}{(2 \pi)^d},
\end{equation}
and $\tilde\phi_a(-p) = \tilde\phi_a^\ast(p)$ for real scalar fields. To save on notation, we won't continue to indicate spacetime dependence nor Fourier transforms $\tilde\phi$ explicitly and take $\phi$ and $J$ index-free as vectors in $N$-dimensional field space.

Expectation values and correlation functions are obtained from $Z[J]$ through functional differentiation,
\begin{align}
	&\varphi
	= \langle\phi\rangle
	= \frac{1}{Z} \frac{\delta Z}{\delta J}
	= \frac{1}{Z} \int \mathcal{D} \phi \, \phi \, e^{-S[\phi] + J \cdot \phi},\\
	&\bigl\langle\phi^n\bigr\rangle
	= \frac{1}{Z} \frac{\delta^n Z}{\delta^n J}
	= \frac{1}{Z} \int \mathcal{D} \phi \, \phi^n \, e^{-S[\phi] + J \cdot \phi},
\end{align}
earning $Z[J]$ the name generating functional. A more efficient description is possible in terms of only the connected correlation functions. These in turn are generated by the Schwinger functional
\begin{equation}
	W[J]
	= \ln Z[J].
\end{equation}
For instance, the connected two point correlator - a.k.a. the propagator - is given by
\begin{mleqn}\label{eqn:propagator}
	G
	&= \frac{\delta^2 W[J]}{\delta^2 J}
	= \frac{\delta}{\delta J} \biggl(\frac{1}{Z} \frac{\delta Z}{\delta J}\biggr)\\
	&= \frac{1}{Z} \frac{\delta^2 Z}{\delta^2 J} - \frac{1}{Z^2} \frac{\delta Z}{\delta J} \frac{\delta Z}{\delta J}\\[1ex]
	&= \bigl\langle\phi \, \phi\bigr\rangle - \varphi \, \varphi
	\equiv \bigl\langle\phi \, \phi\bigr\rangle_c.
\end{mleqn}
$G_{ab}(x,y)$ is an $N \times N$ matrix correlating the field $\phi_a$ at spacetime point $x$ with $\phi_b$ at $y$.

We now modify the Schwinger functional by introducing a renormalization scale-dependent cutoff term $\Delta S_k$ that vanishes in the infrared,
\begin{equation}\label{eqn:flowing schwinger}
	W_k[J]
	= \ln Z_k[J]
	= \ln \int \mathcal{D} \phi \, e^{-S[\phi] + J \cdot \phi - \Delta S_k[\phi]},
\end{equation}
where the renormalization scale $k$ has units of inverse length and can be intuitively understood to specify at which scale we probe a theory. Small $k$ correspond to large distances, large $k$ to small distances. $\Delta S_k[\phi]$ is a quadratic functional of the field $\phi$
\begin{equation}
	\Delta S_k[\phi]
	= \frac{1}{2} \, \phi \cdot R_k \cdot \phi
	= \frac{1}{2} \int_{x,y} \, \phi_a(x) \, R_{k,ab}(x,y) \, \phi_b(y).
\end{equation}
with $R_k$ acting as a momentum-dependent mass. We will see that $R_k$ serves both as an infrared \textit{and} ultraviolet regulator in our description. For an $O(N)$-symmetric scalar theory, it is diagonal both in momentum space and with respect to field indices,
\begin{equation}
	R_{k,ab}(x,y)
	= \delta_{ab} \, \delta(x-y) \, R_k(- \partial_x^2).
\end{equation}
Since the scale dependence of $W_k[J]$ stems solely from $\Delta S_k$, it's $k$-derivative (at fixed source $J$) is
\begin{mleqn}\label{eqn:schwinger flow}
	\partial_k W_k[J]\bigr|_J
	&\crefrel{eqn:flowing schwinger}{=} -\frac{1}{Z_k} \int \mathcal{D} \phi \, \bigl(\partial_k \Delta S_k[\phi]\bigr) e^{-S[\phi] + J \cdot \phi - \Delta S_k[\phi]}\\
	&= -\frac{1}{2} \bigl\langle\phi \cdot \partial_k R_k \cdot \phi\bigr\rangle
	\crefrel{eqn:propagator}{=} -\frac{1}{2} \bigl(\langle\phi \cdot \phi\rangle_c + \varphi \cdot \varphi\bigr) \cdot \partial_k R_k.
\end{mleqn}
For the connected part $\langle\phi \cdot \phi\rangle_c$ we can insert the functional derivative \labelcref{eqn:propagator},
\begin{equation}\label{eqn:flowing propagator}
	\langle\phi \cdot \phi\rangle_c
	\equiv W_k^{(2)}
	= \frac{\delta^2 W_k}{\delta^2 J}
	= \frac{\delta \varphi}{\delta J}
\end{equation}
to rewrite \labelcref{eqn:schwinger flow} as Polchinski's equation \cite{polchinski1984renormalization},
\begin{equation}\label{eqn:polchinski eqn}
	\partial_k W_k[J]\bigr|_J
	= -\frac{1}{2} \Tr\bigl[W_k^{(2)} \, \partial_k R_k\bigr] - \frac{1}{2} \, \varphi \cdot (\partial_k R_k) \cdot \varphi,
\end{equation}
where $\Tr$ integrates over position (or momentum\footnote{\label{ftn:momentum trace}In momentum space $\Tr = \sum_a \int \dif^d p/(2 \pi)^d$, as appropriate for the unit matrix $\mathbf{1} = (2 \pi)^d \delta_{ab} \, \delta(p - q)$.}) space and sums over the field indices $a$, $b$,
\begin{equation}
	\Tr\Bigl[(\partial_k R_k) \, W_k^{(2)}\Bigr]
	= \int_{x,y} \, W_{k,ab}^{(2)}(x,y) \, \partial_k R_{k,ab}(x,y).
\end{equation}

We can construct the flowing action $\Gamma_k[\varphi]$ from the modified Schwinger functional by subtracting from its Legendre transform
\begin{equation}\label{eqn:legendre trafo}
	\tilde\Gamma_k[\varphi]
	= \sup_J\bigr(J \cdot \varphi - W_k[J]\bigr)
	\qquad \mathrlap{\text{where }
	\varphi
	= \frac{\delta W_k}{\delta J},}
\end{equation}
the same cutoff term we added to $W_k$,
\begin{equation}\label{eqn:flowing action}
	\Gamma_k[\varphi]
	= \tilde\Gamma_k[\varphi] - \Delta S_k[\varphi].	
\end{equation}
$\Gamma_k[\varphi]$ is also known as the average action \cite{tetradis1994critical} because it provides an effective description of physics at distance scales $\gtrsim k^{-1}$ for fields $\varphi_a = \langle\phi_a\rangle$ averaged over a volume $k^{-d}$. Moreover, it enables a formulation of quantum theory even more economic than the Schwinger functional. In perturbation theory, it acts as the generating functional for only the one-particle irreducible correlation functions, while still encoding all properties of the underlying quantum fields.

Upon functional differentiation with respect to the average field $\varphi_a$, the Legendre transform $\tilde\Gamma_k$ yields the (scale-dependent) field equation
\begin{equation}
	\frac{\delta}{\delta \varphi} \, \tilde\Gamma_k
	= \tfjm_k.
\end{equation}
Comparing with \labelcref{eqn:flowing propagator}, we identify
\begin{equation}
	\tilde\Gamma_k^{(2)}
	= \frac{\delta^2 \tilde\Gamma_k}{\delta^2 \varphi}
	= \frac{\delta \tfjm_k}{\delta \varphi}
\end{equation}
as the inverse propagator,
\begin{equation}
	\Bigl(\tilde\Gamma_k^{(2)} \cdot W_k^{(2)}\Bigr)_{ab}(x,y)
	= \int_z \, \frac{\delta \tfjm_c(z)}{\delta \varphi_a(x)} \, \frac{\delta \varphi_b(y)}{\delta \tfjm_c(z)}
	= \frac{\delta \varphi_b(y)}{\delta \varphi_a(x)}
	= \delta_{ab} \, \delta(x-y).
\end{equation}
Thus
\begin{equation}\label{eqn:w gamma relation}
	W_k^{(2)}
	= \bigl(\tilde\Gamma_k^{(2)}\bigr)^{-1}
	= \bigl(\Gamma_k^{(2)} + R_k\bigr)^{-1}.
\end{equation}
The $k$-derivative of $\tilde\Gamma_k$ (at fixed average field) reads
\begin{equation}\label{eqn:legendre flow}
	\partial_k \tilde\Gamma_k\bigr|_\varphi
	\crefrel{eqn:legendre trafo}{=} \biggl(\varphi - \frac{\delta W_k}{\delta J}\biggr) \, \partial_k J - \partial_k W_k\bigr|_J
	= -\partial_k W_k\bigr|_J.
\end{equation}
At fixed $\varphi$, $J$ becomes scale-dependent, accounting for the second term in \labelcref{eqn:legendre flow}. The third is due to the scale dependence of $R_k$ in $W_k$ while $J$ is held fixed. Inserting \labelcref{eqn:legendre flow} into the $k$-derivative of \labelcref{eqn:flowing action} gives
\begin{equation}
	\partial_k \Gamma_k[\varphi]
	= -\partial_k W_k\bigr|_J - \frac{1}{2} \, \varphi \cdot (\partial_k R_k) \cdot \varphi.
\end{equation}
Using \labelcref{eqn:polchinski eqn,eqn:w gamma relation}, we arrive at Wetterich's equation
\begin{equation}\label{eqn:wetterich eqn}
	\partial_k \Gamma_k[\varphi]
	\crefrel{eqn:polchinski eqn}{=} \frac{1}{2} \Tr\bigl[W_k^{(2)} \, \partial_k R_k\bigr]
	\crefrel{eqn:w gamma relation}{=} \frac{1}{2} \Tr\Bigl[\bigl(\Gamma_k^{(2)} + R_k\bigr)^{-1} \partial_k R_k\Bigr].
\end{equation}
\labelcref{eqn:wetterich eqn} is a non-linear functional integro-differential equation of one-loop structure that determines the scale-dependence of the flowing action $\Gamma_k$ in terms of fluctuations of the fully-dressed regularized propagator $\smash{[\Gamma_k^{(2)} + R_k]^{-1}}$. \labelcref{eqn:wetterich eqn} admits a simple diagrammatic representation as a one-loop equation,
\begin{mleqn}\label{eqn:graphical wetterich eqn}
	\partial_k \Gamma_k
	= \frac{1}{2} \sum_{i,j=1}^N \int_{\mathrlap{p_1,p_2}} \;
	\begin{aligned}
		\begin{tikzpicture}[pin edge={shorten <=5*\lrad}]
			\def\lrad{1}
			\def\mrad{0.175*\lrad}
			\def\srad{0.15*\lrad}
		    \draw[loop/.list={{0.25}{1},{0.75}{2}}] (0,0) circle (\lrad);
		    \draw[cross] (-\lrad,0) circle (\srad) node[left=2pt] {$\partial_k R_{k,ij}(p_1,p_2)$};
		    \draw[dressed] (\lrad,0) circle (\mrad) node[right=2pt] {$\bigl[\Gamma_k^{(2)} + R_k\bigr]_{ji}^{-1}(p_2,p_1)$};
		\end{tikzpicture}
	\end{aligned},
\end{mleqn}
(Since $\partial_k R_{k,ab}(p,q) = \partial_k R_k(p) \, (2 \pi)^d \, \delta_{ab} \, \delta(p - q)$, the trace in \labelcref{eqn:wetterich eqn} effectively sums over just one index $i$ and integrates over one loop momentum $p$, as stated in \cref{ftn:momentum trace}.) This one-loop structure is important when it comes to practical calculations: only one integral has to be computed. In a rotationally invariant setting, it is even one-dimensional. Compared to perturbation theory where we have to sum a potentially non-convergent series of diagrams in which each $n$-loop diagram requires us to compute $n$ integrals, this amounts to a considerable reduction in complexity \cite{delamotte2012introduction}.

It is worth taking a moment to appreciate the significance of \labelcref{eqn:wetterich eqn}. Had we simply applied perturbation theory to the microscopic action $S$, we would have obtained a structurally very similar equation. Indeed, up to one loop, the perturbative expansion of $\Gamma_k$ reads
\begin{equation}\label{eqn:one-loop gamma}
	\Gamma_k[\varphi]\bigr|_\text{1-loop}
	= S[\varphi] + \frac{1}{2} \Tr\ln\bigl(S^{(2)}[\varphi] + R_k\bigr),
\end{equation}
which upon differentiation with respect to $k$ yields
\begin{equation}\label{eqn:one-loop flow eqn}
	\partial_k \Gamma_k[\varphi]\bigr|_\text{1-loop}
	= \frac{1}{2} \Tr\Bigl[\bigl(S^{(2)}[\varphi] + R_k\bigr)^{-1} \partial_k R_k\Bigr].
\end{equation}
Despite looking almost identical to \labelcref{eqn:wetterich eqn}, replacing $S^{(2)}[\varphi]$ with the fully dressed 2-point function $\Gamma_k^{(2)}$ turns the perturbative one-loop expression \labelcref{eqn:one-loop flow eqn} into an exact identity that incorporates effects of arbitrarily high loop order as well as genuinely non-perturbative effects \cite{wetterich2001effective}! Further noteworthy features of \labelcref{eqn:wetterich eqn} include \cite{wetterich2001effective,berges2002non}.
\begin{enumerate}
	\item\label{itm:hierarchy} Exact flow equations for arbitrarily high $n$–point functions follow from \labelcref{eqn:wetterich eqn} by functional differentiation. For instance, the 2-point function is given by
	\begin{mleqn}\label{eqn:hierarchy}
		\partial_k \Gamma_k^{(2)}
		= \partial_k \, \frac{\delta^2 \Gamma_k}{\delta^2 \phi}
		&= -\frac{1}{2} \Tr\biggl[\partial_k R_k \, \frac{\delta}{\delta \phi} \Bigl(\bigl[\Gamma_k^{(2)} + R_k\bigr]^{-1} \, \Gamma_k^{(3)} \, \bigl[\Gamma_k^{(2)} + R_k\bigr]^{-1}\Bigr)\biggr]\\
		&= \frac{1}{2} \Tr\biggl[\partial_k R_k \, \bigl[\Gamma_k^{(2)} + R_k\bigr]^{-1} \, \Gamma_k^{(3)} \, \bigl[\Gamma_k^{(2)} + R_k\bigr]^{-1} \, \Gamma_k^{(3)} \, \bigl[\Gamma_k^{(2)} + R_k\bigr]^{-1}\\
		&\hphantom{{}=\frac{1}{2} \Tr\biggl[}+ \bigl[\Gamma_k^{(2)} + R_k\bigr]^{-1} \, \Gamma_k^{(3)} \, \partial_k R_k \, \bigl[\Gamma_k^{(2)} + R_k\bigr]^{-1} \, \Gamma_k^{(3)} \, \bigl[\Gamma_k^{(2)} + R_k\bigr]^{-1}\\
		&\hphantom{{}=\frac{1}{2} \Tr\biggl[}-\partial_k R_k \, \bigl[\Gamma_k^{(2)} + R_k\bigr]^{-1} \, \Gamma_k^{(4)} \, \bigl[\Gamma_k^{(2)} + R_k\bigr]^{-1}\biggr].
	\end{mleqn}
	Represented diagrammatically \labelcref{eqn:hierarchy} reads
	\begin{equation}\label{eqn:graphical hierarchy}
		\partial_k \Gamma_k^{(2)}
		= \frac{1}{2} \Tr\Biggl(
		\begin{aligned}
			\begin{tikzpicture}
				\def\unit{0.7}
			    \draw[thick] (0,0) circle (\unit);
			    \draw (-2*\unit,0) -- (-\unit,0) (\unit,0) -- (2*\unit,0);
			    \draw[fill=white,cross] (0,\unit) circle (0.15*\unit) node[below] {$\partial_k R_k$};
			    \draw[fill=white,postaction={pattern=north east lines}] (\unit,0) circle (0.2*\unit) node[above right] {$\Gamma_k^{(3)}$} (-\unit,0) circle (0.2*\unit) node[above left] {$\Gamma_k^{(3)}$};
			\end{tikzpicture}
		\end{aligned}
		+ \begin{aligned}
			\begin{tikzpicture}
				\def\unit{0.7}
			    \draw[thick] (0,0) circle (\unit);
			    \draw (-2*\unit,0) -- (-\unit,0) (\unit,0) -- (2*\unit,0);
			    \draw[fill=white,cross] (0,-\unit) circle (0.15*\unit) node[above] {$\partial_k R_k$};
			    \draw[fill=white,postaction={pattern=north east lines}] (\unit,0) circle (0.2*\unit) node[above right] {$\Gamma_k^{(3)}$} (-\unit,0) circle (0.2*\unit) node[above left] {$\Gamma_k^{(3)}$};
			\end{tikzpicture}
		\end{aligned}
		- \begin{aligned}
			\begin{tikzpicture}
				\def\unit{0.7}
			    \draw[thick] (0,0) circle (\unit);
			    \draw (-2*\unit,-\unit) -- (2*\unit,-\unit);
			    \draw[fill=white,cross] (0,\unit) circle (0.15*\unit) node[below] {$\partial_k R_k$};
			    \draw[fill=white,postaction={pattern=north east lines}] (0,-\unit) circle (0.2*\unit) node[above] {$\Gamma_k^{(4)}$};
			\end{tikzpicture}
		\end{aligned}\Biggr).
	\end{equation}
	This alludes to a general property of flow equations for $n$-point functions: they form a hierarchy; the flow of $\Gamma_k^{(n)}$ depends on $\Gamma_k^{(n+1)}$ and $\Gamma_k^{(n+2)}$.
	
	\item To obtain a scaling form of the evolution equation, we may replace $\partial_k$ on both sides of \labelcref{eqn:wetterich eqn} by a partial derivative with respect to the logarithmic scale $t = \ln(k/\Lambda)$ (also referred to as RG time),
	\begin{equation}
		\partial_t
		= \frac{\partial}{\partial \ln(k/\Lambda)}
		= \frac{k}{\Lambda} \frac{\partial}{\partial (k/\Lambda)}
		= k \, \partial_k.
	\end{equation}
	
	\item The presence of the cutoff function $R_k$ renders the momentum integration in $\Tr$ both infrared and ultraviolet finite. In particular, for $p^2 \ll k^2$ $R_k$ serves as an additional mass–like term $R_k \sim k_2$ that prevents the propagator $\smash{\bigl[\Gamma_k^{(2)} + R_k\bigr]^{-1}}$ from becoming singular at $p = 0$. This makes the formalism suitable for dealing with theories plagued by infrared divergencies when treated perturbatively. These include scalar theories in $d < 4$ or at non-zero temperature near a second order phase transition as well as non-abelian gauge theories \cite{tetradis1994critical}. For instance, \labelcref{eqn:wetterich eqn} can be applied to systems with spontaneously broken $O(N)$ symmetry despite the appearance of massless Goldstone bosons if $N > 1$. Their standard loop expansion is highly infrared divergent, making these massless excitations notoriously difficult to treat with other methods \cite{adams1995solving}.
	
	\item\label{item:decoupling} Since $\partial_k R_k(p)$ appears in the numerator of \labelcref{eqn:wetterich eqn}, its fast decay for $p^2 \gg k^2$ results in UV finiteness of the momentum integration that is part of the trace $\Tr$. Together with the IR regulating properties of $R_k(p)$ in the denominator, this means that only momenta $p^2 \lesssim k^2$ of the order of or smaller than the renormalization scale contribute substantially to the flow at scale $k$. The divergent loop diagrams of perturbation theory are thus avoided.
	
	An important consequence is the decoupling of massive modes $M$ at low energies. Once $k^2 \ll M^2$, fluctuations of massive modes are strongly suppressed by $\partial_k R_k(p) \approx 0$. They were integrated out during earlier stages of the flow where $M \approx k < \Lambda$, resulting in renormalized couplings for the low-energy theory. If $k$ is lowered further, there will be essentially no change in $\Gamma_k$ due to fluctuations of these modes \cite{floerchinger2012analytic}. In this way, the flow equations automatically lead to the emergence of effective theories for the low-energy degrees of freedom \cite{berges2002non,wetterich2001effective}!
	
	Unfortunately, this also means that given a low-energy theory, we cannot know whether the underlying fundamental theory (at scales larger than $M$) involves massive excitations or not. Below the scale $M$ there would remain no signal of such a mode.
		
	
	\item The crucial requirement for practical application of \labelcref{eqn:wetterich eqn} to non-perturbative systems is the availability of sufficiently simple and yet physically relevant truncation schemes. Determining which terms in an expansion of $\Gamma_k$ can safely be discarded and which operators must be kept in order to capture important behavior requires sophisticated physical insight into a model.\footnote{Of course, if we make use of prior knowledge obtained with other methods during this process our formalism looses its claim to being a first-principles-only approach -- at least in practical applications.}
	
	In this context the close resemblance of \labelcref{eqn:wetterich eqn} to a perturbative expression turns out to be of great use. We can benefit from the fact that for many situations of interest the propagator $\smash{(\Gamma_k^{(2)})^{-1}}$ is approximately known, allowing us to devise a simple form for $\Gamma_k^{(2)}$ that depends on only a handful of scale-dependent parameters (typically masses, couplings, decay widths and wave function renormalizations) while still describing the relevant physics. (The success of the entire method ultimately depends on a clever guess for the exact propagator, which in turn can depend on the proper choice of degrees of freedom \cite{tetradis1994critical}.) We can then project the flow \labelcref{eqn:wetterich eqn} of $\Gamma_k$ onto these parameters and obtain a closed and finite set of ordinary coupled non-linear differential equations that is much easier to solve than the flow of $\Gamma_k$ itself.
	
	\item \labelcref{eqn:wetterich eqn} is equivalent to Wilson's exact RG equation \cite{wilson1971renormalization} which describes how the Wilsonian effective action $S_\Lambda^W$ changes with an ultraviolet cutoff $\Lambda$. As we saw in its derivation, Polchinski’s continuum version of Wilson's equation is even related to \labelcref{eqn:wetterich eqn} by a simple Legendre transform, a suitable field redefinition and the association $\Lambda = k$. Although the formal relation is simple, the practical calculation of $S_k^W$ from $\Gamma_k$ (and vice versa) can be quite involved.
\end{enumerate}

\subsection{Average Action}
\label{sec:average action}

The average action $\Gamma_k$ has a number of properties worth mentioning \cite{wetterich2001effective,berges2002non}.
\begin{enumerate}
	\item In perturbation theory, $\Gamma = \lim_{k \to 0} \Gamma_k$ acts as the generating functional of one-particle irreducible correlation functions. Once $\Gamma$ is known, a theory is basically ``solved''. Since it is the result of integrating out fluctuations on \textit{all} momentum scales $0 < k < \Lambda$, it contains effective couplings; physical masses, charges and wave function renormalizations can simply be read off. It also means the effective action is exact at tree level! Instead of having to manage an infinite (often times divergent) series of Feynman diagrams to calculate some physical observable (such as a scattering cross section) it suffices to evaluate tree-level Feynman diagrams \cite{floerchinger2012analytic}.

	\item If the microscopic action $S$ is invariant under some group $G$ and we construct $\Gamma_k$ using an IR cutoff that respects this symmetry, $\Gamma_k$ inherits $G$-invariance from $S$ for all $k$ (assuming the absence of quantum anomalies). In particular, it relays this symmetry to the effective action $\Gamma$ at $k = 0$. For example, this is true for translation and rotation invariance if $R_k$ depends only on the distance $(x-y)$ in position space or $p^2$ in momentum space.
	
	\item The most general form of $\Gamma_k$ is given by an infinite series of \textit{all} field combinations compatible with the given symmetries. Since each term comes with its own scale-dependent coupling (see \cref{sec:truncations}), $\Gamma_k$ in theory contains infinitely many running couplings, making a well-chosen truncation procedure essential.
	
	\item Physical quantities should be independent of the choice of cutoff $R_k$. Scheme independence of final results is a good check for approximations. However, $\Gamma_\Lambda$'s position in and $\Gamma_k$'s flow through theory space \textit{are} scheme-dependent.
	
	\item Despite their similarities, there is a conceptual difference to the Wilsonian effective action \cite{wilson1971renormalization}. $S_W^\Lambda$ describes a set of different actions (parameterized by $\Lambda$) for a single model. In contrast, $\Gamma_k$ acts as effective action for a set of models; for any scale $k$, $\Gamma_k$ is related to the generating functional of 1PI $n$-point functions for a model with a different action $S_k = S + \Delta S_k$. The Wilsonian effective action does not generate the 1PI Green functions.
\end{enumerate}

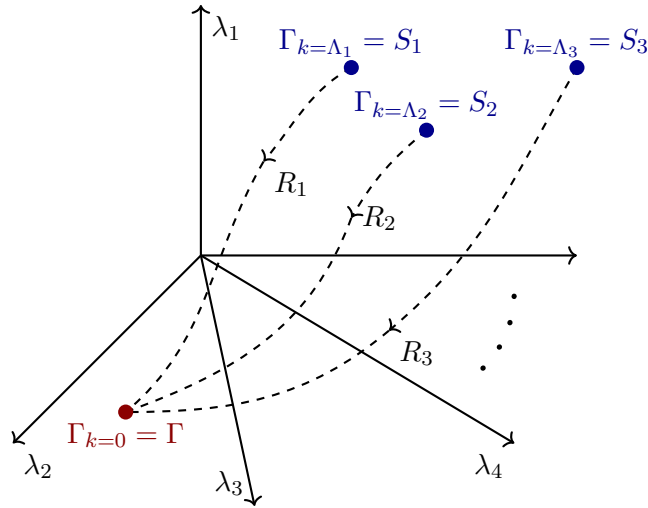
\begin{figure}[htb!]
	\centering
	\begin{tikzpicture}[thick]
		\def\unit{5}
	    
	    \coordinate (qea) at (-1/5*\unit,-5/12*\unit); 
	    \coordinate (ma1) at (2/5*\unit,1/2*\unit); 
	    \coordinate (ma2) at (3/5*\unit,1/3*\unit); 
	    \coordinate (ma3) at (\unit,1/2*\unit); 
	    \coordinate (r1) at (1/6*\unit,1/4*\unit); 
	    \coordinate (r2) at (2/5*\unit,1/10*\unit); 
	    \coordinate (r3) at (1/2*\unit,-1/5*\unit); 
	    
	    \draw[->] (0,0) -- (0,2/3*\unit) node[below right] (l1) {$\lambda_1$};
	    \draw[->] (0,0) -- (-1/2*\unit,-1/2*\unit) node[below right] (l2) {$\lambda_2$};
	    \draw[->] (0,0) -- (1/7*\unit,-2/3*\unit) node[above left] (l3) {$\lambda_3$};
	    \draw[->] (0,0) -- (5/6*\unit,-1/2*\unit) node[below left] (l4) {$\lambda_4$};
	    \draw[->] (0,0) -- (\unit,0);
	    \draw[line width=2,line cap=round,dash pattern=on 0pt off 5\pgflinewidth] (3/4*\unit,-3/10*\unit) edge[bend right=20] (5/6*\unit,-1/10*\unit);
	    
	    \draw[dashed] (ma1) edge[->,in=50,out=210] (r1) (r1) node[below right] {$R_1$} to[out=240,in=40] (qea);
	    \draw[dashed] (ma2) edge[->,in=60,out=220] (r2) (r2) node[right] {$R_2$} to[out=250,in=20] (qea);
	    \draw[dashed] (ma3) edge[->,in=40,out=240] (r3) (r3) node[below right] {$R_3$} to[out=220,in=0] (qea);
	    
	    \fill[DarkRed] (qea) circle (0.1) node[below] {$\Gamma_{k=0} = \Gamma$};
	    \fill[DarkBlue] (ma1) circle (0.1) node[above] {$\Gamma_{k=\Lambda_1} = S_1$};
	    \fill[DarkBlue] (ma2) circle (0.1) node[above] {$\Gamma_{k=\Lambda_2} = S_2$};
	    \fill[DarkBlue] (ma3) circle (0.1) node[above] {$\Gamma_{k=\Lambda_3} = S_3$};
	    
	\end{tikzpicture}
	\caption{Flow of $\Gamma_k$ through infinite-dimensional theory space for different regulators $R_i$. The bare actions $S_i$ obey the same symmetries and thus flow to the same quantum effective action $\Gamma$.}
	\label{fig:theory space}
\end{figure}

This completes the picture we have of the flowing action: in its transition to complexity, $\Gamma_k$ continuously interpolates from the microscopic action at small scales to the effective action at the macroscopic level. It moves through an infinite-dimensional theory space, its path determined by the initial condition $S$ and the choice of regulator $R_k$. Theory space is spanned by the set of all symmetry-compatible operators, e.g. $\phi^2$, $\phi^4$, $(\partial \phi)^2$, etc. in the case of $O(N)$-invariant scalar fields $\phi_a$. (In \cref{fig:theory space} the operators are represented by their couplings $\{\lambda_i\}_{i\in\mathbb{N}}$.) Of course, for a practical treatment to remain manageable requires the truncation of $\Gamma_k$ to but a handful of operators.

\subsection{Regulator}
\label{sec:regulator}

Despite being just a mathematical tool without physical meaning, $R_k$ is a central object in this formulation of quantum field theory. To bestow upon $\Gamma_k$ the property of interpolating between $S$ at $k = \Lambda$ and $\Gamma$ at $k = 0$, it must satisfy
\begin{equation}\label{eqn:regulator props}
	R_k(p)
	\to \begin{cases}
		k^2 & \text{for } p \to 0,\\
		0 & \text{for } p \to \infty,\\
		0 & \text{for } k \to 0,\\
		\infty & \text{for } k \to \Lambda,
	\end{cases}
\end{equation}
where $p$ denotes the internal loop momentum on the r.h.s. of the flow equation.
\begin{itemize}
	\item $R_k(p) \to k^2 > 0$ for $p \to 0$ prevents the propagator $\smash{\bigl[\Gamma_k^{(2)} + R_k\bigr]^{-1}}$ from becoming singular at $p = 0$ and thus regularizes the theory in the infrared.
	
	\item $R_k(p) \to 0$ for $p \to \infty$ ensures a fast decay of $\partial_k R_k$ at high loop-momenta, thus rendering the one-loop flow equation ultraviolet finite.
	
	\item $R_k(p) \xrightarrow{k \to 0} 0$ ensures that the flowing action $\Gamma_k$ approaches the effective action $\Gamma$ at macroscopic scales,
	\begin{equation}
		\lim_{k \to 0} \Gamma_k[\varphi]
		\crefrel{eqn:flowing action}{=} \lim_{k \to 0} \Bigl(\tilde\Gamma_k[\varphi] - \frac{1}{2} \, \varphi \cdot R_k \cdot \varphi\Bigr)
		\crefrel{eqn:legendre trafo}{=} \sup_J\bigr(J \cdot \varphi - W[J]\bigr)
		= \Gamma[\varphi].
	\end{equation}
	\item $R_k(p) \xrightarrow{k \to \Lambda} \infty$ ensures that $\Gamma_k$ flows towards the microscopic action $S$ for $k \to \Lambda$. (This is where the necessity to work with a modified Legendre transform becomes apparent.) We resort again to the functional integral formalism,
	\begin{mleqn}\label{eqn:gammak integral}
		e^{-\Gamma_k[\varphi]}
		&= \exp\Bigl(-\sup_J\bigl(J \cdot \varphi - W_k[J]\bigr) + \frac{1}{2} \, \varphi \cdot R_k \cdot \varphi\Bigr)\\
		&\crefrel{eqn:flowing schwinger}{=} \int \mathcal{D} \phi \, \exp\Bigl(-S[\phi] + J \cdot \phi - \frac{1}{2} \, \phi \cdot R_k \cdot \phi - J \cdot \varphi + \frac{1}{2} \, \varphi \cdot R_k \cdot \varphi\Bigr)\\
		&= \int \mathcal{D} \phi \, \exp\Bigl(-S[\varphi + \phi] - \frac{1}{2} \, \phi \cdot R_k \cdot \phi + \frac{\delta \Gamma_k}{\delta \varphi} \cdot \phi\Bigr),
	\end{mleqn}
	where in the last step we shifted the field $\phi \to \varphi + \phi$, used that $R_{k,ab} \propto \delta_{ab}$ is symmetric so that $\varphi \cdot R_k \cdot \phi = \phi \cdot R_k \cdot \varphi$, and inserted
	\begin{equation}
		\frac{\delta \Gamma_k}{\delta \varphi} \cdot \phi
		\crefrel{eqn:flowing action}{=} J \cdot \phi - \varphi \cdot R_k \cdot \phi.
	\end{equation}
	In the microscopic limit $\lim_{k \to \Lambda} R_k$ diverges and the factor $e^{-\frac{1}{2} \phi \cdot R_k \cdot \phi}$ approaches the limit representation of the functional delta distribution\footnote{The usual normalization includes a prefactor $\delta[\phi] = \lim_{k \to \Lambda} \sqrt{R_k/(2 \pi)} \, e^{-\frac{1}{2} \phi \cdot R_k \cdot \phi}$ which we swept under the rug since it results only in a (divergent) additive constant $-\frac{1}{2} \, \ln[R_\Lambda/(2\pi)]$ to $S$ that doesn't affect the dynamics.} \cite{wetterich2001effective},
	\begin{equation}
		\delta[\phi]
		\sim \lim_{k \to \Lambda} e^{-\frac{1}{2} \, \phi \cdot R_k \cdot \phi},
	\end{equation}
	allowing us to evaluate the path integral in \labelcref{eqn:gammak integral},
	\begin{equation}
		\lim_{k \to \Lambda} e^{-\Gamma_k[\varphi]}
		= \int \mathcal{D} \phi \, \delta[\phi] \, e^{-S[\varphi + \phi] + \frac{\delta \Gamma_k}{\delta \varphi} \cdot \phi}
		= e^{-S[\varphi]}.
	\end{equation}
	Thus
	\begin{equation}\label{eqn:ini cond}
		\lim_{k \to \Lambda} \Gamma_k[\varphi]
		= S[\varphi].
	\end{equation}
	\labelcref{eqn:ini cond} is a useful result since it serves as an initial condition for the flow equation \labelcref{eqn:wetterich eqn}. However, the property $\Gamma_\Lambda = S$ is not essential since we may as well use $\Gamma_\Lambda$ to parametrize the short distance behavior \cite{wetterich2001effective}. When taking $\Gamma_k$ from $\Lambda$ to larger distances, universality ensures that (up to a few relevant renormalized couplings) the precise form of $\Gamma_\Lambda$ is irrelevant in any case. Using $\Gamma_\Lambda$ instead of $S$ can even become necessary in cases where no physical cutoff is present or where a UV cutoff would be in conflict with symmetries as in the case of gauge theories.
\end{itemize}

\subsection{Truncations}
\label{sec:truncations}

Solving a functional differential equation like the Wetterich equation \labelcref{eqn:wetterich eqn} exactly is all but impossible. Fortunately, we can descend from the functional formulation into an infinite system of coupled ordinary differential equations by projecting \labelcref{eqn:wetterich eqn} to the infinite number of couplings $\{\lambda_i|i \in \mathbb{N}\}$ appearing in the most general form of $\Gamma_k$ \cite{tetradis1994critical,wetterich2001effective,berges2002non}. Of course, an exact solution of this infinite system is still impossible \cite{tetradis1994critical}. To make an explicit treatment feasible, we have to heavily restrict the space of action functionals to a finite number of dimensions, meaning we can allow only a handful of relevant couplings in $\Gamma_k$.

This is where (sometimes hard to control) approximations have to be made. Assuming $\Gamma_k[\varphi]$ is invariant under global $O(N)$ transformations, we have several expansion schemes at our disposal.
\begin{description}
	\item[Derivative expansion] The most common way to arrive at a sufficiently simple form of $\Gamma_k[\varphi]$ is to write it as a sum of a few low-order $O(N)$ invariants with order determined by the number of field derivatives $\partial^\mu \varphi_a$. The simplest $O(N)$-invariant $\rho = \frac{1}{2} \phi_a \phi_a$ contains no derivatives and appears at order zero in this classification. By allowing $U_k(\rho)$ to be an arbitrary polynomial of $\rho$, the effective potential covers this order completely.
	
	Of course, $\Gamma_k[\varphi]$ is also constrained by spacetime symmetries. In particular, in a relativistic setting it has to be Lorentz invariant. But $\partial_\mu$ is a Lorentz vector and so for $\Gamma_k$ to include Lorentz invariant dynamics, $\partial_\mu$ needs to be contracted. Unlike gauge fields $A^\mu$, scalar fields don't carry spacetime indices. Neither do they furnish spinor representations of Clifford algebras with spacetime-indexed generators $\gamma^\mu$, like fermions do. Therefore, the only way to include a derivative in $\Gamma_k[\varphi]$ in a Lorentz-invariant fashion is by contracting it with another $\partial_\mu$. Thus, first-order $O(N)$-invariant of scalar fields must already contain two spacetime derivatives. There are two $O(N)$-invariants we can construct in this way,\footnote{By partial integration under the action with vanishing boundary terms, \labelcref{eqn:first order} is equivalent to $\varphi_a \, \partial^2 \varphi_a$ and $\rho \, \partial^2 \rho$.}
	\begin{equation}\label{eqn:first order}
		\partial_\mu \varphi_a \, \partial^\mu \varphi_a
		\quad\text{and}\quad
		\partial_\mu \rho \, \partial^\mu \rho.
	\end{equation}
	By the product rule, acting with derivatives on powers of the fields higher than $\varphi_a$ and $\rho$ just gives sums of these two building blocks. We will, however, include wave function renormalizations $Z_k(\rho)$ and $Y_k(\rho)$ as prefactors to \labelcref{eqn:first order} and these may contain arbitrary order-zero invariants.
	
	Since we mean to truncate our expansion early, we'll stop here. Putting everything together, we get a flowing action of the form \cite{wetterich2001effective}
	\begin{equation}\label{eqn:o(n) flowing action}
		\Gamma_k[\varphi]
		= \int_x \Bigl[U_k(\rho) + \frac{1}{2} \, Z_k \, \partial_\mu \varphi_a \, \partial^\mu \varphi_a + \frac{1}{4} \, Y_k \, \partial_\mu \rho \, \partial^\mu \rho + \mathcal{O}(\partial^4)\Bigr].
	\end{equation}
	with $Z_\Lambda = Y_\Lambda = 1$. \labelcref{eqn:o(n) flowing action} is known as the \textit{derivative expansion}. We can slightly enhance \labelcref{eqn:o(n) flowing action} by generalizing $Z_k$ and $Y_k$ to functions of momentum \cite{floerchinger2012analytic,berges2002non},
	\begin{equation}\label{eqn:o(n) flowing action generalized}
		\Gamma_k[\varphi]
		= \int_x \biggl[U_k(\rho) + \frac{1}{2} \, \partial_\mu \varphi_a \, Z_k(-\partial^2) \, \partial^\mu \varphi_a + \frac{1}{4} \, \partial_\mu \rho \, Y_k(-\partial^2) \, \partial^\mu \rho + \mathcal{O}(\partial^4)\biggr],
	\end{equation}
	\labelcref{eqn:o(n) flowing action generalized} has the advantage that it allows to resolve the propagator's \textit{full} momentum dependence. The effective potential $U_k(\rho)$ can further be expanded into a Taylor series around the location of its minimum,
	\begin{equation}
		U_k(\rho)
		= \frac{\lambda}{2} \, \bigl[\rho - \rho_0(k)\bigr]^2 + \frac{\mu}{6} \bigl[\rho - \rho_0(k)\bigr]^3 + \dots\\[1ex]
	\end{equation}
	$\lambda(k) = \partial_\rho^2 \, U_k(\rho_0)$ and $\mu(k) = \partial_\rho^3 \, U_k(\rho_0)$ are scale-dependent couplings that each span a dimension of theory space. The lowest order in the derivative expansion, known as the local potential approximation, is obtained by setting $Z_k = 1$, $Y_k = 0$ so that $\Gamma_k$ only includes the effective potential and a standard kinetic term,
	\begin{equation}\label{eqn:lpa}
		\Gamma_k[\varphi]
		= \int_x \Bigl[U_k(\rho) + \frac{1}{2} \, \partial_\mu \varphi_a \, \partial^\mu \varphi_a\Bigr].
	\end{equation}
	
	So far the derivative expansion has proven most successful in practical applications of the flow equation \labelcref{eqn:wetterich eqn}. The reason it works so well at capturing relevant physics while still reducing $\Gamma_k$ to but a handful of scale-dependent parameters is quite intuitive. Transformed into Fourier space, \labelcref{eqn:o(n) flowing action generalized} becomes an expansion in powers of loop-momenta $p$ around the scale $k$ which converges rapidly thanks to the separation of momentum scales in the flow equation provided by the fast decay of $\partial_k R_k$ for large $p^2 \gg k^2$ (see \cref{item:decoupling} on \cpageref{item:decoupling}). Only momenta $p^2 \leq k^2$ contribute substantially to the flow and these are captured well by low powers of $p$.
	
	This is why in practice we usually neglect terms higher than quadratic in the momenta. In particular, we tend to work with momentum-independent $Z_k$, $Y_k$ as in \labelcref{eqn:o(n) flowing action} since the main effect of this extra momentum-dependence is to provide an infrared cutoff scale of order $p^2$ which is already provided by $R_k(p)$.
	
	\item[Vertex expansion] We can also expand $\Gamma_k$ in terms of $n$-point functions around some constant field $\varphi_c$ \cite{wetterich2001effective,berges2002non}. This approach known as \textit{vertex expansion} yields
	\begin{equation}\label{eqn:vertex expansion}
		\Gamma_k[\varphi]
		= \sum_{n=0}^\infty \frac{1}{n!} \biggl(\prod_{j=0}^n \int_{x_j} \bigl[\varphi(x_j) - \varphi_c\bigr]\biggr) \Gamma_k^{(n)}(x_1,\dots,x_n).
	\end{equation}
	As mentioned in \cref{itm:hierarchy} on \cpageref{itm:hierarchy}, flow equations for $\Gamma_k^{(n)}$ follow from functional differentiation of \labelcref{eqn:wetterich eqn}. \labelcref{eqn:vertex expansion} makes particularly clear why.
	
	\item[Canonical dimension expansion] A third option very similar to the derivative expansion is to expand in terms of $O(N)$-invariants around some constant background field $\rho_c$ and classify terms not by the number of derivatives but based on their \textit{canonical dimension},
	\begin{mleqn}\label{eqn:canonical dim expansion}
		\Gamma_k[\varphi]
		= \int_x \biggl\{&U_k(\rho_c) + U_k^\prime(\rho_c) (\rho - \rho_c) + \frac{1}{2} U_k^{\prime\prime}(\rho_c) (\rho - \rho_c)^2 + \dots\\
		&-\frac{1}{2} \Bigl(Z_k(\rho_c) + Z_k^\prime(\rho_c) (\rho - \rho_c) + \frac{1}{2} Z_k^{\prime\prime}(\rho_c) (\rho - \rho_c)^2 + \dots\Bigr) \partial_\mu \varphi_a \, \partial^\mu \varphi_a\\
		&+\frac{1}{2} \Bigl(\dot{Z}_k(\rho_c) + \dot{Z}_k^\prime(\rho_c) (\rho - \rho_c) + \dots\Bigr) \varphi_a (\partial_\mu \partial^\mu)^2 \varphi_a\\
		&-\frac{1}{4} Y_k(\rho_c) \rho \partial_\mu \, \partial^\mu \rho\biggr\}.
	\end{mleqn}
	Here, primes denote derivatives with respect to $\rho$ and dots with respect to the scale $t = \ln(k/\Lambda)$. The constant field $\rho_c$ is usually chosen as the minimum of effective potential $\rho_0(k)$. If $\rho_0(k) > 0$ the ground state is not $O(N)$-invariant as can be seen from \cref{fig:mexican hat}, signaling a theory with spontaneous symmetry breaking. Since $U_k^\prime(\rho_0) = 0$ $\rho_0(k)$ then replaces the coupling $U_k^\prime(\rho_0)$ \cite{wetterich2001effective}. In the absence of external sources the value of $\rho_0(k \to 0)$ determines the order parameter $\varphi_0 = \sqrt{2 \rho_0}$.
\end{description}
\begin{figure}[hbt!]
	\centering
	\scalebox{1.5}{
	\begin{tikzpicture}[font=\footnotesize]
	    \begin{axis}[
	        axis lines=center,
	        axis equal,
	        domain=0:360,
	        y domain=0:1.25,
	        y axis line style=stealth-,
	        y label style={at={(0.35,0.18)}},
	        xmax=1.6,zmax=1.3,
	        xlabel = $\varphi_{_1}$,
	        ylabel=$\varphi_{_2}$,
	        zlabel=$U_k(\rho)$,
	        ticks=none
	    ]
	        \addplot3 [surf,shader=flat,draw=black,fill=white,z buffer=sort] ({sin(x)*y}, {cos(x)*y}, {(y^2-1)^2});
	        \coordinate (center) at (axis cs:0,0,1);
	        \coordinate (minimum) at (axis cs:{cos(30)},{sin(30)},0);
	    \end{axis}
	    
	    \fill[DarkBlue] (center) circle (0.1);
	    \fill[DarkRed] (minimum) circle (0.1);
	    \draw (center) edge[shorten <=5,shorten >=5,out=-10,in=150,double,draw=gray,double distance=0.5,-{>[length=2,line width=0.5]}] (minimum);
	\end{tikzpicture}}
	\caption{Shape of the effective potential $U_k(\rho)$ in the presence of spontaneous symmetry breaking. Even if a system starts out in the naive $O(N)$-invariant vacuum (blue dot), quantum fluctuations will quickly push it into the real vacuum (red dot) where $O(N)$ is broken down to $O(N-1)$.}
	\label{fig:mexican hat}
\end{figure}
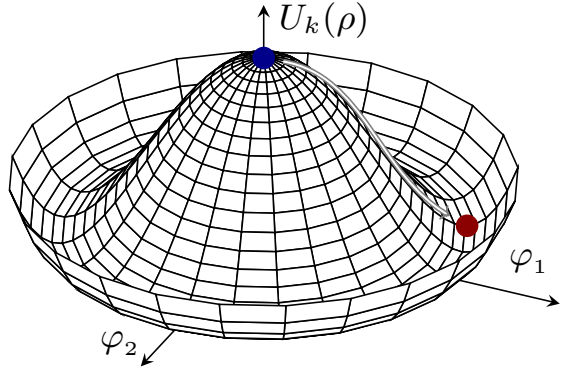

In all three of the above expansion schemes, the basic strategy is to solve the Wetterich equation in a restricted functional space, not as a series expansion in some small parameter. This is why the formalism can be applied to non-perturbative systems \cite{delamotte2012introduction}.

\subsection{Potential Flow}
\label{sec:potential flow}

We consider a set of $O(N)$-invariant scalar fields $\phi_a(x)$ with average action \labelcref{eqn:o(n) flowing action generalized}. When it comes to the ground state, its preserved or spontaneously broken symmetries and the mass spectrum of excitations, the most important quantity is the effective potential $U_k(\rho)$ \cite{wetterich2001effective}. Its scale dependence is two-fold:
\begin{itemize}
	\item one contribution stems from the scale dependence of the fields $\rho = \frac{1}{2} \varphi_a \varphi_a$, which we renormalize according to
	\begin{equation}
		\varphi_r
		= \csqrt{Z_k} \, \varphi,
		\qquad
		\rho_r
		= Z_k \, \rho,
	\end{equation}
	\item the other from the scale dependence of $U_k(\rho)$ itself (via the couplings $\lambda_{k,i}$).
\end{itemize}
The total scale derivative of $U_k$ therefore receives contributions from two terms,
\begin{equation}\label{eqn:eff pot der}
	\frac{\dif U_k}{\dif t}
	= \frac{\partial U_k}{\partial \rho} \frac{\partial \rho}{\partial t}\biggr|_{\mathrlap{\rho_r}} + \partial_t U_k\bigr|_\rho
	= \eta \, \rho \, U_k^\prime + \partial_t U_k\bigr|_\rho,
\end{equation}
where $\bigr|_{\rho_r}$ indicates that $\rho_r$ is held fixed such that $\partial \rho/\partial t\bigr|_{\rho_r} = -Z_k^{-2} \, \rho_r \, \partial_t Z_k = \eta \, \rho$. $\eta = -Z_k^{-1} \, \partial_t Z_k$ denotes the anomalous dimension of the propagator $\smash{\bigl(\Gamma_k^{(2)}\bigr)^{-1}}$ and $\partial_t U_k\bigr|_\rho$ contains the effective potential's inherent scale dependence (via the couplings $\lambda_i$) at fixed $\rho$. Wetterich's equation \labelcref{eqn:wetterich eqn} allows us to derive a flow equation for this contribution.

To that end, we relate $U_k(\rho)$ to $\Gamma_k[\varphi]$ by evaluating the latter for a constant background field $\varphi(x) = \varphi_c \enskip \forall \, x$. ($\varphi_c$ may be any constant field. It is not necessarily related to the minimum of the effective potential $\varphi_0 = \csqrt{2 \rho_0}$. However, in actual calculations, we will often choose $\varphi_c = \varphi_0$.) The derivative terms in $\Gamma_k$ all drop out, leaving us with
\begin{equation}\label{eqn:uniform fields}
	\Gamma_k\bigr|_{\varphi_c}
	= \int_x \, U_k(\rho_c)
	= V_d \, U_k(\rho_c).
\end{equation}
with $\rho_c = \frac{1}{2} \, \varphi_c^2$. Acting on \labelcref{eqn:uniform fields} with a scale derivative yields
\begin{equation}\label{eqn:potential flow}
	\partial_t U_k\bigr|_{\rho_c}
	= V_d^{-1} \, \partial_t \Gamma_k\bigr|_{\varphi_c}
	\crefrel{eqn:wetterich eqn}{=} \frac{1}{2 V_d} \Tr\Biggl[\frac{\partial_t R_k}{\Gamma_k^{(2)}\bigr|_{\varphi_c} + R_k}\Biggr].
\end{equation}
To obtain $\smash{\Gamma_k^{(2)}\bigr|_{\varphi_c}}$ in momentum space, we first expand $\Gamma_k$ in small fluctuations $\chi_a(x)$ around $\varphi_c$. Using $O(N)$ symmetry, we can rotate the fields to have $\varphi_c$ point in, say, $\varphi_1$-direction,
\begin{equation}
	\varphi_a(x)
	= \varphi_c \, \delta_{a1} + \chi_a(x)
	\qquad
	\mathrlap{\text{with } \chi_a(x) \ll \varphi_c \enskip \forall \, a, x.}
\end{equation}
Then $\rho = \frac{1}{2} \varphi_c^2 + \varphi_c \, \chi_1 + \frac{1}{2} \chi_a \chi_a$. Insertion into \labelcref{eqn:o(n) flowing action generalized} yields
\begin{align}
	\Gamma_k
	= \! \int_x &\biggl[U_k(\rho)\Bigr|_{\varphi_c} + \frac{\partial U_k(\rho)}{\partial \chi_a}\biggr|_{\varphi_c} \chi_a + \frac{1}{2} \frac{\partial^2 U_k(\rho)}{\partial \chi_a \partial \chi_b}\biggr|_{\varphi_c} \chi_a \chi_b + \dots\notag\\
	&+ \frac{1}{2} \partial_\mu \chi_a \Bigl[Z_k(\rho_c,-\partial^2) + (\rho - \rho_c) \, Z_k^\prime(\rho_c,-\partial^2) + \dots\Bigr] \partial^\mu \chi_a\\
	&+ \frac{1}{4} \partial_\mu (\varphi_c \chi_1 + \tfrac{1}{2} \chi_a \chi_a) \Bigl[Y_k(\rho_c,\! -\partial^2) + (\rho - \rho_c) \, Y_k^\prime(\rho_c,\! -\partial^2) + \dots\Bigr] \partial^\mu (\varphi_c \chi_1 + \tfrac{1}{2} \chi_a \chi_a) \biggr].\notag
\end{align}
We are only interested in the part $\Gamma_{k,2}$ that is quadratic in the small fluctuations $\chi_a$. Terms less than quadratic drop out when we perform the functional derivative $\smash{\Gamma_k^{(2)} = \delta^2 \Gamma_k/\delta^2 \chi}$ and terms higher than quadratic vanish when we evaluate $\Gamma_k^{(2)}\bigr|_{\varphi_c}$ for constant background $\chi_a(x) = 0 \enskip \forall \, a,x$.
\begin{equation}\label{eqn:quadratic part}
	\Gamma_{k,2}
	= \int_x \biggl[\frac{1}{2} \, m_{ab}^2 \, \chi_a \chi_b + \frac{1}{2} \, \partial_\mu \chi_a \, Z_k(\rho_c,-\partial^2) \, \partial^\mu \chi_a + \frac{1}{4} \, \varphi_c^2 \, \partial_\mu \chi_1 \, Y_k(\rho_c,-\partial^2) \, \partial^\mu \chi_1\biggr],
\end{equation}
where we defined the mass matrix
\begin{mleqn}
	m_{ab}^2
	&= \frac{\partial^2 U_k(\rho)}{\partial \chi_a \partial \chi_b}\biggr|_{\varphi_c}
	= \frac{\partial}{\partial \chi_a} \biggl(\frac{\partial U_k}{\partial \rho} \frac{\partial \rho}{\partial \chi_b}\biggr)\biggr|_{\varphi_c}
	= \frac{\partial}{\partial \chi_a} \Bigl[U_k^\prime(\rho) (\varphi_c \delta_{b1} + \chi_b)\Bigr]\Bigr|_{\varphi_c}\\
	&= U_k^{\prime\prime}(\rho) (\varphi_c \delta_{a1} + \chi_a) (\varphi_c \delta_{b1} + \chi_b) + U_k^\prime(\rho) \delta_{ab}\Bigr|_{\varphi_c}\\[1ex]
	&= 2 \rho_c \, U_k^{\prime\prime}(\rho_c) \, \delta_{a1} \delta_{b1} + U_k^\prime(\rho_c) \, \delta_{ab}.
\end{mleqn}
Expanding $\chi(x)$ into Fourier modes and using the integral representation of the Dirac delta
\begin{equation}\label{eqn:fourier delta}
	\chi(x)
	= \int_p \, \tilde\chi(p) \, e^{i p x},
	\qquad
	\int_x \, e^{i (p - q) x}
	= (2 \pi)^d \, \delta(p - q),
\end{equation}
we can transform \labelcref{eqn:quadratic part} into momentum space (we won't distinguish between $\chi(x)$ and its Fourier transform $\tilde\chi(p)$ in the sequel),
\begin{equation}\label{eqn:fourier quadratic part}
	\Gamma_{k,2}
	= \frac{1}{2} \int_p \, \chi_a(p) \biggl[m_{ab}^2 + p^2 \Bigl[Z_k(\rho_c,p^2) \, \delta_{ab} + \rho_c \, Y_k(\rho_c,p^2) \delta_{a1} \delta_{b1}\Bigr]\biggr] \chi_b(-p).
\end{equation}
The 2-point function for constant fields thus reads
\begin{align}\label{eqn:const field 2-point fct}
	\Gamma_{k,ab}^{(2)}(p,q)\Bigr|_{\varphi_c}
	&= \frac{\delta^2 \Gamma_{k,2}}{\delta \chi_a(p) \, \delta \chi_b(-q)}
	= \frac{1}{(2 \pi)^d} \biggl[m_{ad}^2 + p^2 \Bigl[Z_k(\rho_c,p^2) \, \delta_{ad} + \rho_c \, Y_k(\rho_c,p^2) \delta_{a1} \delta_{d1}\Bigr]\biggr] \frac{\delta \chi_d(-p)}{\delta \chi_b(-q)}\notag\\
	&= \frac{1}{(2 \pi)^d} \biggl[m_{ab}^2 + p^2 \Bigl[Z_k(\rho_c,p^2) \, \delta_{ab} + \rho_c \, Y_k(\rho_c,p^2) \delta_{a1} \delta_{b1}\Bigr]\biggr] \delta(p-q).
\end{align}
Note that the Dirac delta distribution is its own functional inverse,
\begin{equation}
	\int \dif^d k \, \delta(p - k) \, \delta(k - q)
	= \delta(p - q).
\end{equation}
The operator inverse $\bigl[\Gamma_k^{(2)} + R_k\bigr]^{-1}$ is therefore simply given by the algebraic inverse,
\begin{mleqn}\label{eqn:op inv}
	&\bigl[\Gamma_k^{(2)}(p,q)\bigr|_{\varphi_c} + R_k\bigr]_{ab}^{-1}(p,q)\\
	&= \biggl[m_{ab}^2 + p^2 \Bigl[Z_k(\rho_c,p^2) \, \delta_{ab} + \rho_c \, Y_k(\rho_c,p^2) \delta_{a1} \delta_{b1}\Bigr] + R_k(p) \, \delta_{ab}\biggr]^{-1} (2 \pi)^d \, \delta(p-q)
\end{mleqn}
Executing the trace over field indices, we get
\begin{equation}\label{eqn:propagator trace}
	\Bigl[\Gamma_k^{(2)}\bigr|_{\varphi_c} + R_k\Bigr]_{aa}^{-1}(p,q)
	= \biggl[\frac{1}{M_1} + \frac{N-1}{M_0}\biggr] (2 \pi)^d \, \delta(p-q),
\end{equation}
with
\begin{mleqn}
	M_0
	&= Z_k(\rho_c,p^2) \, p^2 + U_k^\prime(\rho_c) + R_k(p),\\
	M_1
	&= \bigl[Z_k(\rho_c,p^2) + \rho_c \, Y_k(\rho_c,p^2)\bigr] \, p^2 + U_k^\prime(\rho_c) + 2 \rho_c \, U_k^{\prime\prime}(\rho_c) + R_k(p).
\end{mleqn}
Momentum conservation requires $p = q$ inside the closed loop on the r.h.s. of the flow equation (cf. \cref{eqn:graphical wetterich eqn}). As can be seen from \labelcref{eqn:fourier delta}, \labelcref{eqn:propagator trace} thus receives a factor $\delta(0) = V_d/(2 \pi)^d$ (with $V_d = \vol(\reals^d)$ the volume of $d$-dimensional Euclidean space) that cancels with the volume factor in \labelcref{eqn:potential flow}, resulting in the flow equation for the effective potential
\begin{equation}\label{eqn:potential flow eqn}
	\partial_t U_k\bigr|_{\rho_c}
	= \frac{1}{2} \int_p \, \partial_t R_k(p) \biggl[\frac{1}{M_1} + \frac{N-1}{M_0}\biggr].
\end{equation}
There are three important things to note here.
\begin{enumerate}
	\item The flow equation for $U_k$ is exact \cite{wetterich1993exact,tetradis1994critical} since our truncation \labelcref{eqn:o(n) flowing action generalized} of $\Gamma_k$ contains the most general terms for quadratic fluctuations around a constant field. As explained above, these are the only ones that contribute to $\Gamma_k^{(2)}$ when evaluated at constant background field.
	
	\item Like \labelcref{eqn:wetterich eqn}, \labelcref{eqn:potential flow eqn} is a partial differential equation containing derivatives of $U_k$ with respect to the independent variables $k$ and $\rho$. But unlike \labelcref{eqn:wetterich eqn} it is no longer functional in nature. In most cases \labelcref{eqn:potential flow eqn} is solved by turning it into an (infinite) set of coupled ordinary differential equations with independent variable $k$ \cite{tetradis1994critical}. This is achieved by expanding $U_k(\rho)$ into a Taylor series around some constant $\rho_c$. If we are interested in excitations close to the vacuum, an expansion around $\rho_0(k)$ is appropriate. In the limit $k \to 0$, $\rho_0(0)$ specifies the macroscopic vacuum and the $\rho$-derivatives of $U_k$ the renormalized masses and couplings of the theory.
	
	\item The term $M_1^{-1}$ in \labelcref{eqn:potential flow eqn} incorporates fluctuations from the massive radial field $\varphi_1$. It contributes most to the flow at sufficiently high temperatures where $\varphi_1$ excitations are not suppressed by their non-zero mass. On the other hand, $M_0^{-1}$ describes fluctuations of the massless Goldstone bosons. It dominates the flow at low temperatures.

	\item As it stands, \labelcref{eqn:potential flow eqn} is not closed. To close it requires flow equations for the $\rho_c$- and $p^2$-dependent wave function renormalizations $Z_k$, $Y_k$ \cite{wetterich2001effective}. In the local potential approximation \labelcref{eqn:lpa} we take
	\begin{equation}
		Z_k(\rho_c,p^2) = Z_k,
		\qquad
		Y_k(\rho_c,p^2) = 0,
	\end{equation}
	so that the only thing needed to close \labelcref{eqn:potential flow eqn} is the flow equation
	\begin{equation}
		\partial_t Z_k
		= -\eta \, Z_k,
	\end{equation}
	or equivalently the anomalous dimension $\eta = -\partial_t \ln(Z_k)$.
\end{enumerate}
Introducing the threshold functions
\begin{mleqn}\label{eqn:threshold}
	\tfie_j(Z,m^2,R)
	&= (\delta_{0j} - j) \int_{\vec p} \frac{\partial_t R}{(Z \, p^2 + m^2 + R)^{j+1}}\\
	&= \tilde\partial_t \int_{\vec p} \begin{cases}
		\ln\bigl(Z \, p^2 + m^2 + R\bigr) & j = 0,\\
		(Z \, p^2 + m^2 + R)^{-j} & j \geq 1,
	\end{cases}
\end{mleqn}
where the cutoff derivative $\tilde\partial_t = \partial_t|_{\Gamma_k^{(2)}}$ targets only the explicit scale dependence of the regulator, we can write the flow equation \labelcref{eqn:potential flow eqn} as
\begin{equation}\label{eqn:potential flow ito i}
	\partial_t U_k\bigr|_{\rho_c}
	= \tfrac{1}{2} \, \tfie_0\bigl(Z_k + \rho_c Y_k,\, U_k^\prime + 2 \rho_c U_k^{\prime\prime},\, R_k\bigr) + \tfrac{1}{2} (N - 1) \, \tfie_0\bigl(Z_k,\, U_k^\prime,\, R_k\bigr).
\end{equation}
The threshold functions \labelcref{eqn:threshold} have the important property that they decay rapidly for $m^2 \gg Z k^2$ \cite{wetterich2001effective,berges2002non}. This implements the decoupling of heavy modes (see \cref{item:decoupling} on \cpageref{item:decoupling}). They also diverge for some negative value of $m^2$ which is related to the fact that the effective potential must become convex for $k \to 0$.

In principle we could attempt to solve \labelcref{eqn:potential flow ito i} as it stands, allowing for a completely general form of $U_k(\rho)$ \cite{floerchinger2012analytic}. This would require solving a two-dimensional partial differential equation numerically. Our investigation is mostly conceptual and qualitative in nature, however, and so we contend ourselves with another restriction to our truncation (and the volume of field space we search for a solution) by Taylor expanding $U_k(\rho)$ around $\rho = \rho_0$ to quartic order in the fields,
\begin{equation}\label{eqn:quartic potential}
	U_k(\rho)
	= U_k(\rho_0) + m^2 (\rho - \rho_0) + \frac{\lambda}{2} (\rho - \rho_0)^2,
\end{equation}
with $m^2 = U_k^\prime(\rho_0)$, $\lambda = U_k^{\prime\prime}(\rho_0)$. In the phase $\rho_0 > 0$ of spontaneously broken $O(N)$, $U_k^\prime(\rho_0)$ vanishes by definition. The term quadratic in the fields is then given by $-\lambda \rho_0 \rho$ which implies $m^2 = -2\lambda \rho_0 < 0$. The potential thus takes the form
\begin{equation}\label{eqn:eff pot exp}
	U_k(\rho)
	= \begin{cases}
		U_k(0) + m^2 \, \rho + \frac{1}{2} \lambda \, \rho^2 & \rho_0 = 0,\\[1ex]
		U_k(\rho_0) + \frac{1}{2} \lambda (\rho - \rho_0)^2 & \rho_0 > 0.
	\end{cases}
\end{equation}

To derive flow equations for the couplings $m^2$, $\lambda$ and the minimum location $\rho_0$, we project these parameters onto the flow of $U_k$. Neglecting the (subleading) $\rho$-dependence of $Z_k$ and $Y_k$\footnote{Terms that would arise from the product rule if we took into account the $\rho$-dependence of $Z_k$ and $Y_k$ are all related to a scale dependence of the kinetic term. They will hence be negligible for small anomalous dimensions \cite{wetterich1993average}.} and using the recursive relation
\begin{equation}\label{eqn:induction}
	\partial_{m^2} \tfie_j
	= (\delta_{0j} - j) \, \tfie_{j+1},
\end{equation}
taking $\rho$-derivatives of \labelcref{eqn:potential flow ito i} evaluated at $\rho_c = \rho_0$ yields
\begin{align}\label{eqn:ukp flow}
	&\partial_t U_k^\prime\bigr|_{\rho_0}
	= \tfrac{1}{2} \bigl(3 U_k^{\prime\prime} + 2 \rho_0 U_k^{(3)}\bigr) \, \tfie_1\bigl(Z_k + \rho_0 Y_k,U_k^\prime + 2 \rho_0 U_k^{\prime\prime},R_k\bigr) + \tfrac{1}{2} (N - 1) \, U_k^{\prime\prime} \, \tfie_1\bigl(Z_k,U_k^\prime,R_k\bigr),\\[1ex]
	&\begin{aligned}\label{eqn:ukpp flow}
		\partial_t U_k^{\prime\prime}\bigr|_{\rho_0}
		&= -\tfrac{1}{2} \bigl(3 U_k^{\prime\prime} + 2 \rho_0 U_k^{(3)}\bigr)^2 \, \tfie_2\bigl(Z_k + \rho_0 Y_k,U_k^\prime + 2 \rho_0 U_k^{\prime\prime},R_k\bigr) - \tfrac{1}{2} (N - 1) \, \bigl(U_k^{\prime\prime}\bigr)^2 \, \tfie_2\bigl(Z_k,U_k^\prime,R_k\bigr)\\
		&\hphantom{{}=}+ \tfrac{1}{2} \bigl(5 U_k^{(3)} + 2 \rho_0 U_k^{(4)}\bigr) \, \tfie_1\bigl(Z_k + \rho_0 Y_k,U_k^\prime + 2 \rho_0 U_k^{\prime\prime},R_k\bigr) + \tfrac{1}{2} (N - 1) \, U_k^{(3)} \, \tfie_1\bigl(Z_k,U_k^\prime,R_k\bigr).
	\end{aligned}
\end{align}
The flow equations for $m^2$ and $\lambda$ follow from \labelcref{eqn:eff pot der} together with \labelcref{eqn:ukp flow,eqn:ukpp flow},
\begin{align}
	&\begin{aligned}
		\mathllap{\partial_t m^2}
		&= \frac{\partial}{\partial \rho} \frac{\dif U_k}{\dif t}\biggr|_{\rho_0=0}
		= \frac{\partial}{\partial \rho} \biggl(\eta \, \rho \, U_k^\prime + \partial_t U_k\bigr|_{\rho_0}\biggr)\biggr|_{\rho_0=0}\\
		&= \eta \, \bigl[U_k^\prime + \rho \, U_k^{\prime\prime}\bigr] + \partial_t U_k^\prime\Bigr|_{\rho_0=0}\\
		&\crefrel{eqn:ukp flow}{=} \eta \, U_k^\prime(0) + \tfrac{3}{2} \lambda \, \tfie_1\bigl(Z_k,\, U_k^\prime,\, R_k\bigr) + \tfrac{1}{2} (N - 1) \, \lambda \, \tfie_1\bigl(Z_k,\, U_k^\prime,\, R_k\bigr)\\
		&= \eta \, m^2 + \frac{\lambda}{2} \, (N + 2) \, \tfie_1\bigl(Z_k,\, m^2,\, R_k\bigr),
	\end{aligned}\\[1ex]
	&\begin{aligned}
		\mathllap{\partial_t \lambda}
		&= \frac{\partial^2}{\partial^2 \rho} \frac{\dif U_k}{\dif t}\biggr|_{\mathrlap{\rho_0}}
		= \frac{\partial}{\partial \rho} \biggl(\eta \, \bigl[U_k^\prime + \rho \, U_k^{\prime\prime}\bigr] + \partial_t U_k^\prime\bigr|_{\rho_0}\biggr)\biggr|_{\rho_0}\\
		&= \eta \, \bigl[2 U_k^{\prime\prime}(\rho_0) + \rho_0 \, U_k^{(3)}(\rho_0)\bigr] + \partial_t U_k^\prime(\rho_0)\Bigr|_{\rho_0}\\
		&\crefrel{eqn:ukpp flow}{=} 2 \eta \, \lambda - \frac{\lambda^2}{2} \Bigl[9 \, \tfie_2\bigl(Z_k + \rho_0 Y_k,\, 2 \rho_0 \lambda,\, R_k\bigr) + (N - 1) \, \tfie_2\bigl(Z_k,0,\, R_k\bigr)\Bigr].
	\end{aligned}
\end{align}
To obtain a flow equation for $\rho_0$, we take the total scale derivative of $U_k^\prime$ evaluated at $\rho_0$,
\begin{equation}\label{eqn:minimum cond}
	0
	= \frac{\dif U_k^\prime(\rho_0)}{\dif t}
	= U_k^{\prime\prime}(\rho_0) \, \partial_t \rho_0 + \partial_t U_k^\prime(\rho_0)\bigr|_{\rho_0},
\end{equation}
which vanishes because the $\rho$-derivative of the effective potential is zero at its minimum. Solving \labelcref{eqn:minimum cond} for $\partial_t \rho_0$ and inserting the $\rho$-derivative of \labelcref{eqn:eff pot der}, we get
\begin{mleqn}\label{eqn:rho flow}
	\partial_t \rho_0
	&= -\frac{1}{\lambda} \partial_t U_k^\prime(\rho_0)
	\crefrel{eqn:eff pot der}{=} -\frac{1}{\lambda} \Bigl(\eta \, \rho_0 \, U_k^{\prime\prime}(\rho_0) + \partial_t U_k^\prime(\rho_0)\bigr|_{\rho_0}\Bigr)\\
	&\crefrel{eqn:ukp flow}{=} -\eta \, \rho_0 - \frac{1}{2} \Bigl[3 \tfie_1\bigl(Z_k + \rho_0 Y_k,\, U_k^\prime + 2 \rho_0 U_k^{\prime\prime},\, R_k\bigr) + (N - 1) \, \tfie_1\bigl(Z_k,\, U_k^\prime,\, R_k\bigr)\Bigr].
\end{mleqn}
This concludes our introductory section on Euclidean functional renormalization. In the next two sections, we will analytically continue flow equations to extend the formalism to Minkowski space.

\section{Functional Renormalization in Minkowski Space}
\label{sec:fr in ms}

So far, the functional renormalization group in its formulation due to Wetterich \cite{wetterich1993average} has been applied mainly in Euclidean space to either static, classical statistical field theories (where the fields depend on spatial position only) or quantum field theories in the Matsubara formalism where time and frequency become imaginary \cite{floerchinger2012analytic}.

While significant progress has been made with this setup over the past 20 years, it is only applicable to static systems and imaginary-time quantities. In nature, actual dynamical processes take place in Minkowski space. It thus stands to reason that our understanding of physics, not to mention the renormalization group itself, particularly where real-time properties such as propagator residues and decay widths are concerned, would greatly benefit from an extension of the formalism to this new domain.

Of course, we expect a number of challenges. Most singular structures become visible only in Minkowski space. (Euclidean space propagators feature singularities too, but only for massless particles at $p = 0$ or at Fermi surfaces \cite{floerchinger2012analytic}.) Singularities are difficult to treat numerically, making it convenient to work in Euclidean space where they are fewer. However, singularities in correlation functions are physical and have crucial repercussions on the behavior of the particles they describe. For instance, a pole in the propagator corresponds to a stable particle, a branch cut to a resonance, i.e. an unstable particle.

If we are to fully understand the real-time dynamics of particle propagation and decay on a fundamental level, we must be able to cope with these analytic structures. Fortunately, functional renormalization has the potential to do that and do it well. In the following we develop an analytic implementation of the FRG that takes poles and branch cuts of the propagator into account in a fully self-consistent manner.

\subsection{Methodology}

Different strategies for performing the analytic continuation are conceivable \cite{floerchinger2012analytic,pawlowski2017finite,pawlowski2015real,kamikado2014real,gasenzer2010far,schoeller2009perturbative,keil2001real,korb2007real,tripolt2014spectral,tripolt2014spectral2,tripolt2013finite,strodthoff2017self}.
\begin{enumerate}
	\item\label[opt]{item:radical} The most radical approach reconstructs the formalism from the ground up in Minkowski space by analytically continuing the Feynman path integral itself (the starting point of our derivation of the Wetterich equation in \cref{sec:wetterich equation}). The advantage of such an approach is its applicability to even far-from-equilibrium dynamics. Unfortunately, we are immediately faced with severe technical complications. Factors of $i$ appear at various places, most importantly in the exponent of the integrand $e^{S_\text{M}} = e^{i S_\text{E}}$, spoiling its interpretation as a weighting factor. Moreover, this approach requires the technically involved Schwinger-Keldysh closed time contour.

	\item\label[opt]{item:modest} A more modest attempt would be to stick to the Euclidean functional integral, work with the formalism as derived in \cref{sec:theoretical foundations} exclusively in Euclidean space and use analytic continuation only on the final result after taking the flow down to $k = 0$. This method has in fact been successfully pursued \cite{dupuis2009infrared,sinner2009spectral,haas2014gluon}. The advantage of this procedure is that we use the formalism in a setting where it is comparatively transparent and well understood \cite{floerchinger2012analytic}.
	
	The disadvantage lies in the analytic continuation itself. It can turn out rather difficult in practice since the Euclidean propagator is known only numerically and only at isolated points along the imaginary axis, the so-called Matsubara frequencies $i \omega_n = 2 \pi i T n$, $n \in \mathbb{N}$. Numerical reconstruction based on Padé approximants or the maximum entropy method require information from many points. As a result, the computational effort gets quite large.
	
	Besides this practical issue, there are some systemic shortcomings. First, knowledge about spectral properties does not enable us to improve the renormalization group running. Second, only linear response properties are accessible.

	\item A third possibility also keeps the Euclidean space functional integral but performs the analytic continuation already on the flow equations rather than the final result at $k = 0$. From an innovation standpoint, i.e. how much new formalism needs to be developed, it is situated somewhere between \cref{item:radical,item:modest}. This is the approach we pursue in our work. It offers a number of advantages \cite{floerchinger2012analytic}.
	\begin{enumerate}[label=\roman*)]
		\item Because the flow equations for objects such as the effective potential or the propagator are available in analytic form, we can do the analytic continuation by hand instead of having to resort to involved numerical techniques.

		\item Real-time properties such as quasi-particle decay widths can be inserted in a self-consistent manner on the r.h.s. of flow equations. This should notably improve the performance of truncations. Particularly properties not directly related to the propagator (such as thermodynamic quantities) are expected to gain enhanced accuracy.

		\item All the usual space-time symmetries, i.e. translational as well as Lorentz (or Galilei) invariance are manifest. (A convenient choice of the infrared regulator due to Flörchinger \cite{floerchinger2012analytic} will nevertheless allow us to perform the Matsubara summation in loop expressions analytically, leading to well behaved expressions on the right hand side of flow equations where at most an integral over spatial momenta remains to be done numerically.)

		\item Compared to the Schwinger-Keldysh contour, this method is significantly less involved.

		\item Since we derive all of our flow equations in Euclidean space where functional renormalization is best understood and has progressed the farthest, we can benefit from existing expertise. For example, it is known how the flowing action approaches the microscopic action for large cutoff scales (this is not obvious in Minkowski space due to the indefiniteness of $p^2 = -p_0^2 + \vec{p}^2 \gtrless 0$) or how to construct useful regulators.
	\end{enumerate}
	The biggest drawback, on the other hand, is that (like \cref{item:modest}) this approach is based on linear response theory. It is hence restricted to close-to-equilibrium physics. Even though it can be applied to weakly non-linear regimes \cite{floerchinger2016variational}, strongly non-linear responses as they dominate far from equilibrium are beyond its scope.
\end{enumerate}

\subsection{Matsubara Formalism}
\label{sec:matsubara formalism}

In the Matsubara or imaginary time formalism, quantum fields at non-zero temperature live on a generalized torus $\mathcal{M}_{d+1} = S^1 \times \reals^d$ with circumference $\beta = 1/T$ in the imaginary time direction $\tau = -i t$. We will refer to this topology as Matsubara space. To understand why time becomes imaginary, compact and periodic at non-zero temperature, we recall some basic concepts of statistical mechanics \cite{das2000topics}. An equilibrium ensemble at temperature $T = 1/\beta$ can be described by its partition function
\begin{equation}
	Z(\beta)
	= \Tr \rho(\beta)
	= \Tr e^{-\beta \mathcal{H}},
\end{equation}
where the density operator $\rho(\beta)$ determines the occupation number of every possible state at a given temperature and $\mathcal{H}$ is a Hamiltonian that specifies the type of system we are dealing with. (If $\mathcal{H} = H$, where $H$ is the Hamiltonian that appears in the unitary time evolution operator $U = e^{-i H t}$, the ensemble is canonical, i.e. it has a fixed particle number but variable energy due to heat exchange with a bath. If instead $\mathcal{H} = H - \mu N$ with $N$ the number operator and $\mu$ the chemical potential, the ensemble is grand canonical and can exchange energy with a bath as well as particles with a reservoir.)

The important observables in a statistical setting are ensemble averages $\langle\mathcal{O}\rangle_\beta$ defined as
\begin{equation}\label{eqn:ensemble avr}
	\langle\mathcal{O}\rangle_\beta
	= \frac{1}{Z(\beta)} \, \Tr \mathcal{O} \, e^{-\beta \mathcal{H}}.
\end{equation}
for any measurable quantity $\mathcal{O}$. Cyclicity of the trace renders such averages periodic under imaginary time evolution,
\begin{mleqn}\label{eqn:kms}
	\langle \mathcal{O}(t)\rangle_\beta
	&= \frac{1}{Z(\beta)} \, \Tr e^{-\beta \mathcal{H}} \, \mathcal{O}(t) \, e^{\beta \mathcal{H}} \, e^{-\beta \mathcal{H}}\\
	&= \frac{1}{Z(\beta)} \, \Tr e^{-\beta \mathcal{H}} \, \mathcal{O}(t + i \beta)
	= \langle \mathcal{O}(t + i \beta)\rangle_\beta.
\end{mleqn}
This is known as the Kubo-Martin-Schwinger relation. It is a result of the fact that $e^{-\beta \mathcal{H}}$ acts as a time evolution operator on the compact imaginary time axis $0 \leq \tau = -i t \leq \beta$ with the extent of time determined by the temperature $T = \beta^{-1}$.

The Matsubara formalism is based on the idea (originally due to Bloch \cite{bloch1932theorie} but first implemented perturbatively by Matsubara \cite{matsubara1955new}) that ensemble averages like \labelcref{eqn:ensemble avr} may be written as expectation values in a Euclidean signature quantum field theory. The trace requires that the bosonic (fermionic) fields of such a theory be (anti-)periodic in the imaginary time direction,
\begin{equation}
	\phi(\tau,\vec{x})
	= \pm \phi(\tau + \beta,\vec{x}).
\end{equation}
In momentum space, this leads to the replacement of continuous frequencies by discrete imaginary Matsubara frequencies $i \omega_n = 2 \pi i n T$.

The Matsubara formalism has proven useful in studying the behavior of quantum field theories at non-zero temperature \cite{weinberg1974gauge}. It has been generalized to theories with gauge invariance and was essential in the study of a conjectured deconfining phase transition of Yang-Mills theory \cite{gross1981qcd}.

\subsection{Analytic Structure}
\label{sec:analytic structure}

As stated above, our approach is to derive flow equations for $n$-point functions $\Gamma_k^{(n)}$ in Euclidean space. This yields analytic expressions for $\partial_k \Gamma_k^{(n)}$ at points $i \omega_n$ on the imaginary axis which we can analytically continue to extend them to the entire complex frequency plane with the exception of possible poles and branch cuts along the real axis \cite{floerchinger2012analytic}. This last assertion constitutes a severe restriction to the analytic structure of $n$-point functions and needs to be justified. We will shed light on how it originates for the example of the 2-point function $\Gamma^{(2)} \sim G^{-1}$. Its analytic structure is of particular importance since real-time properties of the propagator $G(p)$ are the main point of interest in this work. Nonetheless, analogous arguments apply also for $n > 2$.

As we saw in \labelcref{eqn:const field 2-point fct}, $\Gamma^{(2)}$ is of the form
\begin{equation}
	\Gamma^{(2)}(p,q)
	= \frac{\delta^2 \Gamma[\varphi]}{\delta \varphi(p) \, \delta \varphi(-q)}
	= (2 \pi)^d \delta(p-q) \, G^{-1}(p),
\end{equation}
with $G(p)$ the Euclidean propagator in momentum space. Enforcing upon $G(p)$ restrictions deriving from Poincaré invariance, unitarity and causality\footnote{Causality requires that the commutator $[\phi(x),\phi(y)]$ vanishes for spacelike separation $(x - y)^2 > 0$.} we obtain the Källen-Lehmann spectral representation \cite{weinberg1995quantum}
\begin{equation}\label{eqn:spectral repr}
	G(p)
	= \int_0^\infty \dif \mu^2 \, \frac{\rho(\mu^2)}{p^2 + \mu^2},
\end{equation}
with real and non-negative spectral weight $\rho(\mu^2) \geq 0$ normalized according to
\begin{equation}
	\int_0^\infty \rho(\mu^2) \, \dif \mu^2
	\overset{!}{=} 1.
\end{equation}
\labelcref{eqn:spectral repr} is interesting for several reasons. First, from a field theoretical standpoint, it decomposes the interacting propagator into a weighted sum of free propagators. Second and more relevant to our analysis, it allows for a very instructive investigation of the analytic structure of $G(p)$.

In Euclidean space $p^2 = p_0^2 + \vec{p}^2 \geq 0$ is positive semi-definite such that the integrand in \labelcref{eqn:spectral repr} is completely regular, rendering $G(p)$ both real and positive for all $p$. In Minkowski space, on the other hand, $p^2 = -p_0^2 + \vec{p}^2 \gtrless 0$ is indefinite. For $p^2 < 0$, $G(p)$ features singularities on the real frequency axis located at\footnote{Although we integrate over $\mu^2$ in \labelcref{eqn:spectral repr}, we are interested in the location of these poles in $p_0$-space rather than $\mu$-space. This is because to solve flow equations, we have to integrate expressions containing $G(p)$ with respect to spatial momentum and frequency. To perform the $p_0$-integration (or Matsubara summation at $T > 0$), we then need to specify an integration contour in the complex frequency plane that avoids the poles. We can either slightly deform the contour away from the real axis at $p_0 = \pm\csqrt{\vec{p}^2 + \mu^2}$ or add infinitesimal $\pm i \epsilon$-terms in the denominator to shift the poles away from the real axis. See \cref{app:prop analytic structure} for details.}
\begin{equation}\label{eqn:real line poles}
	p_0
	= \pm\csqrt{\vec{p}^2 + \mu^2}.
\end{equation}
\begin{figure}[htb!]
	\centering
	\begin{tikzpicture}[thick]
	    \def\xr{5} \def\yr{1}
	    
	    \fill (-\xr/2,0) circle (2pt) node[above] (leftbranch) {$-\sqrt{\vec{p}^2}$} (\xr/2,0) circle (2pt) node[above] (rightbranch) {$\sqrt{\vec{p}^2}$};
	    
	    \draw [->,decorate,decoration={zigzag,segment length=6,amplitude=2}] (-\xr-0.4,0) -- (leftbranch.south) (rightbranch.south) -- (\xr,0) node [above left]  {$\Re(p_0)$};
	    \draw (leftbranch.south) -- (rightbranch.south);
	    \draw [->] (0,-\yr) -- (0,\yr) node[below left=0.1] {$\Im(p_0)$};
	    
	\end{tikzpicture}
	\caption{Propagator branch cuts along the real frequency axis extending from $\pm |\vec{p}|$ out to $\pm \infty$}
	\label{fig:branch cuts}
\end{figure}
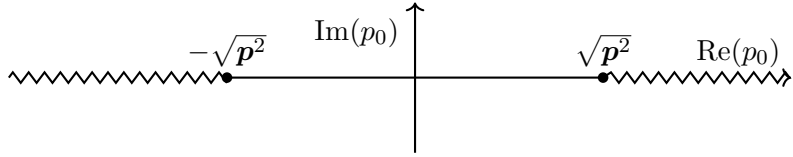%
Since $\mu^2$ is integrated over, \labelcref{eqn:real line poles} actually signals a continuum of singularities, i.e. a branch cut spanning from $\pm |\vec{p}|$ out to infinity in both directions along the real $p_0$-axis as depicted in \cref{fig:branch cuts}. It is typical for $\rho(\mu^2)$ to contain both pole and branch cut contributions from single-particle and bound states, and multi-particle states with continuous energy spectra, respectively. By the Sokhotski-Plemelj theorem, \labelcref{eqn:spectral repr} can be written as (a detailed derivation was relegated to \cref{app:prop analytic structure})
\begin{equation}\label{eqn:sokhotski-plemelj form}
	G(p)
	= \mathcal{P} \biggl(\int_0^\infty \dif \mu^2 \, \frac{\rho(\mu^2)}{p^2 + \mu^2}\biggr) + i \pi \, \sign(\Re p_0 \Im p_0) \, \rho(-p^2),
\end{equation}
where $\mathcal{P}$ denotes the Cauchy principal value. \labelcref{eqn:sokhotski-plemelj form} reveals the branch cut structure in $G(p)$. Since $\sign(\Im p_0)$ abruptly changes sign when crossing the real axis, $G(p)$ features a cut along the real frequency axis at all $p_0$ for which $\rho(-p^2) \neq 0$. \labelcref{eqn:sokhotski-plemelj form} also shows that $G(p)$ is analytic away from the real axis. (Since $\rho(-p^2) = 0 \enskip \forall \, p^2 < 0$, i.e. for all $p_0^2 - \vec{p}^2 \leq p_0^2 <0 \, \Leftrightarrow \, p_0 \in i \reals$, there is no branch cut on the imaginary axis). An important consequence is that also the inverse propagator $G^{-1} \sim \Gamma^{(2)}$ has all its poles, zero-crossings and branch cuts on the real axis as well.

Inverting \labelcref{eqn:sokhotski-plemelj form}, we find that close to the real axis, $\Im(p_0) \approx 0$, the inverse propagator $P(p) = G^{-1}(p)$ is of the form \cite{floerchinger2012analytic}
\begin{equation}\label{eqn:inverse prop}
	P(p)
	= P_1(p^2) - i s(p_0) \, P_2(p^2),
\end{equation}
with $s(p_0) = \sign(\Re p_0 \Im p_0)$ and
\begin{mleqn}
	&P_1(p^2)
	= \frac{\Re G(p)}{\bigl[\Re G(p)\bigr]^2 + \bigl[\Im G(p)\bigr]^2},
	\qquad
	&&P_2(p^2)
	= \frac{\Im G(p)}{\bigl[\Re G(p)\bigr]^2 + \bigl[\Im G(p)\bigr]^2},\\[1ex]
	&\Re G(p)
	= \mathcal{P} \biggl(\int_0^\infty \dif \mu^2 \, \frac{\rho(\mu^2)}{p^2 + \mu^2}\biggr),
	&&\Im G(p)
	= \pi \, \rho(-p^2).
\end{mleqn}
Close to a point $p^2 = -p_0^2 + \vec{p}^2 = -m^2/z$ where $P_1(p^2)$ vanishes (corresponding to a pole in the propagator and thus to a particle), we can expand $P_1$ and $P_2$ as
\begin{equation}\label{eqn:particle expansion}
	P_1(p^2)
	= Z \, (z \, p^2 + m^2) +\dots,
	\qquad
	P_2(p^2)
	= Z \, \gamma^2(p^2) + \dots,
\end{equation}
(with $Z$, $z$, $m^2$ and $\gamma^2$ scale-dependent real and positive quantities) such that the propagator takes the form
\begin{equation}\label{eqn:breit-wigner prop}
	G(p)
	= P^{-1}(p)
	= \frac{1}{Z	} \frac{z \, p^2 + m^2 + i s(p_0) \, \gamma^2}{(z \, p^2 + m^2)^2 + \gamma^4}.
\end{equation}
\labelcref{eqn:breit-wigner prop} describes an unstable particle whose decay is governed by the Breit-Wigner distribution. $\csqrt{Z \, z \, p^2}$ is the center-of-mass energy that produces the resonance, $\sqrt{Z \, m^2}$ the mass of the resonance and $\Gamma = \gamma^2/m$ the decay width (width of the distribution at half-maximum). $\Gamma$ is the inverse of the mean lifetime $\tau = 1/\Gamma$. In the limit of vanishing decay width, $\Gamma \to 0$, the resonance shows up as a delta peak in the spectral function $\rho(p^2)$ and the particle becomes stable. Since this is exactly the type of real-time physics we are interested in, we will continue to employ an inverse propagator of the form \labelcref{eqn:inverse prop} and the expansion \labelcref{eqn:particle expansion} throughout this work.

Before continuing, we explain why we deem it sufficient to know the form of $P(p)$ only close to the real axis $\Im(p_0) \approx 0$. Of course in principle the analytic structure of $(P_k + R_k)^{-1}$ as a function of complex frequency $p_0 \in i \reals$ depends also on the shape of $P(p)$ away from the real axis. However, especially for small $k$, high-energy fluctuations due to virtual particles are strongly suppressed and we expect the propagator $G(p)$ to be dominated by on-shell excitations corresponding to the poles and branch cuts on the real frequency axis. It should be viewed as part of our truncation that possible deviations from this structure at higher scales are neglected. Based on this reasoning, we use \labelcref{eqn:inverse prop,eqn:particle expansion} not only close to the real axis but everywhere in the $p_0$-plane. The coefficients $Z$, $m^2$ and $\gamma^2$ are nonetheless determined by their value at the singularity $p^2 = -m^2/z$ on the real line where $\Re\bigl(P(p)\bigr)$ vanishes.

\section{Analytic Continuation of Flow Equations}\label{sec:analytic continuation of flow equations}

The analytic continuation proceeds differently for different parts of the flowing action. Since the effective potential $U_k$ is momentum-independent, its analytic continuation is trivial. The propagator $G(p)$ and higher-order correlation functions, on the other hand, \textit{are} momentum-dependent. Their flow equations are obtained by expanding the flow of $\Gamma_k$ around a constant background field $\varphi_c$ that fluctuates with a small momentum-dependent part, $\varphi_a(p) = \varphi_c \, \delta_{a1} + \chi_a(p)$, $\chi_a(p) \ll \varphi_c \enskip \forall \, a, p$. In this case, we actually have to work to perform the analytic continuation. Once $G(p)$ is extended to the entire complex plane, however, we have easy access to its real-time properties such as the decay width $\Gamma = \gamma^2/m$ by simply evaluating it for $p_0 \in \reals$.

Working in Minkowski space has some advantages when it comes to devising truncations \cite{floerchinger2012analytic}. By performing the derivative expansion as a Taylor series around on-shell excitations, i.e. around frequencies and momenta corresponding to a pole or branch cut of the propagator, we expect the convergence of the expansion to improve. After all, loop expressions on the r.h.s. of flow equations (as well as on-shell properties of the effective action) are strongly dominated by such singular structures. Compared to an expansion around vanishing frequency in Euclidean space, higher-order terms of the derivative expansion in Minkowski space should therefore be much more strongly suppressed. In most situations, we expect the essential physics to already be well-described by the lowest-order terms.

This has important consequences. In particular, in many situations it may allow us to use an algebraic (as opposed to exponential or Litim-type) regulator $R_k$ even though it exhibits a much milder decay in the ultraviolet. A simple algebraic form of $R_k$ has two major advantages. First, we may construct $R_k$ in Euclidean space and analytically continue it towards the real frequency afterwards. Second, it enables us to decompose the propagator $\smash{(P_k+ R_k)^{-1}}$ in such a way as to perform the summation over Matsubara frequencies analytically!

In this section we assemble the formalism needed to solve flow equations in Minkowski space. First, in \cref{sec:mink truncation} we modify the truncation \labelcref{eqn:o(n) flowing action generalized} of $\Gamma_k$ so as to cope with additional singular structures that arise in Minkowski space. \Cref{sec:mink regulator} introduces the above-mentioned class of regulators. In \cref{sec:feynman rules} we go on to derive the momentum-space Feynman rules for that particular combination of truncation and regulator. With those tools in place, we construct flow equations for parameters of the effective potential and the propagator in \cref{sec:mink potential flow,sec:prop flow}, respectively.

\subsection{Truncation}
\label{sec:mink truncation}

We consider again the $O(N)$-invariant scalar field theory of section \cref{sec:potential flow}, now in $d + 1$ dimensions with time added to the $d$ Euclidean dimensions of space. As we saw in \cref{sec:potential flow}, the spectrum of excitations in the phase with spontaneously broken $O(N)$ consists of a massive radial field $\varphi_1$ and $N - 1$ massless Goldstone bosons.

Due to the term $\sim \varphi_1 \varphi_a^2$ ($a \neq 1$) in $\Gamma_k$ (more precisely in $U_k$), the radial mode can decay into two Goldstone excitations during real-time evolution. This gives rise to a non-vanishing decay width $\gamma_1^2$ for the radial mode which makes it a quasi-particle with finite lifetime. The ordered phase $\rho_0 \neq 0$ thus corresponds to a non-zero density of quasi-particles (which we expect to disperse if we increase the temperature and vacate the ground state).

Except for this new decay channel, the system is very similar to the one treated in \cref{sec:potential flow}. We can therefore employ a similar truncation written in terms of unrenormalized fields $\bar\varphi_a(x)$ as\footnote{Up to this point, $\varphi$ and $\rho$ denoted unrenormalized quantities. We now change notation $\varphi \to \bar\varphi$, $\rho \to \bar\rho$.}
\begin{equation}\label{eqn:minkowski space ansatz}
	\Gamma_k[\varphi]
	= \pmint_x \biggl[\bar{U}_k(\bar\rho) + \frac{1}{2} \, \bar\varphi_a \, Q_k(-\partial^2) \, \bar\varphi_a + \frac{1}{4} \, \bar\rho \, S_k(-\partial^2) \, \bar\rho\biggr],
\end{equation}
where the sum over $a \in \{1,\dots,N\}$ is implied and we introduced the shorthand notation
\begin{equation}
	\pmint_x
	= \int_0^\beta \dif \tau \int_{\reals^d} \! \dif^d x
\end{equation}
to denote integration over Matsubara space $\mathcal{M}_{d+1} = S^1 \times \reals^d$. ($\mathcal{M}_{d+1}$ is a $d + 1$-dimensional generalized torus spanned by the Cartesian product of $d$ Euclidean dimensions of space $\reals^d$ and a circle $S_1$ of temperature-dependent circumference $\beta = 1/T$ for the cyclic dimension of imaginary time $\tau = i t$.) In momentum space this corresponds to 
\begin{equation}\label{eqn:mmint def}
	\mmint_p
	= T \sum_{p_0} \int_{\vec p}
	= T \sum_{p_0} \int_{\reals^d} \frac{\dif^d \vec p}{(2 \pi)^d},
\end{equation}
where $\sum_{p_0}$ sums over the discrete imaginary Matsubara frequencies $p_0 \in \{i \omega_n = 2 \pi i T n | n \in \integers\}$.

In \labelcref{eqn:minkowski space ansatz} we made the crucial assumption that the momentum-dependent parts of the inverse propagator $Q_k(p^2)$ and $S_k(p^2)$ are of the same analytic structure as the inverse propagator \labelcref{eqn:inverse prop} expanded as in \labelcref{eqn:particle expansion} \cite{floerchinger2012analytic}. Removing the momentum-independent mass from \labelcref{eqn:particle expansion}, this amounts to
\begin{equation}
	Q_k(p)
	= Z_k(p^2) \, p^2 - i s(p_0) \, \gamma_k^2(p^2),
	\qquad
	S_k(p)
	= Y_k(p^2) \, p^2 - i s(p_0) \, \delta_k^2(p^2).
\end{equation}
Note that we are neglecting here possible alterations in the analytic structure of the scale-dependent propagator due to the frequency dependence of the regulator $R_k(p)$ used to construct $\Gamma_k$. Since $R_k(p) \to 0$ for $k \to 0$, any such alterations will disappear at $k = 0$. However, they will be present at non-zero $k$ and might even have non-negligible effects at intermediate stages of the flow. It should be viewed as part of our truncation that we disregard these modifications here.

The functions $\smash{\gamma_k^2(p^2)}$, $\smash{\delta_k^2(p^2)}$ determine the size of the jump at the branch cut discontinuity in \cref{fig:branch cuts}. Since $G(p)$ is completely regular for $p^2 > 0$, $\smash{\gamma_k^2(p^2)}$ and $\smash{\delta_k^2(p^2)}$ are non-zero only for $p^2 < 0$. Physically, this ensures causality of the decay process since it requires time-like, i.e. negative $p^2$.

To derive the flow equation for $U_k(\rho)$, we again expand $\Gamma_k$ around a constant background,
\begin{equation}
	\bar\varphi_a(x)
	= \bar\varphi_c \, \delta_{a1} + \bar\chi_a(x),
	\qquad
	\bar\rho_c
	= \frac{1}{2} \, \bar\varphi_c^2.
\end{equation}
Only the part $\Gamma_{k,2}$ that is quadratic in the fluctuating fields $\bar\chi_a(x)$ contributes to the flow of $U_k(\rho)$. In momentum space it reads
\begin{equation}\label{eqn:gammak2}
	\Gamma_{k,2}
	= \frac{1}{2} \mmint_p \, \bar\chi_a(p) \biggl[\delta_{ab} \Bigl(Q_k(p^2) + \bar{U}_k^\prime(\bar\rho_c)\Bigr) + \delta_{a1} \, \delta_{b1} \Bigl(\bar\rho_c \, S_k(p^2) + 2 \bar\rho_c \, \bar{U}_k^{\prime\prime}(\bar\rho_c)\Bigr)\biggr] \bar\chi_b(-p).
\end{equation}
Just like in \labelcref{eqn:particle expansion}, we further expand \labelcref{eqn:gammak2} around zero-crossings of
\begin{equation}\label{eqn:zero crossings}
	\Re\Bigl[Q_k(p) + \bar\rho_c S_k(p) + \bar{U}_k^\prime(\bar\rho_c) + 2 \bar\rho_c \bar{U}_k^{\prime\prime}(\bar\rho_c)\Bigr]
	\qquad\text{and}\qquad
	\Re\Bigl[Q_k(p) + \bar{U}_k^\prime(\bar\rho_c)\Bigr],
\end{equation}
corresponding to the point on the real frequency axis where the propagators $G_1(p)$ and $G_a(p)$ ($a \neq 1$) become singular. These are the on-shell excitations of the radial field and the Goldstone bosons, respectively. Strictly speaking, the location of the zero-crossings depend on the value of the background field $\bar\rho_c$. For the regimes we will study, however, it suffices to expand \labelcref{eqn:zero crossings} around some $p^2 = -m^2$ such that the expressions \labelcref{eqn:zero crossings} vanish at the minimum $\rho_0$. Since
\begin{equation}
	m_{ab}^2
	= \bar{U}_k^\prime(\bar\rho_0) \, \delta_{ab} + 2 \rho_0 \bar{U}_k^{\prime\prime}(\bar\rho_0) \, \delta_{a1} \, \delta_{b1}
	\qquad\text{with}\enskip
	\bar{U}_k^\prime(\bar\rho_0) = 0,
\end{equation}
the Goldstone bosons ($a = b > 1$) are massless at the minimum $\bar\rho_0$. This puts their expansion point at $p = 0$. The discontinuity along the real frequency axis vanishes here, $\gamma^2(0) = 0$, so that
\begin{equation}\label{eqn:goldstone expansion}
	Q_k(p)
	= Z_k(0) \, p^2,
	\mathrlap{\qquad (\text{for } a = b > 1).}
\end{equation}
The massive radial field ($a = b = 1$) we expand around $p^2 = -m_1^2 = -\bigl[2 \bar\rho_0 \, \bar{U}_k^{\prime\prime}(\bar\rho_0)\bigr]^2$, where $\gamma^2(-m_1^2)$ is non-zero such that
\begin{equation}\label{eqn:radial expansion}
	Q_k(p) + \bar\rho_0 \, S_k(p)
	\,=\, \bigl[Z_k(-m_1^2) + \bar\rho_0 \, Y_k(-m_1^2)\bigr] \, p^2 - i s(p_0) \bigl[\gamma_k^2(-m_1^2) + \bar\rho_0 \, \delta_k^2(-m_1^2)\bigr],
\end{equation}
Introducing the abbreviations
\begin{equation}\label{eqn:z1 gamma1}
	Z_1
	= \frac{1}{Z_k} \bigl[Z_k(-m_1^2) + \bar\rho_0 \, Y_k(-m_1^2)\bigr],
	\qquad
	\gamma_1^2
	= \frac{1}{Z_k} \bigl[\gamma_k^2(-m_1^2) + \bar\rho_0 \, \delta_k^2(-m_1^2)\bigr],
\end{equation}
where we set $Z_k = Z_k(0)$, an expression analogous to \labelcref{eqn:gammak2} but in terms of renormalized fields
\begin{equation}
	\varphi_a
	= \csqrt{Z_k} \, \bar\varphi_a,
	\qquad
	\rho
	= Z_k \, \bar\rho,
	\qquad
	\bar{U}_k(\bar\rho)
	= U_k(\rho),
\end{equation}
can be written as
\begin{equation}\label{eqn:ren gammak2}
	\Gamma_{k,2}
	= \frac{1}{2} \mmint_p \, \chi_a(p) \biggl\{\delta_{ab} \Bigl[p^2 + U_k^\prime(\rho_c)\Bigr] + \delta_{a1} \delta_{b1} \Bigl[(Z_1 - 1) \, p^2 - i s(p_0) \, \gamma_1^2 + 2 \rho_c U_k^{\prime\prime}(\rho_c)\Bigr]\biggr\} \chi_b(-p),
\end{equation}
where we pulled a factor $Z_k$ out of every term and absorbed it into the fluctuating fields $\chi_a = \csqrt{Z_k} \, \bar\chi_a$. By evaluating \labelcref{eqn:ren gammak2} for $\rho_c = \rho_0$, we can directly read off the radial mode's renormalized mass and decay width,
\begin{equation}
	m_1 = \csqrt{2 \rho_0 U_k^{\prime\prime}(\rho_0)/Z_1},
	\qquad
	\Gamma_1 = \gamma_1^2/(Z_1 m_1) = \gamma_1^2/\csqrt{2 \rho_0 \, U_k^{\prime\prime}(\rho_0) \, Z_1}
\end{equation}

\subsection{Regulator}
\label{sec:mink regulator}

Our next goal is to find a regulator $R_k$ that allows for analytic continuation of $(P_k + R_k)^{-1}$ in truncations where close to the real frequency axis, $\Im(p_0) \approx 0$, $P_k$ is well approximated by \labelcref{eqn:inverse prop} expanded according to \labelcref{eqn:particle expansion}, i.e.
\begin{equation}\label{eqn:inverse prop expanded}
	P
	= Z \Bigl[z \, p^2 + m^2 - i s(p_0) \, \gamma^2\Bigr].
\end{equation}
The problem we face is that given the indefiniteness of $p^2 = -p_0^2 + \vec{p}^2$ in Minkowski space, it is unclear which modes correspond to the infrared and which to the ultraviolet part of the spectrum. Some high frequency $p_0 \approx \Lambda$ could join with an equally large momentum $|\vec{p}| \approx \Lambda$ to produce a vanishing $p^2$. Yet $R_k$ still needs to suppress fluctuations from these modes during late stages of the flow (i.e. at small $k$) if the derivative expansion is to have any chance at convergence. This might not seem like such a difficult problem until we recall that we cannot split up $p^2$ and simply implement the decay for high $p_0$ and high $\vec{p}$ separately if we wish to keep rotational and Lorentz invariance. At this point it is still unclear which requirements $R_k$ must fulfill in order to act as an effective infrared and ultraviolet regulator in Minkowski space.

In Euclidean space by contrast, $p^2 \geq 0$ establishes an unambiguous order relation for all modes in the spectrum. Constructing a regulator with the desired properties becomes a simple matter. Our approach will therefore be to construct a regulator in Euclidean space and use analytic continuation to extend it to Minkowski space. There are some caveats to this method, however. A function that is smooth and regular on the imaginary frequency axis may nevertheless feature poles and discontinuities in other regions of the complex plane. In fact, it stands to reason that this is even unavoidable if we require $R_k$ to decay rapidly for large imaginary values of $p_0$. We thus expect analytic continuation to be difficult for most choices of $R_k$ that have proven useful in Euclidean space. For that reason, we adopt here a special class of regulators due to Flörchinger \cite{floerchinger2012analytic} which is particularly suited to analytic continuation,
\begin{equation}\label{eqn:algebraic regulator}
	R_k(p)
	= \frac{Z \, k^2}{\sum_{j=0}^\infty c_j \bigl(\frac{p^2}{k^2}\bigr)^j}
	= \frac{Z \, k^2}{c_0 + c_1 \, \frac{p^2}{k^2} + c_2 \, \Bigl(\frac{p^2}{k^2}\Bigr)^2 + \dots}.
\end{equation}
The coefficient $Z$ can be chosen for convenience. We will identify it with the wave function renormalization $Z_k$ but it could be any real, positive function of $k$.

When only a few coefficients $c_j$ are non-zero, \labelcref{eqn:algebraic regulator} has a comparatively mild algebraic decay in the ultraviolet. We still expect it to provide adequate separation of momentum modes due to the improved convergence of the derivative expansion in Minkowski space. \labelcref{eqn:algebraic regulator} has all desired properties for Euclidean argument $p_0^2 + \vec{p}^2 \geq 0$ if the coefficients $c_j$ are real and positive. Regularization of the ultraviolet improves if some $c_j$ with large $j$ are non-zero. On the other hand, calculations simplify if only a few $c_j$ with small $j$ are non-zero. The simplest, non-trivial choice is $c_0 = 1$, $c_1 = c > 0$, $c_j = 0 \; \forall \, j > 1$. Then
\begin{equation}\label{eqn:choice of rk}
	R_k(p)
	= \frac{Z \, k^2}{1 + c \, \frac{p^2}{k^2}},
	\qquad
	\partial_t R_k(p)
	= k \, \partial_k R_k(p)
	= \frac{2 k^2 \, Z + k^2 \, \partial_t Z}{1 + c \, \frac{p^2}{k^2}} + \frac{2 c \, Z \, p^2}{\bigl(1 + c \, \frac{p^2}{k^2}\bigr)^2}.
\end{equation}
This will be our setup in the sequel.

\subsection{Feynman Rules}
\label{sec:feynman rules}

In the truncation \labelcref{eqn:ren gammak2} the (unrenormalized) propagator $\bar G_k = \bigl[\bar\Gamma_k^{(2)} + R_k\bigr]^{-1}$ reads
\begin{align}\label{eqn:feynman rule 1}
	&\begin{aligned}
		\begin{tikzpicture}
			\draw (-2.25,0) node[left] {$\varphi_a$} -- (2.25,0) node[right] {$\varphi_b$};
			\draw[->,yshift=5pt] (-2,0) -- (-1,0) node[midway,above] {$p_1$};
			\draw[->,yshift=5pt] (1,0) -- (2,0) node[midway,above] {$p_2$};
			\draw[fill=white,postaction={pattern=north east lines}] (0,0) circle (0.25) node[above=5pt] {$G_{k,ab}(p_1,p_2)$};
		\end{tikzpicture}
	\end{aligned}\notag
	\quad (p_1 = p_2)\\
	&\quad= \bigl[\bar\Gamma_k^{(2)} + R_k\bigr]_{ab}^{-1}(p_1,p_2)\bigr|_{\varphi_c}
	= \biggl[\frac{\delta^2 \Gamma_k[\varphi]}{\delta \bar\chi_a(p_1) \, \delta \bar\chi_b(-p_2)}\biggr|_{\varphi_c} + R_k(p_1) \, (2 \pi)^{d+1} \, \delta_{ab} \, \delta(p_1 - p_2)\biggr]^{-1}\\
	&\quad= \frac{\delta(p_1 - p_2)}{(2 \pi)^{d+1} \, Z_k} \begin{cases}
		\bigl[Z_1 \, p_1^2 - i s(p_{0,1}) \, \gamma_1^2 + U_k^\prime(\rho_c) + 2 \rho_c \, U_k^{\prime\prime}(\rho_c) + k^2/(1 + c \, p^2/k^2)\bigr]^{-1} & a = b = 1,\\[1ex]
		\bigl[p_1^2 + U_k^\prime(\rho_c) + k^2/(1 + c \, p^2/k^2)\bigr]^{-1} & a = b > 1,
	\end{cases}\notag
\end{align}
where we defined the Matsubara space Dirac delta as
\begin{equation}
	\delta(p - q)
	\equiv \delta_\text{M}^{(d+1)}(p - q)
	= \frac{T}{2 \pi} \, \delta_{m,n} \, \delta^{(d)}(\vec p - \vec q).
\end{equation}
In the zero-temperature limit, we have $\sum_{n \in \mathbb{Z}} \frac{T}{2 \pi} \xrightarrow{T \to 0} \int_{-\infty}^\infty \dif p_0$ and $\delta_{m,n} \xrightarrow{T \to 0} \delta(p_0 - q_0)$ so that
\begin{equation}
	\lim_{T \to 0} \delta_\text{M}^{(d+1)}(p - q)
	= \delta(p_0 - q_0) \, \delta^{(d)}(\vec p - \vec q).
\end{equation}
Since we will evaluate $n$-point functions for constant fields, we consider again the field configuration
\begin{equation}
	\varphi_a(p)
	= \varphi_c \, \delta_{a1} + \chi_a(p),
	\qquad
	\text{with } \chi_a(p) \ll \varphi_c \enskip \forall \, a,p.
\end{equation}
in which $\delta \varphi_a(p) = \delta \chi_a(p)$ and $\rho = \frac{1}{2} \sum_{a=1}^N \bigl(\varphi_c \, \delta_{a1} + \chi_a(p)\bigr)^2$. The 3-point function is then given by
\begin{align}\label{eqn:3pt function}
	&\begin{aligned}
		\begin{tikzpicture}
			\draw (-2,0) node[left] {$\varphi_a$} -- (0,0) -- (1.5,1.5) node[above right] {$\varphi_b$} (0,0) -- (1.5,-1.5) node[below right] {$\varphi_c$};
			\draw[->,yshift=5pt] (-1.7,0) -- (-0.7,0) node[midway,above] {$p_1$};
			\draw[<-,yshift=5pt] (0.4,0.4) -- (1.2,1.2) node[midway,above left] {$p_2$};
			\draw[<-,xshift=5pt] (0.4,-0.4) --  (1.2,-1.2) node[midway,above right] {$p_3$};
			\draw[fill=white,postaction={pattern=north east lines}] (0,0) circle (0.25) node[right=5pt] {$\Gamma_{k,abc}^{(3)}(p_1,p_2,p_3)$};
		\end{tikzpicture}
	\end{aligned}
	\qquad (p_1 + p_2 + p_3 = 0)\notag\\
	&\quad= \Gamma_{k,abc}^{(3)}(p_1,p_2,p_3)\bigr|_{\varphi_c}
	\overset{\smash[t]{\overset{\mathclap{\text{assuming momentum-independent vertices}}}{\bigl\downarrow}}}{=} \mmint_p \frac{\delta^3 U_k(\rho)}{\delta \chi_a(p_1) \, \delta \chi_b(p_2) \, \delta \chi_c(p_3)}\biggr|_{\varphi_c}\\
	&\quad= \mmint_p \Biggl[U_k^{\prime\prime\prime} \frac{\delta \rho}{\delta \chi_a(p_1)} \frac{\delta \rho}{\delta \chi_b(p_2)} \frac{\delta \rho}{\delta \chi_c(p_3)} + U_k^{\prime\prime} \frac{\delta^2 \rho}{\delta \chi_a(p_1) \, \delta \chi_b(p_2)} \frac{\delta \rho}{\delta \chi_c(p_3)} + U_k^{\prime\prime} \frac{\delta^2 \rho}{\delta \chi_a(p_1) \, \delta \chi_c(p_3)} \frac{\delta \rho}{\delta \chi_c(p_2)}\notag\\
	&\quad\hphantom{{}= \mmint_p \biggl[}+ U_k^{\prime\prime} \frac{\delta^2 \rho}{\delta \chi_b(p_2) \, \delta \chi_c(p_3)} \frac{\delta \rho}{\delta \chi_a(p_1)} + U_k^\prime \frac{\delta^3 \rho}{\delta \chi_a(p_1) \, \delta \chi_b(p_2) \, \delta \chi_c(p_3)}\Biggr]\Biggr|_{\mathrlap{\varphi_c}},\notag
\end{align}
The functional derivatives in \labelcref{eqn:3pt function} are
\begin{align}
	&\frac{\delta \rho\bigl(\chi(p)\bigr)}{\delta \chi_a(p_1)}\biggr|_{\varphi_c}
	= \bigl[\varphi_c \, \delta_{a1} + \chi_a(p)\bigr] \delta(p - p_1)\Bigr|_{\varphi_c}
	= \varphi_c \, \delta_{a1} \, \delta(p - p_1),\\
	&\frac{\delta^2 \rho}{\delta \chi_a(p_1) \, \delta \chi_b(p_2)}\biggr|_{\varphi_c}
	= \delta_{ab} \, \delta(p - p_1) \, \delta(p - p_2),\\
	&\frac{\delta^3 \rho}{\delta \chi_a(p_1) \, \delta \chi_b(p_2) \, \delta \chi_c(p_3)}\biggr|_{\varphi_c}
	= 0.
\end{align}
The Feynman rule for the 3-point function therefore reads
\begin{equation}\label{eqn:3pt function evaluated}
	\Gamma_{k,abc}^{(3)}(p_1,p_2,p_3)\bigr|_{\varphi_c}
	= \Bigl[\varphi_c^3 \, U_k^{\prime\prime\prime} \, \delta_{a1} \, \delta_{b1} \, \delta_{c1} + \varphi_c \, U_k^{\prime\prime} \Bigl(\delta_{a1} \, \delta_{bc} + \delta_{b1} \, \delta_{ac} + \delta_{c1} \, \delta_{ab}\Bigr)\Bigr] \delta(p_1 + p_2 + p_3).
\end{equation}
We can immediately read off two non-zero index combinations,
\begin{equation}
	\Gamma_{k,abc}^{(3)}(p_1,p_2,p_3)\bigr|_{\varphi_c}
	= \delta(p_1 + p_2 + p_3)
	\begin{cases}
		\varphi_c^3 \, U_k^{\prime\prime\prime} + 3 \varphi_c \, U_k^{\prime\prime} & a = b = c = 1,\\
		\varphi_c \, U_k^{\prime\prime} & a = 1,\ b = c \neq 1,\\
		0 & \text{else}.
	\end{cases}
\end{equation}
Else contains the cases
\begin{equation}
	\newcommand{\tpf}[3]{\tikz[inner sep=0,scale=0.25,baseline=-0.7ex] \filldraw (-1,0) node[left] {$#1$} -- (0,0) circle (0.2) -- (50:1) node[above right] {$#2$} (0,0) -- (-50:1) node[below right] {$#3$};}
	\tpf{1}{1}{a}
	\;=\; \tpf{1}{a}{b}
	\;=\; \tpf{a}{a}{a}
	\;=\; \tpf{a}{b}{b}
	\;=\; \tpf{a}{b}{c}
	\;=\; 0,\qquad
	\text{with $a \neq b \neq c \in \{2,\dots,N\}$.}
\end{equation}
Thus our theory features two kinds of 3-point interactions: one purely radial and one between a radial mode and any two identical Goldstone bosons. All other 3-point correlations vanish (within our approximation). We will see in \cref{sec:prop flow} that despite the hierarchy of flow equations (see \cref{itm:hierarchy} on \cpageref{itm:hierarchy}), higher vertices won't be necessary to compute the flow of parameters of the 2-point function under the assumption of momentum-independent vertices.

\subsection{Potential Flow}
\label{sec:mink potential flow}

Except for an additional discrete frequency dimension and the presence of the discontinuity $\gamma_1^2$, \cref{eqn:fourier quadratic part,eqn:ren gammak2} are strikingly similar. We can follow exactly the same steps as in \cref{sec:potential flow} to derive the flow of the effective potential. This time around, the result is
\begin{equation}\label{eqn:real time potential flow}
	\partial_t U_k\bigr|_{\rho_c}
	= \frac{1}{2} \mmint_p \frac{\partial_t R_k(p)}{Z_k} \biggl[\frac{1}{\bar M_1} + \frac{N-1}{\bar M_0}\biggr],
\end{equation}
with
\begin{equation}
	\bar M_1
	= Z_1 \, p^2 - i s(p_0) \, \gamma_1^2 + U_k^\prime + 2 \rho_c \, U_k^{\prime\prime} + \frac{R_k}{Z_k},
	\qquad
	\bar M_0
	= p^2 + U_k^\prime + \frac{R_k}{Z_k}.
\end{equation}
Note that \labelcref{eqn:goldstone expansion,eqn:radial expansion} are expansions around points on the real frequency axis. Thus \labelcref{eqn:real time potential flow} looses all validity when taken too far from the real axis. In particular, we should not evaluate it for large imaginary values $p_0 = 2 \pi i T n$ with $n \gg 1$. Fortunately, performing the Matsubara summation $\sum_{p_0}$ in \labelcref{eqn:real time potential flow} relies on contour integration methods. Due to the analytic structure of the propagator $G(p)$ discussed in \cref{sec:analytic structure}, this means we only need to evaluate residues and integrals along branch cuts on the real frequency axis (or close to it for $k > 0$) where \labelcref{eqn:real time potential flow} is valid.

By modifying the threshold functions $\tfie_j$ in \cref{eqn:threshold} for renormalized quasi-particles in $d + 1$ dimensions \cite{floerchinger2012analytic,floerchinger2017unpublished},
\begin{mleqn}\label{eqn:real time threshold fcts}
	\tfim_j(Z,z,m^2,\gamma^2,R)
	&= (\delta_{j0} - j) \mmint_p \frac{1/Z \, \partial_t R}{(z \, p^2 + m^2 - i s(p_0) \, \gamma^2 + R/Z)^{j+1}}\\
	&= \tilde\partial_t \mmint_p \begin{cases}
		\ln\bigl(z \, p^2 + m^2 - i s(p_0) \, \gamma^2 + R/Z\bigr) & j = 0,\\
		\bigl(z \, p^2 + m^2 - i s(p_0) \, \gamma^2 + R/Z\bigr)^{-j} & j \geq 1,
	\end{cases}
\end{mleqn}
we can rewrite \labelcref{eqn:real time potential flow} as in \labelcref{eqn:potential flow ito i},
\begin{equation}\label{eqn:potential flow ito ibar}
	\partial_t U_k\bigr|_{\rho_c}
	= \tfrac{1}{2} \tfim_0\bigl(Z_k,Z_1,\,U_k^\prime + 2 \rho_c U_k^{\prime\prime},\,\gamma_1^2,R_k\bigr) + \tfrac{1}{2} (N - 1) \, \tfim_0\bigl(Z_k,1,\,U_k^\prime,\,0,\, R_k\bigr).
\end{equation}
Neglecting again the (subleading) $\rho$-dependence of $Z_1$ and $\gamma_1^2$, the flow equations for the derivatives of $U_k(\rho)$ expanded as in \labelcref{eqn:eff pot exp} and evaluated at $\rho_c = \rho_0$ read
\begin{align}
	&\begin{aligned}
		\partial_t U_k^\prime\bigr|_{\rho_0}
		&= \tfrac{1}{2} \bigl(3 U_k^{\prime\prime} + 2 \rho_0 U_k^{(3)}\bigr) \, \tfim_1\bigl(Z_k,\, Z_1,\, U_k^\prime + 2 \rho_0 U_k^{\prime\prime},\, \gamma_1^2,\, R_k\bigr)\\
		&\hphantom{{}=}+ \tfrac{1}{2} (N - 1) \, U_k^{\prime\prime} \, \tfim_1\bigl(Z_k,\, 1,U_k^\prime, 0,\, R_k\bigr),
	\end{aligned}\\
	&\begin{aligned}
		\partial_t U_k^{\prime\prime}\bigr|_{\rho_0}
		&= -\tfrac{1}{2} \bigl(3 U_k^{\prime\prime} + 2 \rho_0 U_k^{(3)}\bigr)^2 \, \tfim_2\bigl(Z_k,\, Z_1,\, U_k^\prime + 2 \rho_0 U_k^{\prime\prime},\, \gamma_1^2,\, R_k\bigr)\\
		&\hphantom{{}=}- \tfrac{1}{2} (N - 1) \, \bigl(U_k^{\prime\prime}\bigr)^2 \, \tfim_2\bigl(Z_k,\, 1,U_k^\prime, 0,\, R_k\bigr)\\
		&\hphantom{{}=}+ \tfrac{1}{2} \bigl(5 U_k^{(3)} + 2 \rho_0 U_k^{(4)}\bigr) \, \tfim_1\bigl(Z_k,\, Z_1,\, U_k^\prime + 2 \rho_0 U_k^{\prime\prime},\, \gamma_1^2,\, R_k\bigr)\\
		&\hphantom{{}=}+ \tfrac{1}{2} (N - 1) \, U_k^{(3)} \, \tfim_1\bigl(Z_k,\, 1,U_k^\prime, 0,\, R_k\bigr).
	\end{aligned}
\end{align}
from which we obtain by the same steps taken in \cref{sec:potential flow} the following flow equations for the mass squared $m^2$, the location of the minimum $\rho_0(k)$ and the quartic coupling $\lambda$,
\begin{align}
	\label{eqn:mk mink flow}
	\partial_t m^2
	&= \eta \, m^2 + \frac{\lambda}{2} \, (N + 2) \, \tfim_1\bigl(Z_k,\, 1,\, m^2,\, 0,\, R_k\bigr),\\[1ex]
	\partial_t \rho_0
	\label{eqn:rhok mink flow}
	&= -\eta \, \rho_0 - \frac{1}{2} \, \Bigl[3 \,\tfim_1\bigl(Z_k,\, Z_1,\, 2 \rho_0 \lambda,\, \gamma_1^2,\, R_k\bigr) + (N - 1) \, \tfim_1\bigl(Z_k,\, 1,0,0,\, R_k\bigr)\Bigr],\\[1ex]
	\label{eqn:lambdak mink flow}
	\partial_t \lambda
	&= 2 \eta \, \lambda - \frac{\lambda^2}{2} \, \Bigl[9 \, \tfim_2\bigl(Z_k,\, Z_1,\, 2 \rho_0 \lambda,\, \gamma_1^2,\, R_k\bigr) + (N - 1) \, \tfim_2\bigl(Z_k,\, 1,0,0,\, R_k\bigr)\Bigr].
\end{align}

\subsection{Propagator Flow}
\label{sec:prop flow}

To close the system of differential equations \crefrange{eqn:mk mink flow}{eqn:lambdak mink flow}, we need three additional flow equations for the remaining scale-dependent parameters $Z_k$, $\gamma_1^2$ and $Z_1$ that are incorporated self-consistently in our truncation. Our course of action will be to calculate the flow of $\smash{\Gamma_k^{(2)}}$ and then project it to $\partial_t Z_k$, $\partial_t \gamma_1^2$ and $\partial_t Z_1$.

To derive the unrenormalized 2-point function $\bar\Gamma_k^{(2)} = \Gamma_k^{(2)}/Z_k$, we expand $\Gamma_k$ in powers of $\bar\chi_a(p)$, take two functional derivatives with respect to the unrenormalized fluctuating fields $\bar\chi_a(p)$ and evaluate the result for constant fields $\varphi(x) = \varphi_c$, i.e. $\chi_a(p) = 0 \enskip \forall \, a$, resulting in\footnote{The two functional derivatives strip away all terms less than quadratic in the fluctuating fields. Setting $\chi_a(p) = 0$ afterwards removes any terms higher than quadratic. So we only need to consider the quadratic fluctuations \cref{eqn:ren gammak2} to obtain the most general form of $\smash{\Gamma_k^{(2)}\bigr|_{\varphi_c}}$ (cf. \cref{eqn:fourier quadratic part,eqn:const field 2-point fct})}
\begin{align}
	&\bar\Gamma_{k,ab}^{(2)}(p,q)\Bigr|_{\varphi_c}
	= \frac{\delta^2 \Gamma_{k,2}}{\delta \bar\chi_a(p) \, \delta \bar\chi_b(-q)}\\
	&\crefrel{eqn:ren gammak2}{=} Z_k \biggl\{\delta_{ab} \Bigl[p^2 + U_k^\prime(\rho_c)\Bigr] + \delta_{a1} \delta_{b1} \Bigl[(Z_1 - 1) \, p^2 - i s(p_0) \, \gamma_1^2 + 2 \rho_c U_k^{\prime\prime}(\rho_c)\Bigr]\biggr\} \frac{\delta(p - q)}{(2 \pi)^{d+1}}\notag
\end{align}
The $a = b = 1$-component governs propagation of the radial mode $\varphi_1$ (equal incoming and outgoing momenta ensure momentum conservation),
\begin{equation}\label{eqn:radial 2-point fct}
	\bar\Gamma_{k,11}^{(2)}(q,q)\Bigr|_{\varphi_c}
	= \bar{P}_r(q) \, \frac{T}{(2 \pi)^d} \, \delta(0),
\end{equation}
where the inverse propagator of the radial mode is given by
\begin{equation}\label{eqn:inv radial prop}
	\bar{P}_r(q)
	= \bar{G}_r^{-1}(q)
	= Z_k \Bigl[Z_1 \, q^2 - i s(q_0) \, \gamma_1^2 + U_k^\prime(\rho) + 2 \rho \, U_k^{\prime\prime}(\rho)\Bigr].
\end{equation}
To project onto $Z_1$, we take the $q^2$-derivative of \labelcref{eqn:inv radial prop},
\begin{equation}\label{eqn:z1}
	Z_1
	= \frac{1}{Z_k} \, \partial_{q^2} \, \bar{P}_r(q).
\end{equation}
($Z_1$ and $\gamma_1^2$ are $q$-independent within our approximation since we expanded the inverse propagator around the radial mode's on-shell energy $q^2 = -m_1^2$, cf. \cref{eqn:z1 gamma1}.) The $t$-derivative of \labelcref{eqn:z1} yields
\begin{equation}\label{eqn:z1 scale dep}
	\partial_t Z_1
	= -\frac{1}{Z_k^2} \, (\partial_t Z_k) \, \partial_{q^2} \, \bar{P}_r(q)\ + \frac{1}{Z_k} \, \partial_t \, \partial_{q^2} \, \bar{P}_r(q)
	= \eta \, Z_1 + \frac{1}{Z_k} \, \partial_t \, \partial_{q^2} \, \bar{P}_r(q).
\end{equation}
Similarly, we can project onto $\gamma_1^2$ \cite{floerchinger2012analytic},
\begin{equation}\label{eqn:g1}
	\gamma_1
	= \frac{1}{Z_k} \, \disc_{q_0} \, \bar{P}_r(q)
\end{equation}
and so
\begin{equation}\label{eqn:g1 scale dep}
	\partial_t \gamma_1
	= \eta \, \gamma_1 + \frac{1}{Z_k} \, \partial_t \, \disc_{q_0} \, \bar{P}_r(q),
\end{equation}
where the discontinuity projector $\disc_{q_0}$ is defined as
\begin{equation}
	\disc_{q_0} f(x)
	= \frac{i}{2} \, \sign(q_0) \, \smashoperator{\lim_{\epsilon \to 0^+}} \, \Bigl[f(q_0 + i \epsilon) - f(q_0 - i \epsilon)\Bigr],
\end{equation}
Propagation of the Goldstone bosons is governed by components of $\Gamma_k^{(2)}$ with $a = b > 1$,
\begin{equation}\label{eqn:goldstone 2-point fct}
	\bar\Gamma_{k,22}^{(2)}(q,q)\Bigr|_{\varphi_c}
	= \, \bar{P}_g(q) \, \frac{T}{(2 \pi)^d} \, \delta(0),
\end{equation}
with inverse Goldstone propagator
\begin{equation}\label{eqn:inv goldstone prop}
	\bar{P}_g(q)
	= \bar G_g^{-1}(q)
	= Z_k \bigl[q^2 + U_k^\prime(\rho)\bigr].
\end{equation}
so that
\begin{equation}\label{eqn:zk scale dep}
	\partial_t Z_k
	= \partial_t \, \partial_{q^2} \bar{P}_g(q).
\end{equation}
The flow equations for $Z_1$, $\gamma_1^2$ and $Z_k$ can therefore be derived by using Wetterich's equation to determine an algebraic expression for $\smash{\partial_t \Gamma_k^{(2)}}$ and inserting that into \labelcref{eqn:z1 scale dep,eqn:g1 scale dep,eqn:zk scale dep}. A step by step prescription for the construction of flow equations for $n$-point functions goes as follows \cite{berges2002non}:
\begin{enumerate}
	\item Write down all one-loop Feynman diagrams obtained by taking $n$ functional derivatives of \labelcref{eqn:wetterich eqn}. These diagrams incorporate all quantum fluctuations contributing to the scale dependence of $\smash{\Gamma^{(n)}}$. Alternatively, since functional and scale derivatives commute, we can perform the functional derivatives after casting \labelcref{eqn:wetterich eqn} into a form particularly suited for taking derivatives,
	\begin{equation}\label{eqn:wetterich eqn cutoff undone}
		\partial_t \Gamma_k
		= \frac{1}{2} \Tr \tilde\partial_t \ln(\Gamma_k^{(2)} + R_k)
		= \tilde\partial_t \Gamma_k[\varphi]\bigr|_\text{1-loop},
	\end{equation}
	with cutoff derivative $\tilde\partial_t = \partial_t|_{\Gamma_k^{(2)}}$ and renormalization group-improved one-loop contribution to the flowing action $\Gamma_k[\varphi]\bigr|_\text{1-loop} = S[\varphi] + \frac{1}{2} \Tr\ln(\Gamma_k^{(2)} + R_k)$ (cf. \cref{eqn:one-loop gamma}).
		
	\item\label{itm:vertices} Insert $n$th functional derivatives of $\Gamma_k$ for all $n$-point functions with $n \geq 3$. It is important to be aware that such vertices may be momentum-dependent even if they were constant on the classical level (i.e. as functional derivatives of the microscopic action $S = \Gamma_\Lambda$). Unless prohibited by symmetries there will also be contributions from higher vertices that are absent in $S$. This is a consequence of the partial integration of momentum modes with $q^2 > k^2$ in $\Gamma_k$.
	
	\item Insert for all propagator lines the full regularized propagator $\smash{G_k = (\Gamma_k^{(2)} + R_k)^{-1}}$ (evaluated at fixed field $\varphi_c$).
	
	\item If we used \labelcref{eqn:wetterich eqn cutoff undone}, reapply the $\tilde\partial_t$-derivative which acts on the integrand of the one-loop momentum integrals. This will increase the number of diagrams since it generates multiple diagrams with identical topology but regulator insertions attached to different internal lines \cite{huber2012algorithmic}. It also renders momentum integrals both UV and IR finite. The resulting exact flow equation for $\smash{\Gamma_k^{(n)}}$ is therefore fully regularized.
\end{enumerate}
An example of this procedure, the case $n = 2$ for a scalar theory, was presented in \cref{eqn:hierarchy,eqn:graphical hierarchy} using a simplified notation that suppresses momentum arguments and field indices. To gain a more detailed understanding of the type of fluctuations that generate the flow of $\smash{\Gamma_k^{(2)}}$, we rederive the result here with indices and momenta reinstated. For the one-point function we get\footnote{A primed integration variable indicates that it does not include the factor $T/(2 \pi)^d$ as in \cref{eqn:mmint def}, but rather $\mmint_{p^\prime} = \sum_{p_0} \int_{\reals^d} \dif^d p$ as appropriate for the functional chain rule.}
\begin{mleqn}\label{eqn:one-point flow}
	\partial_t \Gamma_{k,a}^{(1)}(q)
	&= \frac{\delta}{\delta \varphi_a(q)} \, \frac{1}{2} \Tr\Biggl[\frac{\partial_t R_k}{\Gamma_k^{(2)} + R_k}\Biggr]= \frac{\delta}{\delta \varphi_a(q)} \, \frac{1}{2} \sum_{i,j=1}^N \mmint_{p_1,p_2} \, \frac{\partial_t R_{k,ij}(p_1,p_2)}{\Gamma_{k,ji}^{(2)}(p_2,p_1) + R_{k,ji}(p_2,p_1)}\\
	&= -\frac{1}{2} \sum_{\substack{i,j\\k,l}}^N \mmint_{\substack{p_1,p_2\\p_3^\prime,p_4^\prime}} \frac{\partial_t R_{k,ij}(p_1,p_2)}{\Gamma_{k,jk}^{(2)}(p_2,p_3) + R_{k,jk}(p_2,p_3)} \, \frac{\Gamma_{k,akl}^{(3)}(q,p_3,p_4)}{\Gamma_{k,li}^{(2)}(p_4,p_1) + R_{k,li}(p_4,p_1)}.
\end{mleqn}
The integrand in \labelcref{eqn:one-point flow} corresponds to the diagram
\begin{mleqn}
\begin{tikzpicture}[pin edge={shorten <=5*\lrad}]
	\def\lrad{1}
	\def\mrad{0.175*\lrad}
	\def\srad{0.15*\lrad}
    
    \draw[loop/.list={{0.125}{2},{0.375}{3},{0.625}{4},{0.875}{1}}] (0,0) circle (\lrad);
    \draw[cross] (\lrad,0) circle (\srad) node[right=6pt] {$\partial_k R_{k,ij}(p_1,p_2)$};
	\draw[dressed] (0,\lrad) circle (\srad) node[above=3pt] {$G_{k,jk}(p_2,p_3)$};
	\draw[dressed] (0,-\lrad) circle (\srad) node[below=3pt] {$G_{k,li}(p_4,p_1)$};
        
    \draw (-2*\lrad,0) -- (-\lrad,0) node[pos=0.4,below] {$\varphi_a$};
	\draw[momentum] (-2*\lrad,0) -- (-1.25*\lrad,0) node[midway,above] {$q$};
	
	\node (Gkakl) at (-2*\lrad,\lrad) {$\Gamma_{k,akl}^{(3)}(q,p_3,-p_4)$};
	\draw[label] (Gkakl.-30) -- (-\lrad,0);
	\draw[dressed] (-\lrad,0) circle (\mrad);
    
\end{tikzpicture}
\end{mleqn}
The sign as well as the additional index summation over $k$, $l$ and momentum integration over $p_3$, $p_4$ in \labelcref{eqn:one-point flow} stem from the functional chain rule,
\begin{mleqn}
	\frac{\delta \bigl[\Gamma_{k,bc}^{(2)}(q_2,q_3)\bigr]^{-1}}{\delta \varphi_a(q_1)}
	&= \smashoperator[l]{\sum_{i,j=1}^N} \mmint_{p_1^\prime,p_2^\prime} \frac{\delta [\Gamma_{k,bc}^{(2)}(q_2,q_3)]^{-1}}{\delta \Gamma_{k,ij}^{(2)}(p_1,p_2)} \frac{\delta \Gamma_{k,ij}^{(2)}(p_1,p_2)}{\delta \varphi_a(q_1)}\\
	&= -\sum_{\substack{i,j\\k,l}}^N \mmint_{\substack{p_1^\prime,p_2^\prime\\p_3^\prime,p_4^\prime}} \frac{1}{\Gamma_{k,bk}^{(2)}(q_2,p_3)} \frac{\delta \Gamma_{k,kl}^{(2)}(p_3,p_4)}{\delta \Gamma_{k,ij}^{(2)}(p_1,p_2)} \frac{1}{\Gamma_{k,lc}^{(2)}(p_4,q_3)} \, \Gamma_{k,aij}^{(3)}(q_1,p_1,p_2)\\
	&= -\smashoperator[l]{\sum_{i,j=1}^N} \mmint_{p_1^\prime,p_2^\prime} \frac{\Gamma_{k,aij}^{(3)}(q_1,p_1,p_2)}{\Gamma_{k,bi}^{(2)}(q_2,p_1) \, \Gamma_{k,cj}^{(2)}(p_2,q_3)}.
\end{mleqn}
Taking a further derivative $\delta/\delta \varphi_b(q_2)$ of \labelcref{eqn:one-point flow} yields the flow equation for the 2-point function,
\begin{align}\label{eqn:2-point flow}
	\partial_t &\Gamma_{k,ab}^{(2)}(q_1,q_2)
	= \frac{1}{2} \, \smashoperator[l]{\sum_{\substack{i,j,k\\l,m,n}}^N} \smash{\smashoperator[r]{\mmint_{\substack{p_1,p_2,p_3^\prime\\p_4^\prime,p_5^\prime,p_6^\prime}}}} \partial_t R_{k,ij}(p_1,p_2)\\
	\label{eqn:2-point t1}
	&\times \biggl(G_{k,jk}(p_2,p_3) \, \Gamma_{k,akl}^{(3)}(q_1,p_3,-p_4) \, G_{k,lm}(p_4,p_5) \, \Gamma_{k,bmn}^{(3)}(-q_2,p_5,-p_6) \, G_{k,ni}(p_6,p_1)\\
	\label{eqn:2-point t2}
	&\hphantom{\times \biggl[}+ G_{k,jm}(p_2,p_5) \, \Gamma_{k,bmn}^{(3)}(-q_2,p_5,-p_6) \, G_{k,nk}(p_6,p_3) \, \Gamma_{k,akl}^{(3)}(q_1,p_3,-p_4) \, G_{k,li}(p_4,p_1)\\
	\label{eqn:2-point t3}
	&\hphantom{\times \biggl[}- G_{k,jk}(p_2,p_3) \, \Gamma_{k,abkl}^{(4)}(q_1,-q_2,p_3,-p_4) \, G_{k,li}(p_4,p_1)\biggr),
\end{align}
where
\begingroup\allowdisplaybreaks
\begin{align}
	&\labelcref{eqn:2-point t1}
	= \begin{aligned}
		\begin{tikzpicture}
		    \def\lrad{5/4}
			\def\mrad{0.15*\lrad}
			\def\srad{0.1*\lrad}
		    \draw[loop/.list={{0.0625}{6},{0.1875}{1},{0.3125}{2},{0.4375}{3},{0.625}{4},{0.875}{5}}] (0,0) circle (\lrad);
		    \draw[cross] (0,\lrad) circle (\srad) node[above=5pt] {$\partial_k R_{k,ij}(p_1,p_2)$};
			\draw[dressed] (135:\lrad) circle (\srad) node[above left] {$G_{k,jk}(p_2,p_3)$};
			\draw[dressed] (45:\lrad) circle (\srad) node[above right] {$G_{k,ni}(p_6,p_1)$};
			\draw[dressed] (0,-\lrad) circle (\srad) node[below=3pt] {$G_{k,lm}(p_4,p_5)$};
		    \draw (-2*\lrad,0) coordinate (xl) -- (-\lrad,0) node[pos=0.4,below] {$\varphi_a$};
		    \draw[momentum] (-2*\lrad,0) -- (-1.25*\lrad,0) node[midway,above] {$q_1$};
		    \draw (\lrad,0) -- (2*\lrad,0) coordinate (xr) node[pos=0.6,below] {$\varphi_b$};
		    \draw[momentum] (1.25*\lrad,0) -- (2*\lrad,0) node[midway,above] {$q_2$};
		    \node at (-1.6*\lrad,-\lrad) (Gkail) {$\Gamma_{k,akl}^{(3)}(q_1,p_3,-p_4)$};
		    \draw[label] (Gkail) -- (-\lrad,0);
			\draw[dressed] (-\lrad,0) circle (\mrad);
		    \node at (2*\lrad,-\lrad) (Gkbde) {$\Gamma_{k,bmn}^{(3)}(-q_2,p_5,-p_6)$};
		    \draw[label] (Gkbde.150) -- (\lrad,0);
			\draw[dressed] (\lrad,0) circle (\mrad);
		\end{tikzpicture}
	\end{aligned},\\
	&\labelcref{eqn:2-point t2}
	= \begin{aligned}
		\begin{tikzpicture}
		    \def\lrad{5/4}
			\def\mrad{0.15*\lrad}
			\def\srad{0.1*\lrad}
		    \draw[loop/.list={{0.125}{6},{0.375}{3},{0.5625}{4},{0.6875}{1},{0.8125}{2},{0.9375}{5}}] (0,0) circle (\lrad);
		    \draw[cross] (0,-\lrad) circle (\srad) node[below=5pt] {$\partial_k R_{k,ij}(p_1,p_2)$};
			\draw[dressed] (-45:\lrad) circle (\srad) node[below right] {$G_{k,jm}(p_2,p_5)$};
			\draw[dressed] (-135:\lrad) circle (\srad) node[below left] {$G_{k,li}(p_4,p_1)$};
			\draw[dressed] (0,\lrad) circle (\srad) node[above=3pt] {$G_{k,nk}(p_6,p_3)$};
		    \draw (-2*\lrad,0) coordinate (xl) -- (-\lrad,0) node[pos=0.4,below] {$\varphi_a$};
		    \draw[momentum] (-2*\lrad,0) -- (-1.25*\lrad,0) node[midway,above] {$q_1$};
		    \draw (\lrad,0) -- (2*\lrad,0) coordinate (xr) node[pos=0.6,below] {$\varphi_b$};
		    \draw[momentum] (1.25*\lrad,0) -- (2*\lrad,0) node[midway,above] {$q_2$};
		    \node at (-1.8*\lrad,\lrad) (Gkail) {$\Gamma_{k,akl}^{(3)}(q_1,p_3,-p_4)$};
		    \draw[label] (Gkail) -- (-\lrad,0);
			\draw[dressed] (-\lrad,0) circle (\mrad);
		    \node at (1.8*\lrad,\lrad) (Gkbde) {$\Gamma_{k,bmn}^{(3)}(-q_2,p_5,-p_6)$};
		    \draw[label] (Gkbde) -- (\lrad,0);
			\draw[dressed] (\lrad,0) circle (\mrad);
		\end{tikzpicture}
	\end{aligned},\\
	\label{eqn:2-point d3}
	&\labelcref{eqn:2-point t3}
	= \begin{aligned}
		\begin{tikzpicture}
		    \def\lrad{5/4}
			\def\mrad{0.15*\lrad}
			\def\srad{0.1*\lrad}
		    \draw[loop/.list={{0.125}{1},{0.375}{2},{0.625}{3},{0.875}{4}}] (0,0) circle (\lrad);
		    \draw[cross] (0,\lrad) circle (\srad) node[above=5pt] {$\partial_k R_{k,ij}(p_1,p_2)$};
		    \draw[dressed] (-\lrad,0) circle (\srad) node[left=2pt] {$G_{k,jk}(p_2,p_3)$};
			\draw[dressed] (\lrad,0) circle (\srad) node[right=2pt] {$G_{k,li}(p_4,p_1)$};
		    \draw (-2.2*\lrad,-\lrad) -- (2.2*\lrad,-\lrad) node[pos=0.1,below] {$\varphi_a$} node[pos=0.9,below] {$\varphi_b$};
			\draw[momentum] (-2*\lrad,-\lrad) -- (-\lrad,-\lrad) node[midway,above] {$q_1$};
			\draw[momentum] (\lrad,-\lrad) -- (2*\lrad,-\lrad) node[midway,above] {$q_2$};
			\draw[dressed] (0,-\lrad) circle (\mrad) node[below] {$\Gamma_{k,abkl}^{(4)}(q_1,-q_2,p_3,-p_4)$};
		\end{tikzpicture}
	\end{aligned}.
\end{align}
\endgroup
(Sums over indices and integrals over momenta apply only to diagrams in which they appear, i.e. $m$, $n$, $p_5$ and $p_6$ are not traced in the third diagram.)

By undoing the cutoff derivative, \labelcref{eqn:2-point flow} reduces to
\begingroup\allowdisplaybreaks
\begin{align}\label{eqn:2-point flow undone cutoff}
	\partial_t \Gamma_{k,ab}^{(2)}(q_1,q_2)
	&= \frac{1}{2} \sum_{\substack{i,j\\k,l}}^N \mmint_{\substack{p_1,p_2\\p_3^\prime,p_4^\prime}} \tilde\partial_t \biggl[G_{k,ij}(p_1,p_2) \, \Gamma_{k,abji}^{(4)}(q_1,-q_2,p_1,-p_2)\notag\\
	&\hphantom{{}=}- G_{k,ij}(p_1,p_2) \, \Gamma_{k,akl}^{(3)}(q_1,p_2,-p_3) \, G_{k,kl}(p_3,p_4) \, \Gamma_{k,bli}^{(3)}(-q_2,-p_1,p_4)\biggr]\\
	&= \frac{1}{2} \sum_{\substack{i,j\\k,l}}^N \mmint_{\substack{p_1,p_2\\p_3^\prime,p_4^\prime}} \tilde\partial_t \Biggl(\hspace{-4ex} \raisebox{-10.5ex}{
	\begin{tikzpicture}
		\def\lrad{1}
		\def\mrad{0.175*\lrad}
		\def\srad{0.15*\lrad}
		\draw[loop/.list={{0}{1},{0.5}{2}}] (0,0) circle (\lrad);
		\draw[dressed] (0,\lrad,0) circle (\srad) node[above=2pt] {$G_{k,ij}(p_1,p_2)$};
		\draw (-2*\lrad,-\lrad) node[left] {$\varphi_a$} -- (2*\lrad,-\lrad) node[right] {$\varphi_b$};
		\draw[momentum] (-2*\lrad,-\lrad) -- (-\lrad,-\lrad) node[midway,above] {$q_1$};
		\draw[momentum] (\lrad,-\lrad) -- (2*\lrad,-\lrad) node[midway,above] {$q_2$};
		\draw[dressed] (0,-\lrad) circle (\mrad) node[below] {$\Gamma_{k,abji}^{(4)}(q_1,-q_2,-p_1,p_2)$};
	\end{tikzpicture}
	}\hspace{-6ex} - \hspace{-5ex}\raisebox{-10ex}{
	\begin{tikzpicture}
		\def\lrad{1}
		\def\mrad{0.175*\lrad}
		\def\srad{0.15*\lrad}
		\draw[loop/.list={{0.125}{1},{0.375}{2},{0.625}{3},{0.875}{4}}] (0,0) circle (\lrad);
		\draw[dressed] (0,\lrad) circle (\srad) node[above=2pt] {$G_{k,ij}(p_1,p_2)$};
		\draw[dressed] (0,-\lrad) circle (\srad) node[below=3pt] {$G_{k,kl}(p_3,p_4)$};
		\draw (-2*\lrad,0) coordinate (xl) -- (-\lrad,0) node[pos=0.4,below] {$\varphi_a$};
		\draw[momentum] (-2*\lrad,0) -- (-1.25*\lrad,0) node[midway,above] {$q_1$};
		\draw (\lrad,0) -- (2*\lrad,0) coordinate (xr) node[pos=0.6,below] {$\varphi_b$};
		\draw[momentum] (1.25*\lrad,0) -- (2*\lrad,0) node[midway,above] {$q_2$};
		\node at (-2.1*\lrad,\lrad) (Gkajk) {$\Gamma_{k,ajk}^{(3)}(q_1,p_2,-p_3)$};
		\draw[label] (Gkajk.-30) -- (-\lrad,0);
		\draw[dressed] (-\lrad,0) circle (\mrad);
	    \node at (2.1*\lrad,\lrad) (Gkbli) {$\Gamma_{k,bli}^{(3)}(-q_2,-p_1,p_4)$};
		\draw[label] (Gkbli.-150) -- (\lrad,0);
		\draw[dressed] (\lrad,0) circle (\mrad);
	\end{tikzpicture}
	}\hspace{-9ex}\Biggr).\notag
\end{align}
\endgroup
Despite what we said in \cref{itm:vertices} on \cpageref{itm:vertices}, we will now make the simplifying assumption of momentum-independent vertices. This has two important consequences.
\begin{enumerate}
	\item The effective vertices $\Gamma_k^{(n)}$ follow more easily as field derivatives not of the entire effective action $\Gamma_k$ but only its momentum-independent part, i.e. the effective potential $U_k(\rho)$.
	\item\label{itm:q-indep diagrams} Since we are working with amputated diagrams, neglecting the momentum-dependence of vertices renders \labelcref{eqn:2-point d3} and the first diagram in \labelcref{eqn:2-point flow undone cutoff} $q$-independent. As a result, it drops out of all flow equations upon projecting with $\partial_{q^2}$ or $\disc_{q_0}$ so we don't need to consider it (nor the 4-point function) further.
\end{enumerate}
Using the Feynman rules derived in \cref{sec:feynman rules}, we can assemble an algebraic expression for the r.h.s. of \labelcref{eqn:2-point flow undone cutoff}. For $a = b = 1$ and $q_1 = q_2 = q$ (as appropriate for quantum fluctuations generating the scale dependence of $Z_1$ and $\gamma_1^2$), the second diagram evaluates to
\begin{mleqn}\label{eqn:a1b1 flow}
	&-\frac{1}{2} \sum_{\substack{i,j\\k,l}}^N \mmint_{\substack{p_1,p_2\\p_3^\prime,p_4^\prime}} \hspace{-9ex}\raisebox{-10ex}{
	\begin{tikzpicture}
		\def\lrad{1}
		\def\mrad{0.175*\lrad}
		\def\srad{0.15*\lrad}
		\draw[loop/.list={{0.125}{1},{0.375}{2},{0.625}{3},{0.875}{4}}] (0,0) circle (\lrad);
		\draw[dressed] (0,\lrad) circle (\srad) node[above=2pt] {$G_{k,ij}(p_1,p_2)$};
		\draw[dressed] (0,-\lrad) circle (\srad) node[below=3pt] {$G_{k,kl}(p_3,p_4)$};
		\draw (-2*\lrad,0) coordinate (xl) -- (-\lrad,0) node[pos=0.4,below] {$\varphi_a$};
		\draw[momentum] (-2*\lrad,0) -- (-1.25*\lrad,0) node[midway,above] {$q_1$};
		\draw (\lrad,0) -- (2*\lrad,0) coordinate (xr) node[pos=0.6,below] {$\varphi_b$};
		\draw[momentum] (1.25*\lrad,0) -- (2*\lrad,0) node[midway,above] {$q_2$};
		\node at (-2.1*\lrad,\lrad) (Gkajk) {$\Gamma_{k,ajk}^{(3)}(q_1,p_2,-p_3)$};
		\draw[label] (Gkajk.-30) -- (-\lrad,0);
		\draw[dressed] (-\lrad,0) circle (\mrad);
	    \node at (2.1*\lrad,\lrad) (Gkbli) {$\Gamma_{k,bli}^{(3)}(-q_2,-p_1,p_4)$};
		\draw[label] (Gkbli.-150) -- (\lrad,0);
		\draw[dressed] (\lrad,0) circle (\mrad);
	\end{tikzpicture}}
	(a = b = 1, \enskip q_1 = q_2 = q)\\
	&= -\frac{1}{2} \Bigl(\varphi_c^3 \, U_k^{\prime\prime\prime} + 3 \varphi_c \, U_k^{\prime\prime}\Bigr)^2 \mmint_p G_r(p) \, G_r(q + p) - \frac{1}{2} (N - 1) \varphi_c^2 \, \bigl(U_k^{\prime\prime}\bigr)^2 \mmint_p G_g(p) \, G_g(q + p),
\end{mleqn}
where $G_r$ and $G_g$ denote the (renormalized) radial and Goldstone propagators,
\begin{mleqn}
	G_r(p)
	&= \frac{1}{Z_1 p^2 + U_k^\prime(\rho_c) + 2 \rho_c \, U_k^{\prime\prime}(\rho_c) - i s(p_0) \, \gamma_1^2 + R_k(p)/Z_k},\\
	G_g(p)
	&= \frac{1}{p^2 + U_k^\prime(\rho_c) + R_k(p)/Z_k}
\end{mleqn}
The index summation and momentum integration in \labelcref{eqn:a1b1 flow} were carried out separately, the former yielding
\begin{mleqn}
	\smashoperator{\sum_{i,j,k,l=1}^N} \; &G_{ij} \Bigl[\varphi_c^3 \, U_k^{\prime\prime\prime} \, \delta_{11} \, \delta_{j1} \, \delta_{k1} + \varphi_c \, U_k^{\prime\prime} \Bigl(\delta_{11} \, \delta_{jk} + \delta_{j1} \, \delta_{k1} + \delta_{k1} \, \delta_{j1}\Bigr)\Bigr]\\
	&\times G_{kl} \Bigl[\varphi_c^3 \, U_k^{\prime\prime\prime} \, \delta_{11} \, \delta_{i1} \, \delta_{l1} + \varphi_c \, U_k^{\prime\prime} \Bigl(\delta_{11} \, \delta_{il} + \delta_{i1} \, \delta_{l1} + \delta_{i1} \, \delta_{l1}\Bigr)\Bigr]\\
	&= \Bigl(\varphi_c^3 \, U_k^{\prime\prime\prime} + 3 \varphi_c \, U_k^{\prime\prime}\Bigr)^2 G_r^2 + (N - 1) \varphi_c^2 \, \bigl(U_k^{\prime\prime}\bigr)^2 G_g^2,
\end{mleqn}
while for the latter, three of the four momentum integrals are trivial,
\begin{mleqn}\label{eqn:integrating p2-p4}
	\smash{\mmint_{\substack{p_1,p_2\\p_3^\prime,p_4^\prime}}} \delta(q + p_2 - p_3) \, G(p_1) \, \delta(p_1 - p_2) \, \delta(-q - p_1 + p_4) \, G(p_3) \, \delta(p_3 - p_4)
	= \frac{T \, \delta(0)}{(2 \pi)^d} \mmint_p G(p) \, G(q + p).
\end{mleqn}
Notice the $T \, \delta(0)/(2 \pi)^d$ which cancels with the one in \cref{eqn:radial 2-point fct,eqn:goldstone 2-point fct}. Thus the flow equations for $Z_1$ and $\gamma_1$ read
\begin{mleqn}\label{eqn:z1 g1 flow}
	\begin{rcases}
		\partial_t Z_1\\
		\partial_t \gamma_1^2
	\end{rcases}
	= \left\{\begin{aligned}
		&\eta \, Z_1 + \partial_{q^2}\\
		&\eta \, \gamma_1^2 + \disc_{q_0}
	\end{aligned}\right\} \frac{1}{Z_k} \, \tilde\partial_t \biggl[&-\frac{1}{2} \Bigl(\varphi_c^3 \, U_k^{\prime\prime\prime} + 3 \varphi_c \, U_k^{\prime\prime}\Bigr)^2 \mmint_p G_r(p) \, G_r(q + p)\\
	&- \frac{1}{2} (N - 1) \, \varphi_c^2 \, \bigl(U_k^{\prime\prime}\bigr)^2 \mmint_p G_g(p) \, G_g(q + p)\biggr].
\end{mleqn}
Using $\varphi_c = \csqrt{2 \, \rho_c}$, the prefactor of the radial propagator arising from the momentum-independent vertices can be expanded into
\begin{mleqn}
	\frac{1}{2} \Bigl(\varphi_c^3 \, U_k^{\prime\prime\prime} + 3 \varphi_c \, U_k^{\prime\prime}\Bigr)^2
	&= 4 \rho_c^3 \, \bigl(U_k^{\prime\prime\prime}\bigr)^2 + 12 \rho_c^2 \, U_k^{\prime\prime} \, U_k^{\prime\prime\prime} + 9 \rho_c \, \bigl(U_k^{\prime\prime}\bigr)^2.
\end{mleqn}
For our choice of a quartic effective potential of the form \labelcref{eqn:eff pot exp} this reduces to
\begin{equation}
	4 \rho_c^3 \, \bigl(U_k^{\prime\prime\prime}\bigr)^2 + 12 \rho_c^2 \, U_k^{\prime\prime} \, U_k^{\prime\prime\prime} + 9 \rho_c \, \bigl(U_k^{\prime\prime}\bigr)^2
	= 9 \rho_c \, \lambda^2.
\end{equation}
Moving on to derive the flow of $Z_k$, we set $a = b = 2$. The index summation now yields
\begin{mleqn}
	\smashoperator{\sum_{i,j,k,l=1}^N} \; &G_{ij} \Bigl[\varphi_c^3 \, U_k^{\prime\prime\prime} \, \delta_{21} \, \delta_{j1} \, \delta_{k1} + \varphi_c \, U_k^{\prime\prime} \Bigl(\delta_{21} \, \delta_{jk} + \delta_{j1} \, \delta_{k2} + \delta_{k1} \, \delta_{j1}\Bigr)\Bigr]\\
	&G_{kl} \Bigl[\varphi_c^3 \, U_k^{\prime\prime\prime} \, \delta_{11} \, \delta_{i1} \, \delta_{l1} + \varphi_c \, U_k^{\prime\prime} \Bigl(\delta_{11} \, \delta_{il} + \delta_{i1} \, \delta_{l1} + \delta_{i1} \, \delta_{l1}\Bigr)\Bigr]\\
	&= 2 \varphi_c^2 \, \bigl(U_k^{\prime\prime}\bigr)^2 \, G_r \, G_g.
\end{mleqn}
The structure of Dirac deltas remains the same. Thus the flow equation for $Z_k$ is
\begin{equation}\label{eqn:zk flow}
	\partial_t Z_k
	= -\frac{1}{2} \partial_{q^2} \, \tilde\partial_t \, 2 \varphi_c^2 \, \bigl(U_k^{\prime\prime}\bigr)^2 \mmint_p G_r(p) \, G_g(q + p).
\end{equation}
We now introduce the threshold functions $\tilde\partial_t \tfjm_{ab}^{ij} = \tilde\partial_t \mmint_p G_a^i(p) \, G_b^j(q + p)$, where $a,b \in \{r,g\}$ specify the type of fields running in the loop (radial or Goldstone mode). In our truncation they take the following explicit forms,
\begin{mleqn}
	\tfjm_{11}^{ij}
	&= \tfjm_{11}^{ij}(q,z_1,z_2,m_1^2,m_2^2,\gamma_1^2,\gamma_2^2,R)\\
	&= \mmint_p \frac{1}{\bigl(z_1 p^2 + m_1^2 - i s(p_0) \, \gamma_1^2 + R\bigr)^i} \frac{1}{\bigl(z_2 (p + q)^2 + m_2^2 - i s(p_0 + q_0) \, \gamma_2^2 + R\bigr)^j},\\
	\tfjm_{22}^{ij}
	&= \tfjm_{22}^{ij}(q,1,1,0,0,0,0,R)
	= \mmint_p \frac{1}{\bigl(p^2 + R\bigr)^i} \frac{1}{\bigl((p + q)^2 + R\bigr)^j},\\
	\tfjm_{12}^{ij}
	&= \tfjm_{12}^{ij}(q,z_1,1,m_1^2,0,\gamma_1^2,0,R)
	= \mmint_p \frac{1}{\bigl(z_1 p^2 + m_1^2 - i s(p_0) \, \gamma_1^2 + R\bigr)^i} \frac{1}{\bigl((p + q)^2 + R\bigr)^j}.
\end{mleqn}
Using the shorthand notation $\tfjm_{ab}^{11} = \tfjm_{ab}$, we can write the flow equations \labelcref{eqn:z1 g1 flow,eqn:zk flow} for $\rho_c = \rho_0$ very compactly,
\begin{align}\label{eqn:z1 flow at rho0}
	&\partial_t Z_1
	= \eta \, Z_1 - \partial_{q^2} \tilde\partial_t \, \rho_0 \, \lambda^2 \bigl[9 \, \tfjm_{11} + (N - 1) \, \tfjm_{22}\bigr],\\[1ex]
	\label{eqn:g1 flow at rho0}
	&\partial_t \gamma_1^2
	= \eta \, \gamma_1^2 - \disc_{q_0} \tilde\partial_t \, \rho_0 \, \lambda^2 \, \bigl[9 \, \tfjm_{11} + (N - 1) \, \tfjm_{22}\bigr],\\[1ex]
	\label{eqn:zk flow at rho0}
	&\partial_t Z_k
	= -\partial_{q^2} \, \tilde\partial_t \, 2 \rho_0 \, \lambda^2 \, \tfjm_{12}.
\end{align}
To describe on-shell excitations of the radial field, \labelcref{eqn:z1 flow at rho0,eqn:g1 flow at rho0} should be evaluated at the external energy $q^2 = 2 \rho_0 \, \lambda^2/Z_1$, whereas Goldstone bosons are on-shell for $q^2 = 0$ which is where we evaluate \labelcref{eqn:zk flow at rho0}. It is worth noting, however, that if we were interested in virtual particles, we would be free to solve the flow equations \labelcref{eqn:z1 flow at rho0,eqn:g1 flow at rho0,eqn:zk flow at rho0} for arbitrary $q$.

\section{Matsubara Summation}
\label{sec:matsubara summation}

Our goal in this section is to analytically perform the summation over Matsubara frequencies $i \omega_n$, $n \in \integers$ which for $T > 0$ is part of the trace $\Tr$ on the r.h.s. of all flow equations. We will carry out the summation on the level of the threshold functions $\tfim_j$ and $\tilde\partial_t \tfjm_{ab}^{ij}$. They provide a unified means of formulating flow equations, allowing us to perform the summation once and apply it to multiple flow equations by virtue of the recursive relation \labelcref{eqn:induction}.

In \cref{sec:prop decomp} we show how the class of regulators introduced in \cref{sec:mink regulator} enables us to decompose the regularized propagator $G_k(p) = (P_k + R_k)^{-1}$ into a sum of free propagators \cite{floerchinger2012analytic}. This will be the crucial ingredient that allows us to perform the Matsubara summations in $\tfim_j$ in \cref{sec:eff pot ms} and in $\tfjm_{ij}$ in \cref{sec:prop ms} analytically.

\subsection{Propagator Decomposition}
\label{sec:prop decomp}

The choices \labelcref{eqn:inverse prop expanded,eqn:choice of rk} for the inverse propagator $P_k$ and regulator $R_k$,
\begin{equation}
	P_k
	= Z_k \Bigl[z \, p^2 + m^2 - i s(p_0) \, \gamma^2\Bigr],
	\qquad
	R_k(p)
	= \frac{Z_k \, k^2}{1 + c \, \frac{p^2}{k^2}}
\end{equation}
enable us to conveniently decompose the regularized propagator $(P_k + R_k)^{-1}$ \cite{floerchinger2012analytic}. A detailed calculation can be found in \cref{app:propagator decomposition}. Here we only quote the final result,
\begin{equation}\label{eqn:prop decomp}
	\frac{1}{P_k + R_k}
	= \frac{1}{Z_k} \biggl(\frac{\beta^+}{p^2 + \alpha^+ k^2} + \frac{\beta^-}{p^2 + \alpha^- k^2}\biggr),
\end{equation}
with dimensionless complex scale-dependent coefficients
\begin{mleqn}
	&\alpha^\pm
	= \frac{1}{2} \biggl(\frac{1}{c} + \frac{\tilde m^2}{z} - i \, s(p_0) \, \frac{\tilde\gamma^2}{z}\biggr) \pm \bigl(A + i \, s(p_0) \, B\bigr),\\
	&\beta^\pm
	= \frac{1}{2 \, z} \pm \bigl(C + i \, s(p_0) \, D\bigr),
\end{mleqn}
where $\tilde m^2 = m^2/k^2$, $\tilde\gamma^2 = \gamma^2/k^2$ and 
\begin{align}
	&A
	= \frac{1}{2 i} \Biggl[\sqrt{\frac{1}{c \, z} - \frac{1}{4} \biggl(\frac{1}{c} - \frac{\tilde m^2}{z} - \frac{i \tilde\gamma^2}{z}\biggr)^2} - \, \sqrt{\frac{1}{c \, z} - \frac{1}{4} \biggl(\frac{1}{c} - \frac{\tilde m^2}{z} + \frac{i \tilde\gamma^2}{z}\biggr)^2}\Biggr],\\
	&B
	= \frac{1}{2} \Biggl[\sqrt{\frac{1}{c \, z} - \frac{1}{4}\biggl(\frac{1}{c} - \frac{\tilde m^2}{z} - \frac{i \tilde\gamma^2}{z} \biggr)^2} + \sqrt{\frac{1}{c \, z} - \frac{1}{4} \biggl(\frac{1}{c} - \frac{\tilde m^2}{z} + \frac{i \tilde\gamma^2}{z} \biggr)^2}\Biggr],
\end{align}
\begin{equation}
	C
	= -\frac{A \Bigl(\frac{1}{c} - \frac{\tilde m^2}{z} \Bigr) + B \,\frac{\tilde\gamma^2}{z}}{4 \, z\, (A^2 + B^2)},
	\qquad
	D
	= \frac{B \Bigl(\tfrac{1}{c} - \frac{\tilde m^2}{z}\Bigr) - A \, \frac{\tilde\gamma^2}{z}}{4 \, z\, (A^2 + B^2)}.
\end{equation}
\labelcref{eqn:prop decomp} closely resembles the sum of two free propagators, which significantly simplifies calculations. Choosing the branch cut of the complex square root along the negative real axis ensures that $A$, $B$, $C$ and $D$ are always real. (Even though we set $c_0 = 1$, $c_1 = c > 0$, $c_j = 0 \; \forall \, j > 1$ to obtain this result, a similar decomposition is possible for the whole class of regulators \labelcref{eqn:algebraic regulator}.)

We showed in \cref{sec:analytic structure} that the propagator $P_k^{-1}(p)$ may exhibit poles and branch cuts only on the real frequency axis. The same cannot be said for the regularized propagator $(P_k+R_k)^{-1}$. Since $R_k$ brings with it its own analytic structure, $(P_k+R_k)^{-1}$ will in general feature singularities away from the real frequency axis \cite{floerchinger2012analytic}. In our case \labelcref{eqn:prop decomp}, for $\tilde\gamma^2/z - B < 0$ and $s(p_0) = 1$ there are poles at $p_0 = \pm\csqrt{\vec{p}^2 + \alpha^+ k^2}$. A Källen-Lehmann spectral representation of the form \labelcref{eqn:spectral repr} is therefore not possible for $(P_k+R_k)^{-1}$. Although a proof is still pending, this is believed to be a generic feature of cutoff functions that serve as effective UV regulators in Minkowski space \cite{floerchinger2012analytic}. Notably, $(P_k + R_k)^{-1}$ also has a branch cut. However, assuming all integrals along this branch cut are dominated by nearby poles on the different Riemann sheets, it will not inhibit our analytic treatment.

\subsection{Effective Potential}
\label{sec:eff pot ms}

We now perform the Matsubara summation for the simplest threshold function $\tfim_0$ \cite{floerchinger2017unpublished}. We can afterwards extend the result to higher-orders $\tfim_j$ with $j \geq 1$ by using the relation \labelcref{eqn:induction}, $\partial_{m^2} \tfim_j = (\delta_{j0} - j) \, \tfim_{j+1}$. $\tfim_0$ was defined as (cf. \cref{eqn:real time threshold fcts})
\begin{mleqn}
	\tfim_0(Z_k,z,m^2,\gamma^2,R_k)
	&= T \sum_{p_0} \int_{\vec p} \frac{\partial_t R_k/Z_k}{z \, p^2 + m^2 - i s(p_0) \, \gamma^2 + R_k/Z_k}\\
	&= \mmint_p \frac{\partial_t R_k}{P_k + R_k}
	= \tilde\partial_t \mmint_p \ln(P_k + R_k).
\end{mleqn}
We denote the difference between $\tfim_0$ evaluated at some intermediate renormalization scale $k$ and in the ultraviolet $\Lambda$ as $\Delta \tfim_0$,\footnote{This step serves as a type of implicit Pauli-Villars regularization with the heavy mass term replaced by the momentum-dependent $R_\Lambda(p)$ in the inverse propagator \cite{berges2002non}.}
\begin{align}\label{eqn:delta i0}
	\Delta \tfim_0
	&= \int_\Lambda^k \tilde\dif t \, \tfim_0
	= \mmint_p \bigl[\ln(P_k + R_k) - \ln(P_\Lambda + R_\Lambda)\bigr]\\
	&= -\mmint_p \biggl\{\ln\frac{1}{Z_k} \biggl[\frac{\beta^+}{p^2 + \alpha^+ k^2} + \frac{\beta^-}{p^2 + \alpha^- k^2}\biggr] - \ln\frac{1}{Z_\Lambda} \biggl[\frac{\beta^+}{p^2 + \alpha^+ \Lambda^2} + \frac{\beta^-}{p^2 + \alpha^- \Lambda^2}\biggr]\biggr\},\notag
\end{align}
where we inserted the decomposition \labelcref{eqn:prop decomp}. The tilde on $\tilde\dif t$ indicates that we integrate with respect to the explicit scale dependence in $R_k$ only. Since the second term in $\Delta \tfim_0$ is $k$-independent, we can at any time easily recover $\tfim_0$ by taking the $\tilde\partial_t$ derivative of $\Delta \tfim_0$.

\begin{figure}[htb!]
	\centering
	\begin{tikzpicture}[thick,scale=0.8]
	    \def\xr{3}\def\yr{3}
	    
	    \draw [->] (-\xr-1,0) -- (\xr+1,0) node [above left]  {$\Re(p_0)$};
	    \draw [->] (0,-\yr-0.7) -- (0,\yr+0.7) coordinate [below left = 0.3 and 0.1] (y-axis);
	    \node (y-label) at ([xshift=-50]y-axis) {$\Im(p_0)$};
	    \draw[ultra thin,gray] (y-axis) -- (y-label);
	    
	    \foreach \n in {-\yr,...,-1,1,2,...,\yr}{%
	        \draw[fill] (0,\n) circle (1pt) node [right,font=\scriptsize] {$i \omega_{_{\!\n}}$};}
	    \draw[fill] (0,0) circle (1pt) node [above right] {0};
	    
	    \draw[DarkBlue,decoration={markings,mark=between positions 0 and 1 step 0.28 with \arrow{>}},postaction={decorate}] (1,-\yr) -- (1,\yr) node [below right] {$C$} arc (0:180:1) (-1,\yr) -- (-1,-\yr) arc (180:360:1);
	    
	    \node (poles) at (3,1.5) {poles of $\smash{\frac{1}{-p_0^2 + x^2}}$};
		\draw[fill]
		(2,2.5) coordinate [circle,fill,inner sep=1pt,label=right:$p_1$] (p1)
		(-2,-2.5) coordinate [circle,fill,inner sep=1pt,label=below:$p_2$] (p2);
		\begin{scope}[on background layer]
			\draw[ultra thin,gray]
			(poles) -- (p1)
			(poles) -- (p2);
		\end{scope}
	\end{tikzpicture}
	\caption{Counterclockwise path $C$ around the imaginary $p_0$-axis but excluding poles of $(-p_0^2 + x^2)^{-1}$}
	\label{fig:contour 1}
\end{figure}
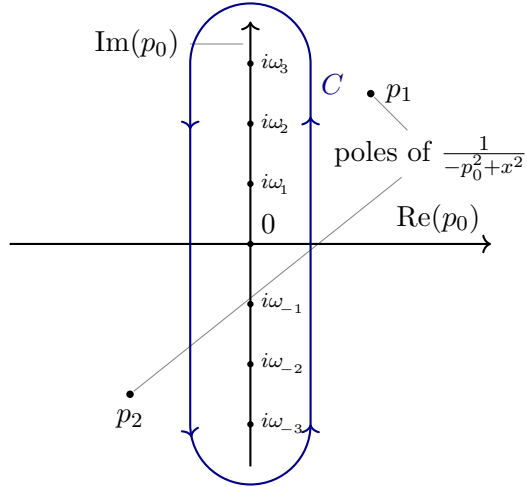

Using $Z_\Lambda = 1 \approx Z_k$, $\beta^+ + \beta^- = \frac{1}{z}$ and $\alpha^+ \, \beta^- + \alpha^- \, \beta^+ = \frac{1}{c \, z}$ (see \labelcref{eqn:ABCD relations}), we can write
\begin{mleqn}
	-\ln\biggl(\frac{\beta^+}{p^2 + \alpha^+ k^2} + \frac{\beta^-}{p^2 + \alpha^- k^2}\biggr)
	&= \ln\bigl[(p^2 + \alpha^+ k^2) (p^2 + \alpha^- k^2)\bigr]\\
	&\hphantom{{}={}}- \ln\bigl[\beta^+ (p^2 + \alpha^- k^2) + \beta^- (p^2 + \alpha^+ k^2)\bigr]\\[1ex]
	&= \ln(p^2 + \alpha^+ k^2) + \ln(p^2 + \alpha^- k^2) - \ln\bigl(\tfrac{1}{z} \, p^2 + \tfrac{1}{c z} \, k^2\bigr),
\end{mleqn}
(and similarly for the $\Lambda$-term) so that $\Delta \tfim_0$ can also be written
\begin{mleqn}\label{eqn:delta i0 simplified}
	\Delta \tfim_0
	&= \mmint_p \Bigl[\ln(p^2 + \alpha^+ k^2) + \ln(p^2 + \alpha^- k^2) - \ln(p^2 + \tfrac{1}{c} \, k^2)\\
	&\hphantom{{}= \mmint_p \Bigl[}- \ln(p^2 + \alpha^+ \Lambda^2) - \ln(p^2 + \alpha^- \Lambda^2) + \ln(p^2 + \tfrac{1}{c} \, \Lambda^2)\Bigr],
\end{mleqn}
The terms in \labelcref{eqn:delta i0 simplified} can be pairwise combined into integrals,
\begin{equation}\label{eqn:x integral}
	T \sum_{p_0} \ln(p^2 + \alpha^+ k^2) - \ln(p^2 + \alpha^+ \Lambda^2)
	= \int_{\vec{p}^2 + \alpha^+ \Lambda^2}^{\vec{p}^2 + \alpha^+ k^2} \dif x^2 \, T \sum_{n \in \integers} \frac{1}{\omega_n^2 + x^2},
\end{equation}
and similarly for the other four terms. (Recall that $p^2 = -p_0^2 + \vec p^2$ where $p_0 \in \{i \omega_n = 2 \pi i T n|n \in \integers\}$.) The Matsubara summation can then be expressed as the complex contour integral,
\begin{equation}\label{eqn:matsubara summation}
	T \sum_{n \in \integers} \frac{1}{\omega_n^2 + x^2}
	= \oint_C \frac{\dif p_0}{2 \pi i} \, \frac{1}{-p_0^2 + x^2} \bigl[n_\text{B}(p_0) + \tfrac{1}{2}\bigr],
\end{equation}
where
\begin{equation}\label{eqn:be dist}
	n_\text{B}(p_0)
	= \frac{1}{e^{p_0/T} - 1}
\end{equation}
is the Bose-Einstein distribution and $C$ is the path shown in \cref{fig:contour 1}.

In \labelcref{eqn:matsubara summation}, we used that $n_\text{B}(p_0)$ has only simple poles located at all Matsubara frequencies $i \omega_n$, $n \in \integers$. At $p_0 = i \, \omega_n = i 2 \pi T n$, we have $e^{p_0/T} = e^{2 \pi i \, n} = 1$ and so the denominator in \labelcref{eqn:be dist} vanishes. Using l'Hôpital's rule, we find that the residue at all poles is the temperature $T$,
\begin{equation}
	\lim_{p_0 \to i \, \omega_n} \frac{p_0 - i \, \omega_n}{e^{p_0/T} - 1}
	= \lim_{p_0 \to i \, \omega_n} \frac{T}{e^{p_0/T}}
	= T
	\quad \forall \, n \in \integers.
\end{equation}
The $+\frac{1}{2}$ was merely added to antisymmetrize $n_\text{B}(-p_0) + \frac{1}{2} = -\bigl[n_\text{B}(p_0) + \frac{1}{2}\bigr]$. Applying \labelcref{eqn:matsubara summation} to all terms in $\Delta \tfim_0$ we get
\begin{equation}
	\Delta \tfim_0
	= \int_{\vec p} \Biggl(\int_{\vec{p}^2 + \alpha^+ \Lambda^2}^{\vec{p}^2 + \alpha^+ k^2} + \int_{\vec{p}^2 + \alpha^- \Lambda^2}^{\vec{p}^2 + \alpha^- k^2} + \int_{\vec{p}^2 + \Lambda^2/c}^{\vec{p}^2 + k^2/c}\Biggr) \dif x^2 \, \oint_C \frac{\dif p_0}{2 \pi i} \, \frac{n_\text{B}(p_0) + \tfrac{1}{2}}{-p_0^2 + x^2}.
\end{equation}

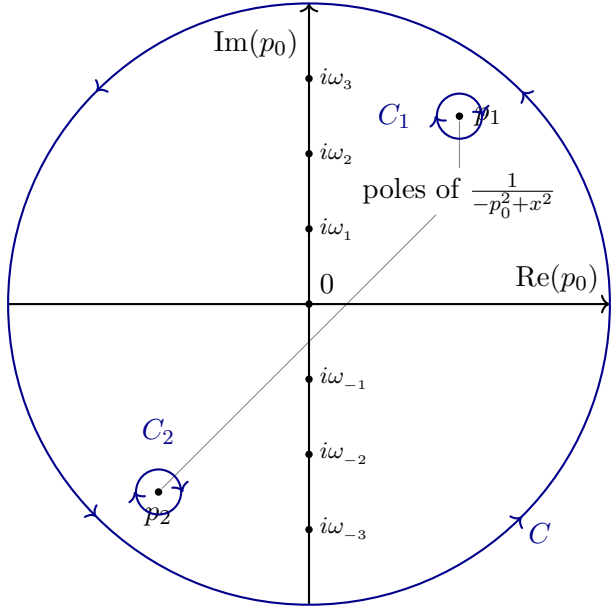
\begin{figure}[htb!]
	\centering
	\begin{tikzpicture}[thick]
	    \def\xr{3} \def\yr{3}
	    
	    \draw [->] (-\xr-1,0) -- (\xr+1,0) node [above left]  {$\Re(p_0)$};
	    \draw [->] (0,-\yr-1) -- (0,\yr+1) node [below left=0.2 and 0] {$\Im(p_0)$};
	    
	    \foreach \n in {-\yr,...,-1,1,2,...,\yr}{%
	        \draw[fill] (0,\n) circle (1pt) node [right,font=\footnotesize] {$i \omega_{_{\n}}$};}
	    \draw[fill] (0,0) circle (1pt) node [above right] {0};
	    
	    \draw[DarkBlue,decoration={markings,mark=between positions 0.125 and 0.875 step 0.25 with \arrow{>}},postaction={decorate}] circle (\yr+1) node [below right=0.925*\xr and 0.925*\yr] {$C$};
	    
	    \node (poles) at (2,1.5) {poles of $\smash{\frac{1}{-p_0^2 + x^2}}$};
		\draw[fill]
		(2,2.5) coordinate [circle,fill,inner sep=1pt,label=right:$p_1$] (p1)
		(-2,-2.5) coordinate [circle,fill,inner sep=1pt,label=below:$p_2$] (p2);
		\begin{scope}[on background layer]
			\draw[ultra thin,gray]
			(poles) -- (p1)
			(poles) -- (p2);
		\end{scope}
	    
	    \draw[DarkBlue,decoration={markings,mark=between positions 0.03 and 1.03 step 0.25 with \arrow{<}},postaction={decorate}] (p1) circle (0.3) node [left=0.5] {$C_1$} (p2) circle (0.3) node [above=0.5] {$C_2$};
	\end{tikzpicture}
	\caption{Sum of contours equivalent to $C$ in \cref{fig:contour 1}}
	\label{fig:contour 2}
\end{figure}

We now deform the contour $C$ into a circle and take the radius to infinity. This will enclose the poles of $(-p_0^2 + x^2)^{-1}$ scattered throughout the complex plane. Their contribution is removed again by enclosing them in clockwise contours as shown in \cref{fig:contour 2}.

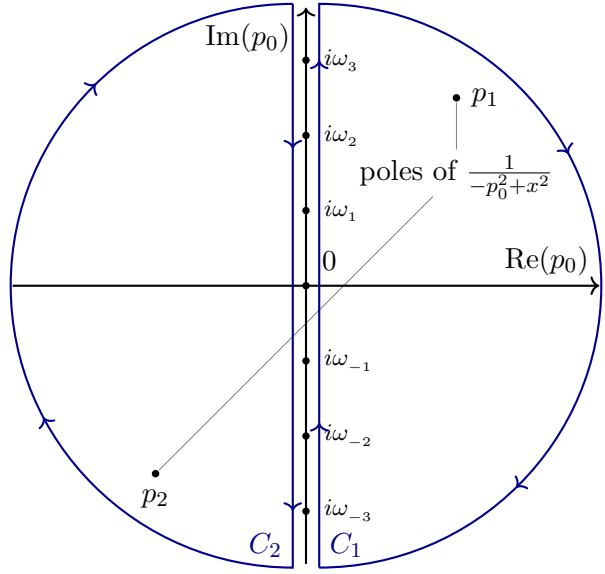
\begin{figure}[htb!]
	\centering
	\begin{tikzpicture}[thick]
	    \def\xr{3.5} \def\yr{3}
	    
	    \draw [->] (-\xr-0.4,0) -- (\xr+0.4,0) node [above left]  {$\Re(p_0)$};
	    \draw [->] (0,-\yr-0.7) -- (0,\yr+0.7) node[below left=0.1] {$\Im(p_0)$};
	    
	    \foreach \n in {-\yr,...,-1,1,2,...,\yr}{%
	        \draw[fill] (0,\n) circle (1pt) node [right=0.1,font=\footnotesize] {$i \omega_{_{\n}}$};}
	    \draw[fill] (0,0) circle (1pt) node [above right=0.1] {0};
	    
	    \draw[xshift=5,DarkBlue,decoration={markings,mark=between positions 0.1 and 1 step 0.25 with \arrow{>}},postaction={decorate}] (0,-\yr-0.75) node [above right] {$C_1$} -- (0,\yr+0.75) arc (90:-90:\yr+0.75);
	    
	    \draw[xshift=-5,DarkBlue,decoration={markings,mark=between positions 0.1 and 1 step 0.25 with \arrow{>}},postaction={decorate}] (0,\yr+0.75) -- (0,-\yr-0.75) node [above left] {$C_2$} arc (270:90:\yr+0.75);
	    
	    \node (poles) at (2,1.5) {poles of $\smash{\frac{1}{-p_0^2 + x^2}}$};
		\draw[fill]
		(2,2.5) coordinate [circle,fill,inner sep=1pt,label=right:$p_1$] (p1)
		(-2,-2.5) coordinate [circle,fill,inner sep=1pt,label=below:$p_2$] (p2);
		\begin{scope}[on background layer]
			\draw[ultra thin,gray]
			(poles) -- (p1)
			(poles) -- (p2);
		\end{scope}
	\end{tikzpicture}
	\caption{Contour enclosing the entire plane except for the imaginary axis}
	\label{fig:contour 3}
\end{figure}

Since the integrand \labelcref{eqn:matsubara summation} falls off faster than $1/p_0$, contributions from the circle $C$ at infinity vanish. This contour is thus equivalent to the one in \cref{fig:contour 3} where we discarded $C$ and blew up $C_1$ and $C_2$ to enclose the entire $p_0$-plane save for the imaginary axis.

It is important in this context that $\alpha^\pm$ depends on $s(p_0)$ as this implies $x = x\bigl(s(p_0)\bigr)$. The contour in \labelcref{eqn:matsubara summation} therefore encloses both poles and branch cuts. If there were only poles, we could simply invoke the residue theorem to turn the infinite sum over Matsubara frequencies \labelcref{eqn:matsubara summation} into a finite sum over the residues at the poles $p_0 = \pm x$ of $(-p_0^2 + x^2)^{-1}$. This would give
\begin{equation}
	T \sum_{n \in \integers} \frac{1}{\omega_n^2 + x^2}
	= -\smashoperator{\sum_{p_0 = \pm x}} \Res \frac{n_\text{B}(p_0) + \tfrac{1}{2}}{-p_0^2 + x^2}
	= -\frac{n_\text{B}(-x) - n_\text{B}(x)}{2 x}
	= \frac{\coth(x/2 T)}{2 x}.
\end{equation}
(The residues appear with a sign due to the clockwise contour.) Evaluating the contribution from the branch cuts, however, requires additional work. Using
\begin{equation}\label{eqn:pfd}
	\frac{1}{-p_0^2 + x^2}
	= \frac{1}{2 x} \biggl(\frac{1}{p_0 + x} - \frac{1}{p_0 - x}\biggr)
\end{equation}
and substituting $\dif x^2 = 2 x \, \dif x$, \labelcref{eqn:x integral} becomes
\begin{mleqn}\label{eqn:partial delta i0 integral}
	&T \sum_{n \in \integers} \ln(p^2 + \alpha^+ k^2) - \ln(p^2 + \alpha^+ \Lambda^2)\\
	&= \smash{\int_{\csqrt{\vec{p}^2 + \alpha^+ \Lambda^2}}^{\csqrt{\vec{p}^2 + \alpha^+ k^2}}} \dif x \oint_{C_1+C_2} \frac{\dif p_0}{2 \pi i} \biggl[\frac{1}{p_0 + x} - \frac{1}{p_0 - x}\biggr] \bigl[n_\text{B}(p_0) + \tfrac{1}{2}\bigr].
\end{mleqn}
Note that the integral boundaries receive a square root when substituting $\dif x^2 \to \dif x$.

\begin{figure}[htb!]
	\centering
	\begin{tikzpicture}
	    \begin{axis}[
	        xlabel=$\Re(p_0)$,
	        ylabel=$\Im(p_0)$,
	        zlabel=$s(p_0)$,
	        domain=-1:1,
	        surf,shader=flat,
	        xtick distance=1,
	        ytick distance=1,
	        ztick distance=1,
	        tickwidth=0pt
	    ]
	    
	    \addplot3[blue!20] coordinates {
			(-1,1,-1) (0,1,-1)
			
			(-1,0,-1) (0,0,-1)
		};
	    
	    \addplot3[blue!20] coordinates {
			(1,-1,-1) (0,-1,-1)
			
			(1,0,-1) (0,0,-1)
		};
	
	    \addplot3[
	        gray,opacity=0.1,
	        samples=2,
	    ]{0};
	    
	    \addplot3[orange!80] coordinates {		
			(0,0,1) (1,0,1)
			
			(0,1,1) (1,1,1)
		};
	    
	    \addplot3[orange!80] coordinates {		
			(0,0,1) (-1,0,1)
			
			(0,-1,1) (-1,-1,1)
		};
	    
	    \end{axis}
	\end{tikzpicture}
	\caption{Structure of the sign function $s(p_0) = \sign(\Re p_0 \, \Im p_0)$}
	\label{fig:sign}
\end{figure}
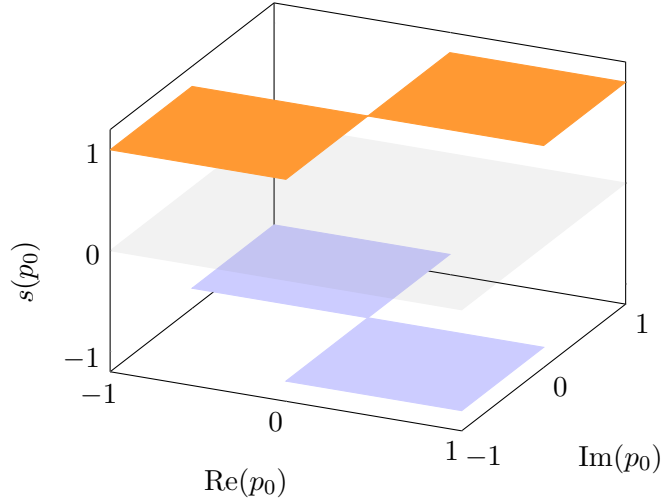

We can now use the structure of $s(p_0)$ plotted in \cref{fig:sign} to split the contour integral into two parts, the first being branch-cut free and the second with branch cut structure manifest.

\begin{align}\label{eqn:integral decomposition}
	\labelcref{eqn:partial delta i0 integral}
	&= \frac{1}{2} \int_{C_1+C_2} \frac{\dif p_0}{2 \pi i} \Biggl[\int_{\csqrt{\vec{p}^2 + \alpha^{++} \Lambda^2}}^{\csqrt{\vec{p}^2 + \alpha^{++} k^2}} + \int_{\csqrt{\vec{p}^2 + \alpha^{+-} \Lambda^2}}^{\csqrt{\vec{p}^2 + \alpha^{+-} k^2}}\Biggr] \dif x \biggl[\frac{1}{p_0 + x} - \frac{1}{p_0 - x}\biggr] \bigl[n_\text{B}(p_0) + \tfrac{1}{2}\bigr]\\
	&\hphantom{=}+ \frac{1}{2} \int_{C_1+C_2} \frac{\dif p_0}{2 \pi i} \, s(p_0) \Biggl[\int_{\csqrt{\vec{p}^2 + \alpha^{++} \Lambda^2}}^{\csqrt{\vec{p}^2 + \alpha^{++} k^2}} - \int_{\csqrt{\vec{p}^2 + \alpha^{+-} \Lambda^2}}^{\csqrt{\vec{p}^2 + \alpha^{+-} k^2}}\Biggr] \dif x \biggl[\frac{1}{p_0 + x} - \frac{1}{p_0 - x}\biggr] \bigl[n_\text{B}(p_0) + \tfrac{1}{2}\bigr].\notag
\end{align}
Observe that in the first and third quadrant of the complex plane where $s(p_0) = 1$, the first and third $x$-integrals in \labelcref{eqn:integral decomposition} combine to give \labelcref{eqn:partial delta i0 integral} while the second and fourth cancel. Conversely, in the second and fourth quadrant $s(p_0) = -1$ and so the first and third $x$-integrals cancel while the second and fourth sum to \labelcref{eqn:partial delta i0 integral}. This split up is beneficial in two ways. Not only does it reveal the branch cut structure in $\Delta \tfim_0$, it also separates the $x$-integration into parts where $s(p_0)$ takes a definite sign as reflected in the shorthand notation $\alpha^{\pm\pm} = \alpha^\pm\bigl(s(p_0) = \pm 1\bigr)$.

We now perform the $p_0$-integration in the upper and lower line of \labelcref{eqn:integral decomposition} separately. The former contains only (simple) poles at $p_0 = \pm x$. The residues can be computed by the limit formula,
\begin{equation}
	\Res_{p_0 = \mp x} \frac{1}{p_0 \pm x}
	= \lim_{p_0 \to \mp x} (p_0 \pm x) \, \frac{1}{p_0 \pm x}
	= 1.
\end{equation}
The first line thus integrates to
\begin{equation}\label{eqn:simple poles eval}
	\Biggl[\int_{\csqrt{\vec{p}^2 + \alpha^{++} \Lambda^2}}^{\csqrt{\vec{p}^2 + \alpha^{++} k^2}}	 + \int_{\csqrt{\vec{p}^2 + \alpha^{+-} \Lambda^2}}^{\csqrt{\vec{p}^2 + \alpha^{+-} k^2}}\Biggr] \dif x \bigl[n_\text{B}(x) + \tfrac{1}{2}\bigr]
\end{equation}
Due to $s(p_0)$, the integral on the second line encloses a branch cut along the real $p_0$-axis in addition to the poles at $p_0 = \pm x$. The contribution from the poles alone is
\begin{equation}\label{eqn:pole contrib}
	\Biggl[\int_{\csqrt{\vec{p}^2 + \alpha^{++} \Lambda^2}}^{\csqrt{\vec{p}^2 + \alpha^{++} k^2}} - \int_{\csqrt{\vec{p}^2 + \alpha^{+-} \Lambda^2}}^{\csqrt{\vec{p}^2 + \alpha^{+-} k^2}}\Biggr] \dif x \, s_\text{I}(x) \bigl[n_\text{B}(x) + \tfrac{1}{2}\bigr],
\end{equation}
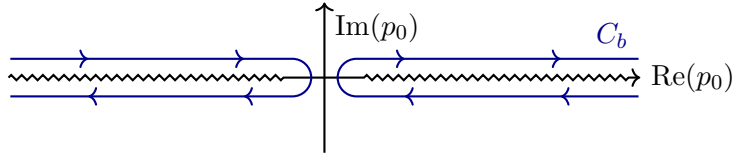
\begin{figure}[htb!]
	\centering
	\begin{tikzpicture}[thick]
	    \def\xr{4} \def\yr{1}
	    
	    \draw [decorate,decoration={zigzag,segment length=4,amplitude=1,post=lineto,post length=15}] (-1.05*\xr,0) -- (0,0);
	    \draw [->,decorate,decoration={zigzag,segment length=4,amplitude=1,pre=lineto,pre length=15,post=lineto,post length=3}] (0,0) -- (1.05*\xr,0) node [right]  {$\Re(p_0)$};
	    \draw [->] (0,-\yr) -- (0,\yr) node [below right] {$\Im(p_0)$};
	    
	    \draw[xshift=5,DarkBlue,decoration={markings,mark=between positions 0.125 and 0.875 step 0.25 with \arrow{>}},postaction={decorate}] (\xr,-\yr/4) -- (\yr/4,-\yr/4) arc (-90:-270:\yr/4) (\yr/4,\yr/4) -- (\xr,\yr/4) node[above left] {$C_b$};
	    
	    \draw[xshift=-5,DarkBlue,decoration={markings,mark=between positions 0.125 and 0.875 step 0.25 with \arrow{>}},postaction={decorate}] (-\xr,\yr/4) -- (-\yr/4,\yr/4) arc (90:-90:\yr/4) (-\yr/4,-\yr/4) -- (-\xr,-\yr/4);
	\end{tikzpicture}
	\caption{Contour $C_b$ suitable to evaluate only branch cut contributions}
	\label{fig:contour 4}
\end{figure}
where $s_\text{I}(x) = \sign(\Im x)$. To isolate the contribution from the branch cut, we integrate along the contour $C_b$ in \cref{fig:contour 4}. $C_b$ amounts to an integral from $0$ to $\infty$ and one from $-\infty$ to $0$. We integrate twice along these stretches in opposite directions. Usually the resulting contributions cancel. In our case, however, $s_\text{I}(p_0) = -1$ along the lower lines so we get a factor of 2,
\begin{mleqn}
	\int_{C_b} \frac{\dif p_0}{2 \pi i}
	&= \int_{-\infty}^0 \frac{\dif p_0}{2 \pi i} \, \underbrace{s_\text{I}(p_0 + i \epsilon)}_1 + \int_0^{-\infty} \frac{\dif p_0}{2 \pi i} \, \underbrace{s_\text{I}(p_0 - i \epsilon)}_{-1}\\
	&\hphantom{{}=}+ \int_0^\infty \frac{\dif p_0}{2 \pi i} \, \underbrace{s_\text{I}(p_0 + i \epsilon)}_1 + \int_\infty^0 \frac{\dif p_0}{2 \pi i} \, \underbrace{s_\text{I}(p_0 - i \epsilon)}_{-1}
	= 2 \int_{-\infty}^0 \frac{\dif p_0}{2 \pi i} + 2 \int_0^\infty \frac{\dif p_0}{2 \pi i},
\end{mleqn}
Exchanging bounds on the first integral and substituting $p_0 \to -p_0$ (which leaves the integrand $(p_0 + x)^{-1} - (p_0 - x)^{-1}$ invariant), the total contribution from the cut is therefore
\begin{equation}\label{eqn:l+r contrib}
	\Biggl[\int_{\csqrt{\vec{p}^2 + \alpha^{++} \Lambda^2}}^{\csqrt{\vec{p}^2 + \alpha^{++} k^2}} - \int_{\csqrt{\vec{p}^2 + \alpha^{+-} \Lambda^2}}^{\csqrt{\vec{p}^2 + \alpha^{+-} k^2}}\Biggr] \dif x \, 2 \int_0^\infty \frac{\dif p_0}{2 \pi i} \biggl[\frac{1}{p_0 + x} - \frac{1}{p_0 - x}\biggr] \bigl[n_\text{B}(p_0) + \tfrac{1}{2}\bigr]
\end{equation}
Away from the imaginary axis the Bose-Einstein distribution plotted in \cref{fig:bose distribution} becomes approximately flat, particularly at sufficiently low temperatures. We assume $|\Re(x)| \gg 0$ (which is satisfied unless $\vec{p}^2 + \Re(\alpha^{\pm\pm}) \, k^2 < 0$ and $\Im(\alpha^{\pm\pm}) \, k^2 \approx 0$) so that the distribution is approximately constant across the width of the poles in \labelcref{eqn:l+r contrib}. Since $p_0$ is integrated from $0$ to $\infty$, these integrals are dominated by the poles at $p_0 = x$ (as opposed to those at $p_0 = -x$) so we make the simplifying replacement $n_\text{B}(p_0) \to n_\text{B}(x)$. This allows us to write
\begin{figure}[htb!]
	\centering
	\begin{tikzpicture}
	    \begin{axis}[
	        xlabel=$\Re(p_0)$,
	        ylabel=$\Im(p_0)$,
	        zlabel=$n_\text{B}(p_0)$,
	        x label style={at={(0.35,0)}},
	        y label style={at={(0.95,0.15)}},
	        shader=flat,
	        tickwidth=0pt
	    ]
	    
	    \def\nB{(e^(2*x) - 2*e^x*cos(deg(y)) + 1)^(-1/2)}
	
	    \addplot3[surf,opacity=0.5,domain=1:10,y domain=-10:10]{\nB};
	
	    \addplot3[surf,opacity=0.5,domain=-10:-2,y domain=-10:10]{\nB};
	
	    \addplot3[surf,opacity=0.5,domain=-2:1,y domain=-10:10,restrict z to domain*=0:2]{\nB};
	
	    \end{axis}
	\end{tikzpicture}
	\begin{tikzpicture}
        \begin{axis}[
            domain = 0:2,ymax = 5,
            xlabel = $\Re(p_0)$,
            ylabel = $n_\text{B}(p_0)$,
            ticks=none,smooth,
            thick,axis lines = left,
            every tick/.style = {thick},
            width=8cm,height=7cm]
            
            \def\nB#1{1/(e^(x/#1) - 1) + 1/2}
            
            \addplot[name path=T1,color=red]{\nB{0.5}};
            
            \addplot[name path=T2,color=yellow]{\nB{1}};
            
            \addplot[name path=T3,color=blue]{\nB{2}};
            
            \addplot [draw=none,name path=aux] {3*x};
            
        \end{axis}
        
        \draw[shorten >=2,shorten <=2,name intersections={of=T1 and aux,name=int1},name intersections={of=T2 and aux,name=int2}] (int1-1) edge[->,bend left] node[midway,below right=-1pt,font=\scriptsize] {$2 \cdot T$} (int2-1);
        
        \draw[shorten >=2,shorten <=2,name intersections={of=T3 and aux,name=int3}] (int2-1) edge[->,bend left] node[midway,below right=-1pt,font=\scriptsize] {$2 \cdot T$} (int3-1);

    \end{tikzpicture}
	\caption{Bose distribution plotted over the complex plane and for different temperatures}
	\label{fig:bose distribution}
\end{figure}
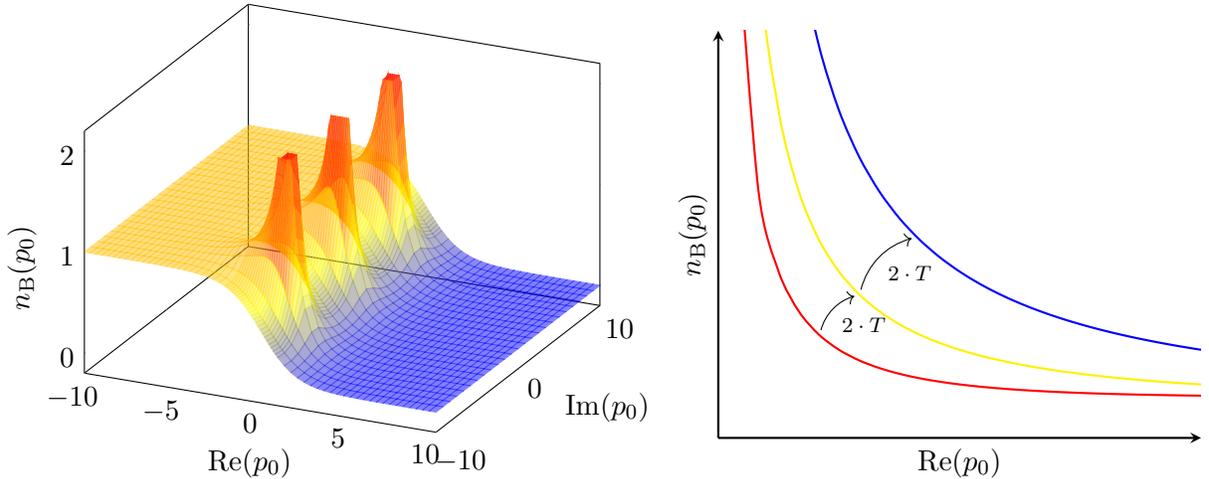
\begin{equation}\label{eqn:si of x}
	2 \int_0^\infty \frac{\dif p_0}{2 \pi i} \biggl[\frac{1}{p_0 + x} - \frac{1}{p_0 - x}\biggr]
	= \int_{-\infty}^\infty \frac{\dif p_0}{2 \pi i} \biggl[\frac{1}{p_0 + x} - \frac{1}{p_0 - x}\biggr]
	= -s_I(x).
\end{equation}
Plugging this back into \labelcref{eqn:l+r contrib}, we find that it precisely cancels with the contribution \labelcref{eqn:pole contrib} from the poles. The result of the Matsubara summation in \labelcref{eqn:partial delta i0 integral} is therefore simply given by \labelcref{eqn:simple poles eval}. Executing the $x$-integration we get \labelcref{eqn:simple poles eval}
\begin{mleqn}\label{eqn:matsubara result}
	\sum_{p_0}& \Bigl[\ln(p^2 + \alpha^+ k^2) - \ln(p^2 + \alpha^+ \Lambda^2)\Bigr]
	= \Biggl(\int_{\csqrt{\vec{p}^2 + \alpha^{++} \Lambda^2}}^{\csqrt{\vec{p}^2 + \alpha^{++} k^2}} + \int_{\csqrt{\vec{p}^2 + \alpha^{+-} \Lambda^2}}^{\csqrt{\vec{p}^2 + \alpha^{+-} k^2}}\Biggr) \dif x \bigl[n_\text{B}(x) + \tfrac{1}{2}\bigr]\\[1ex]
	&= \frac{1}{2} \biggl[\csqrt{\vec{p}^2 + \alpha^{++} k^2} - \csqrt{\vec{p}^2 + \alpha^{++} \Lambda^2} + \csqrt{\vec{p}^2 + \alpha^{+-} k^2} - \csqrt{\vec{p}^2 + \alpha^{+-} \Lambda^2}\biggr]\\
	&\hphantom{{}={}}+ T \Biggl[\ln\Biggl(\frac{e^{\csqrt{\vec{p}^2 + \alpha^{++} k^2}/T} - 1}{e^{\csqrt{\vec{p}^2 + \alpha^{++} \Lambda^2}/T} - 1}\Biggr) + \ln\Biggl(\frac{e^{\csqrt{\vec{p}^2 + \alpha^{+-} k^2}/T} - 1}{e^{\csqrt{\vec{p}^2 + \alpha^{+-} \Lambda^2}/T} - 1}\Biggr)\Biggr],
\end{mleqn}
where we used
\begin{equation}
	\int_a^b \biggl(\frac{1}{e^{x/T} - 1} + \frac{1}{2}\biggr) \dif x
	= \tfrac{1}{2} (b - a) + T \ln\biggl(\frac{e^{b/T} - 1}{e^{a/T} - 1}\biggr).
\end{equation}
The second bracket in \labelcref{eqn:matsubara result} disappears for $T \to 0^+$ so the threshold functions split up into a term that carries the entire temperature dependence and a $T$-independent offset. In total, $\Delta \tfim_0$ after Matsubara summation reads
\begin{equation}\label{eqn:delta i0 result}
	\Delta \tfim_0
	= \int_{\vec p} \sum_{j=1}^{10} w_j \biggl[\frac{1}{2} \csqrt{\vec{p}^2 + \mu_j} + T \ln\Bigl(e^{\csqrt{\vec{p}^2 + \mu_j}/T} - 1\Bigr)\biggr],
\end{equation}
where 
\begin{center}
	\begin{tabular}{ccccccccccc}
	\hline
	$j$ & 1 & 2 & 3 & 4 & 5 & 6 & 7 & 8 & 9 & 10 \\
	$\mu_j$ & $\alpha^{++} k^2$ & $\alpha^{+-} k^2$ & $\alpha^{-+} k^2$ & $\alpha^{--} k^2$ & $\frac{1}{c} \, k^2$ & $\alpha^{++} \Lambda^2$ & $\alpha^{+-} \Lambda^2$ & $\alpha^{-+} \Lambda^2$ & $\alpha^{--} \Lambda^2$ & $\frac{1}{c} \, \Lambda^2$ \\
	$w_j$ & $1$ & $1$ & $1$ & $1$ & $-2$ & $-1$ & $-1$ & $-1$ & $-1$ & $2$ \\ \hline
	\end{tabular}
\end{center}
The terms $\frac{1}{c} \, k^2$ and $\frac{1}{c} \, \Lambda^2$ appear with a factor of $\pm 2$ because they are independent of $s(p_0)$. They therefore give the same contribution twice in \labelcref{eqn:integral decomposition} where we performed the split into $s(p_0) = 1$ and $s(p_0) = -1$. We can recover $\tfim_0$ by taking the derivative with respect to explicit $k$-dependence of \labelcref{eqn:delta i0 result}
\begin{equation}
	\tfim_0
	= \tilde\partial_t \, \Delta \tfim_0
	= \int_{\vec p} \sum_{j=1}^5 \frac{w_j}{2} \frac{\tilde\partial_t \, \mu_j}{\csqrt{\vec{p}^2 + \mu_j}} \biggl[\frac{1}{2} + \frac{1}{1 - e^{-\csqrt{\vec{p}^2 + \mu_j}/T}}\biggr].
\end{equation}
Recall that $\tilde\partial_t$ targets only the explicit scale dependence that was introduced into the regularized propagator $(P_k + R_k)^{-1}$ by the regulator $R_k = Z_k \, k^2/(1 + c \, p^2/k^2)$. The only explicit $k$-dependence in $\alpha^{\pm\pm}$ is contained in $\tilde m^2 = m^2/k^2$ and $\tilde\gamma^2 = \gamma^2/k^2$. By the chain rule $\tilde\partial_t \, \mu_j$ thus evaluates to
\begin{equation}
	\tilde\partial_t \mu_j
	= k \, \tilde\partial_k \mu_j
	= \begin{cases}
		2 k^2 \, \alpha^{\pm\pm} - 2 \tilde m^2 \, \partial_{\tilde m^2} \alpha^{\pm\pm} - 2 \tilde\gamma^2 \, \partial_{\tilde\gamma^2} \alpha^{\pm\pm}
		& j \in \{1,2,3,4\},\\
		2 k^2/c
		& j = 5,
	\end{cases}
\end{equation}
and zero for $j > 5$. Higher orders follow from $\tfim_0$ by taking derivatives with respect to $\tilde m^2$.

\subsection{Propagator}
\label{sec:prop ms}

We now perform the Matsubara summation for the threshold function $\tilde\partial_t \tfjm_{ab}$ in terms of which we formulated the flow equations for $Z_k$, $Z_1$, and $\gamma_1^2$. Consider \cite{floerchinger2017unpublished}
\begin{mleqn}
	\tfjm_{11}
	= \tfjm_{11}(&q,z_1,z_2,m_1^2,m_2^2,\gamma_1^2,\gamma_2^2,R)
	= \mmint_p G_1(p) \, G_2(p+q)\\
	&= \mmint_p \frac{1}{z_1 p^2 + m_1^2 - i s(p_0) \, \gamma_1^2 + R} \, \frac{1}{z_2 (p + q)^2 + m_2^2 - i s(p_0 + q_0) \, \gamma_2^2 + R}.
\end{mleqn}
The two propagators can be decomposed according to \labelcref{eqn:prop decomp},
\begin{align}\label{eqn:prop g1}
	G_1(p)
	&= \frac{\beta_1^+}{-p_0^2 + \vec{p}^2 + \alpha_1^+ k^2} + \frac{\beta_1^-}{-p_0^2 + \vec{p}^2 + \alpha_1^- k^2},\\
	\label{eqn:prop g2}
	G_2(p+q)
	&= \frac{\beta_2^+}{-(p_0 + q_0)^2 + \vec{p}^2 + \alpha_2^+ k^2} + \frac{\beta_2^-}{-(p_0 + q_0)^2 + \vec{p}^2 + \alpha_2^- k^2}.
\end{align}
We set $\vec{q} = 0$ since the external spatial momentum is irrelevant for the Matsubara summation. It affects neither the poles nor the branch cut structure in a qualitative way. We further defined
\begin{mleqn}
	&\alpha_1^\pm = \alpha^\pm\bigl(m_1^2,\gamma_1^2,s(p_0),z_1,c\bigr),
	\qquad
	&&\beta_1^\pm = \beta^\pm\bigl(m_1^2,\gamma_1^2,s(p_0),z_1,c\bigr),\\
	&\alpha_2^\pm = \alpha^\pm\bigl(m_2^2,\gamma_2^2,s(p_0 + q_0),z_2,c\bigr),
	\qquad
	&&\beta_2^\pm = \beta^\pm\bigl(m_2^2,\gamma_2^2,s(p_0 + q_0),z_2,c\bigr).
\end{mleqn}
Multiplying \labelcref{eqn:prop g1,eqn:prop g2} we get four terms,
\begin{mleqn}\label{eqn:j integral}
	\tfjm_{11}
	&= \smashoperator{\sum_{i,j\in\pm}} \mmint_p \frac{\beta_1^i}{-p_0^2 + \vec{p}^2 + \alpha_1^i k^2} \, \frac{\beta_2^j}{-(p_0 + q_0)^2 + \vec{p}^2 + \alpha_2^j k^2}\\
	&= \smashoperator{\sum_{i,j\in\pm}} \mmint_p \frac{\beta_1^i}{2 \csqrt{\vec{p}^2 + \alpha_1^i k^2}} \Biggl[\frac{1}{-p_0 + \csqrt{\vec{p}^2 + \alpha_1^i k^2}} - \frac{1}{-p_0 - \csqrt{\vec{p}^2 + \alpha_1^i k^2}}\Biggr]\\
	&\hphantom{{}=\smashoperator{\sum_{i,j\in\pm}} \mmint_p}\times \frac{\beta_2^j}{2 \csqrt{\vec{p}^2 + \alpha_2^j k^2}} \Biggl[\frac{1}{-(p_0 + q_0) + \csqrt{\vec{p}^2 + \alpha_2^j k^2}} - \frac{1}{-(p_0 + q_0) - \csqrt{\vec{p}^2 + \alpha_2^j k^2}}\Biggr].
\end{mleqn}
In the second step, we used the decomposition \labelcref{eqn:pfd}. \labelcref{eqn:j integral} has four simple poles at
\begin{equation}\label{eqn:poles}
	p_0 = \pm \csqrt{\vec{p}^2 + \alpha_1^i k^2}
	\qquad\text{and}\qquad
	p_0 = -q_0 \pm \csqrt{\vec{p}^2 + \alpha_2^j k^2},
\end{equation}
Due to the presence of $s(p_0)$ and $s(p_0 + q_0)$ in $\alpha_{1/2}^\pm$, $\beta_{1/2}^\pm$, the integrand \labelcref{eqn:j integral} also exhibits branch cuts.\footnote{Our notation suggests $q_0$ and $p_0$ are real. Keep in mind that $p_0$, $q_0$ are analytically continued frequencies. Their original domain, the imaginary axis, was extended to the entire complex plane. This enables $s(p_0)$ and $s(p_0 + q_0)$ to parametrize branch cuts along the real frequency axis.} To deal with these cuts, we use the structure of $s(p_0)$ to decompose the integrand in a manner very similar to \cref{eqn:integral decomposition} into a sum of four terms $\tfjm_{11} = \tfjm_1 + \tfjm_2 + \tfjm_3 + \tfjm_4$, each with a different branch cut structure parametrized by one of the following factors,
\begin{equation}
	1,
	\qquad
	s(p_0),
	\qquad
	s(p_0 + q_0),
	\qquad
	s(p_0) \, s(p_0 + q_0).
\end{equation}
$\tfjm_1$ contains neither $s(p_0)$ nor $s(p_0 + q_0)$ and is thus branch-cut-free. It takes the form
\begin{mleqn}\label{eqn:j1}
	\tfjm_1
	= \frac{1}{4} \sum_{i,j\in\pm} \sum_{r,s\in\pm} \mmint_p \frac{\beta_1^{ir}}{2 \csqrt{\vec{p}^2 + \alpha_1^{i\vphantom{j}r} k^2}} \Biggl[\frac{1}{-p_0 + \csqrt{\vec{p}^2 + \alpha_1^{i\vphantom{j}r} k^2}} - \frac{1}{-p_0 - \csqrt{\vec{p}^2 + \alpha_1^{i\vphantom{j}r} k^2}}\Biggr]\\
	\times \frac{\beta_2^{js}}{2 \csqrt{\vec{p}^2 + \alpha_2^{js} k^2}} \Biggl[\frac{1}{-(p_0 + q_0) + \csqrt{\vec{p}^2 + \alpha_2^{js} k^2}} - \frac{1}{-(p_0 + q_0) - \csqrt{\vec{p}^2 + \alpha_2^{js} k^2}}\Biggr]\mathrlap{.}
\end{mleqn}
The $\tfjm_i$ with $i > 1$ are identical except for additional factors of
\begin{equation}
	\tfjm_2: \enskip r \, s(p_0),
	\qquad
	\tfjm_3: \enskip s \, s(q_0 + p_0),
	\qquad
	\tfjm_4: \enskip r \, s \, s(p_0) \, s(q_0 + p_0),
\end{equation}
to be placed inside the sum and $p_0$-integral. Since $\tfjm_1$ contains no cut, we can immediately carry out the Matsubara summation via contour integration as done previously. This will pick up the residues at the poles in \labelcref{eqn:poles}, resulting in
\begingroup\allowdisplaybreaks
\begin{align}
	\tfjm_1
	&= \frac{1}{4} \sum_{i,j,r,s} \int_{\vec p} \frac{\beta_1^{ir}}{2 \csqrt{\vec{p}^2 + \alpha_1^{i\vphantom{j}r} k^2}} \frac{\beta_2^{js}}{2 \csqrt{\vec{p}^2 + \alpha_2^{js} k^2}}\notag\\
	&\hphantom{{}=} \Biggl\{ \Bigl[n_\text{B}\Bigl(\!\csqrt{\vec{p}^2 + \alpha_1^{i\vphantom{j}r} k^2}\Bigr) + \tfrac{1}{2}\Bigr] \tag{\footnotesize{contribution from poles at $p_0 = \pm \csqrt{\vec{p}^2 + \alpha_1^{i\vphantom{j}r} k^2}$}}\\
	\label{eqn:p0 poles}
	&\hphantom{{}=+}\times \Biggl(\frac{1}{-q_0 - \csqrt{\vec{p}^2 + \alpha_1^{i\vphantom{j}r} k^2} + \csqrt{\vec{p}^2 + \alpha_2^{js} k^2}} - \frac{1}{-q_0 - \csqrt{\vec{p}^2 + \alpha_1^{i\vphantom{j}r} k^2} - \csqrt{\vec{p}^2 + \alpha_2^{js} k^2}}\\
	&\hphantom{{}=+\times\Biggl(}+ \frac{1}{-q_0 + \csqrt{\vec{p}^2 + \alpha_1^{i\vphantom{j}r} k^2} + \csqrt{\vec{p}^2 + \alpha_2^{js} k^2}} - \frac{1}{-q_0 + \csqrt{\vec{p}^2 + \alpha_1^{i\vphantom{j}r} k^2} - \csqrt{\vec{p}^2 + \alpha_2^{js} k^2}}\Biggr)\notag\\[1ex]
	&\hphantom{{}=}+ \Bigl[n_\text{B}\Bigl(\!\csqrt{\vec{p}^2 + \alpha_2^{js} k^2}\Bigr) + \tfrac{1}{2}\Bigr] \tag{\footnotesize{contribution from poles at $p_0 = -q_0 \pm \csqrt{\vec{p}^2 + \alpha_2^{js} k^2}$}}\\
	\label{eqn:p0+q0 poles}
	&\hphantom{{}=+}\times \Biggl(\frac{1}{q_0 - \csqrt{\vec{p}^2 + \alpha_2^{js} k^2} + \csqrt{\vec{p}^2 + \alpha_1^{i\vphantom{j}r} k^2}} - \frac{1}{q_0 - \csqrt{\vec{p}^2 + \alpha_2^{js} k^2} - \csqrt{\vec{p}^2 + \alpha_1^{i\vphantom{j}r} k^2}}\\
	&\hphantom{{}=+\times\Biggl(}+ \frac{1}{q_0 + \csqrt{\vec{p}^2 + \alpha_2^{js} k^2} + \csqrt{\vec{p}^2 + \alpha_1^{i\vphantom{j}r} k^2}} - \frac{1}{q_0 + \csqrt{\vec{p}^2 + \alpha_2^{js} k^2} - \csqrt{\vec{p}^2 + \alpha_1^{i\vphantom{j}r} k^2}}\Biggr)\Biggr\}\notag\\[1ex]
	\label{eqn:p0 p0+q0 poles combined}
	&= \frac{1}{4} \sum_{i,j,r,s} \int_{\vec p} \frac{\beta_1^{ir}}{2 \csqrt{\vec{p}^2 + \alpha_1^{i\vphantom{j}r} k^2}} \frac{\beta_2^{js}}{2 \csqrt{\vec{p}^2 + \alpha_2^{js} k^2}}\\
	&\text{\small$\displaystyle\hphantom{{}=}\times \Biggl[\frac{n_\text{B}\Bigl(\!\csqrt{\vec{p}^2 + \alpha_1^{i\vphantom{j}r} k^2}\Bigr) + n_\text{B}\Bigl(\!\csqrt{\vec{p}^2 + \alpha_2^{js} k^2}\Bigr) + 1}{-q_0 + \csqrt{\vec{p}^2 + \alpha_1^{i\vphantom{j}r} k^2} + \csqrt{\vec{p}^2 + \alpha_2^{js} k^2}} + \frac{n_\text{B}\Bigl(\!\csqrt{\vec{p}^2 + \alpha_1^{i\vphantom{j}r} k^2}\Bigr) - n_\text{B}\Bigl(\!\csqrt{\vec{p}^2 + \alpha_2^{js} k^2}\Bigr)}{-q_0 - \csqrt{\vec{p}^2 + \alpha_1^{i\vphantom{j}r} k^2} + \csqrt{\vec{p}^2 + \alpha_2^{js} k^2}}$}\notag\\
	&\text{\small$\displaystyle\hphantom{{}=\times\Biggl(}
	- \frac{n_\text{B}\Bigl(\!\csqrt{\vec{p}^2 + \alpha_1^{i\vphantom{j}r} k^2}\Bigr) + n_\text{B}\Bigl(\!\csqrt{\vec{p}^2 + \alpha_2^{js} k^2}\Bigr) + 1}{-q_0 - \csqrt{\vec{p}^2 + \alpha_1^{i\vphantom{j}r} k^2} - \csqrt{\vec{p}^2 + \alpha_2^{js} k^2}}
	- \frac{n_\text{B}\Bigl(\!\csqrt{\vec{p}^2 + \alpha_1^{i\vphantom{j}r} k^2}\Bigr) - n_\text{B}\Bigl(\!\csqrt{\vec{p}^2 + \alpha_2^{js} k^2}\Bigr)}{-q_0 + \csqrt{\vec{p}^2 + \alpha_1^{i\vphantom{j}r} k^2} - \csqrt{\vec{p}^2 + \alpha_2^{js} k^2}}\Biggr].$}\notag
\end{align}\endgroup
In \labelcref{eqn:p0 poles}, we used antisymmetry of $n_\text{B}(-p_0) + \frac{1}{2} = -\bigl[n_\text{B}(p_0) + \frac{1}{2}\bigr]$ to pull out an overall factor of $\bigl[n_\text{B}\bigl(\!\csqrt{\vec{p}^2 + \alpha_1^{i\vphantom{j}r} k^2}\bigr) + \tfrac{1}{2}\bigr]$ in front of all four terms, even though the last two terms originate from the pole at $p_0 = -\csqrt{\vec{p}^2 + \alpha_1^{i\vphantom{j}r} k^2}$ and therefore initially appear with a factor $\bigl[n_\text{B}\bigl(\!-\csqrt{\vec{p}^2 + \alpha_1^{i\vphantom{j}r} k^2}\bigr) + \tfrac{1}{2}\bigr]$. This gives rise to a relative sign in front of the last two terms which is compensated by another sign in front of $-\bigl(-p_0 - \csqrt{\vec{p}^2 + \alpha_1^{i\vphantom{j}r} k^2}\bigr)^{-1}$. Likewise, in \labelcref{eqn:p0+q0 poles} we used $n_\text{B}(-p_0 - q_0) + \frac{1}{2} = -\bigl[n_\text{B}(p_0 + q_0) + \frac{1}{2}\bigr]$ which compensates the sign in front of $-\bigl[-(p_0 + q_0) - \csqrt{\vec{p}^2 + \alpha_2^{js} k^2}\bigr]^{-1}$. \labelcref{eqn:p0 p0+q0 poles combined} follows from combining terms in \labelcref{eqn:p0 poles,eqn:p0+q0 poles}.

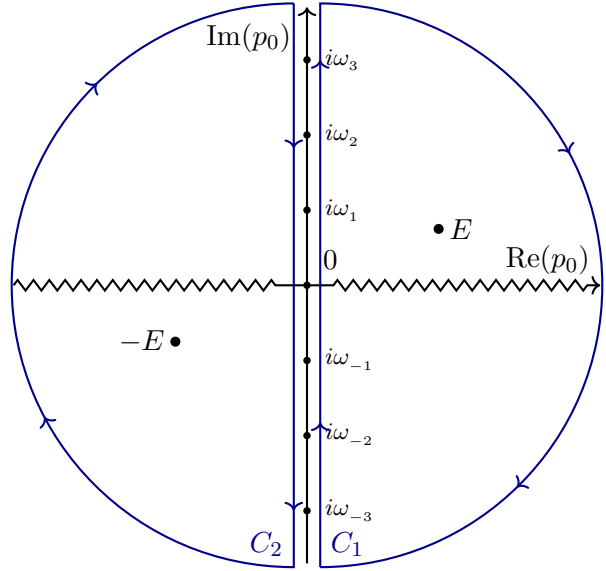
\begin{figure}[htb!]
	\centering
	\begin{tikzpicture}[thick]
	    
	    \def\xr{3.5} \def\yr{3}
	    
	    \draw [decorate,decoration={zigzag,segment length=6,amplitude=2,post=lineto,post length=10}] (-\xr-0.4,0) -- (0,0);
	    \draw [->,decorate,decoration={zigzag,segment length=6,amplitude=2,pre=lineto,pre length=10,post=lineto,post length=3}] (0,0) -- (\xr+0.4,0) node [above left]  {$\Re(p_0)$};
	    \draw [->] (0,-\yr-0.7) -- (0,\yr+0.7) node[below left=0.1] {$\Im(p_0)$};
	    
	    \foreach \n in {-\yr,...,-1,1,2,...,\yr}{%
	        \draw[fill] (0,\n) circle (1pt) node [right=0.1,font=\footnotesize] {$i \omega_{_{\n}}$};}
	    \draw[fill] (0,0) circle (1pt) node [above right=0.1] {0};
	    
	    \draw[xshift=5,DarkBlue,decoration={markings,mark=between positions 0.1 and 1 step 0.25 with \arrow{>}},postaction={decorate}] (0,-\yr-0.75) node [above right] {$C_1$} -- (0,\yr+0.75) arc (90:-90:\yr+0.75);
	    
	    \draw[xshift=-5,DarkBlue,decoration={markings,mark=between positions 0.1 and 1 step 0.25 with \arrow{>}},postaction={decorate}] (0,\yr+0.75) -- (0,-\yr-0.75) node [above left] {$C_2$} arc (270:90:\yr+0.75);
	    
	    \draw[fill] (\xr/2,\yr/4) circle (1.5pt) node [right] {$E$};
	    \draw[fill] (-\xr/2,-\yr/4) circle (1.5pt) node [left] {$-E$};
	    
	\end{tikzpicture}
	\caption{Contour enclosing poles and branch cuts of $\tfjm_{2/3}$ but excluding the Matsubara frequencies}
	\label{fig:contour 5}
\end{figure}

The contributions from $\tfjm_2$ and $\tfjm_3$ vanish to good approximation. The reason is again the same cancellation between pole and branch cut contributions demonstrated for $\Delta \tfim_0$ in \cref{eqn:pole contrib,eqn:l+r contrib,eqn:si of x}. To see this explicitly, consider the integral
\begin{equation}\label{eqn:vanishing contribution}
	\int_{C_1+C_2} \frac{\dif p_0}{2 \pi i} \, s(p_0) \, f(p_0) \biggl(\frac{1}{p_0 - E} - \frac{1}{p_0 + E}\biggr) \bigl[n_\text{B}(p_0) + \tfrac{1}{2}\bigr],
\end{equation}
where $C_1$ and $C_2$ enclose the entire complex plane save the imaginary axis as shown in \cref{fig:contour 5}.
The poles at $p_0 = \pm E$ contribute
\begin{equation}\label{eqn:pole contribution}
	-s(E) \bigl[f(E) + f(-E)\bigr] \bigl[n_\text{B}(E) + \tfrac{1}{2}\bigr],
\end{equation}
where we used $s(E) = s(-E)$ and $n_\text{B}(-E) + \tfrac{1}{2} = -\bigl[n_\text{B}(E) + \tfrac{1}{2}\bigr]$. The overall sign comes from the clockwise contour. The branch cut is evaluated by shrinking $C_1$ and $C_2$ until all poles scattered throughout the complex plane are excluded. We can target only the contribution from the branch cut by again integrating along $C_b$ in \cref{fig:contour 4}. This gives
\begin{equation}\label{eqn:cut contribution}
	2 \Biggl[\int_0^\infty - \int_{-\infty}^0\Biggr] \frac{\dif p_0}{2 \pi i} \, f(p_0) \biggl(\frac{1}{p_0 - E} - \frac{1}{p_0 + E}\biggr) \bigl[n_\text{B}(p_0) + \tfrac{1}{2}\bigr].
\end{equation}
The factor of 2 comes from running back and forth along the real axis. The relative sign derives from $s(p_0)$ since $\sign(\Re p_0) < 0$ for the left half of the contour. For sufficiently large $\Re(E)$ the first integral is strongly dominated by the pole at $p_0 = E$ while the second is dominated by $p_0 = -E$. That is because, as shown in \cref{fig:bose distribution}, $n_\text{B}(p_0)$ is approximately constant away from the real axis, especially at low temperatures. If we assume the same property for $f(p_0)$ (in the case of the $\tfjm_i$, this is even true exactly since they contain no $p_0$-dependence besides the pole structure and the jump at $\Re p_0 = 0$ due to $s(p_0)$), we can replace
\begin{equation}
	\bigl[n_\text{B}(p_0) + \tfrac{1}{2}\bigr]
	\to \bigl[n_\text{B}(E) + \tfrac{1}{2}\bigr],
	\qquad
	f(p_0)
	\to f(E)
\end{equation}
in the first integral in \labelcref{eqn:cut contribution}, and
\begin{equation}
	\bigl[n_\text{B}(p_0) + \tfrac{1}{2}\bigr] 
	\to \bigl[n_\text{B}(-E) + \tfrac{1}{2}\bigr] = -\bigl[n_\text{B}(E) + \tfrac{1}{2}\bigr],
	\qquad
	f(p_0)
	\to f(-E)
\end{equation}
in the second. This gives
\begin{equation}\label{eqn:symmetric integrand}
	2 \Biggl[\int_0^\infty f(E) + \int_{-\infty}^0 f(-E)\Biggr] \frac{\dif p_0}{2 \pi i} \biggl(\frac{1}{p_0 - E} - \frac{1}{p_0 + E}\biggr) \bigl[n_\text{B}(E) + \tfrac{1}{2}\bigr].
\end{equation}
Since the integrand is now symmetric with respect to $p_0 \to -p_0$, we have $2 \int_0^\infty = 2 \int_{-\infty}^0 = \int_{-\infty}^\infty$, and so \labelcref{eqn:symmetric integrand} can be written
\begin{equation}
	\int_{-\infty}^\infty \frac{\dif p_0}{2 \pi i} \, \bigl[f(E) + f(-E)\bigr]\biggl(\frac{1}{p_0 - E} - \frac{1}{p_0 + E}\biggr) \bigl[n_\text{B}(E) + \tfrac{1}{2}\bigr].
\end{equation}
The integral can be closed with a half-circle at infinity and evaluated by means of the residue theorem, resulting in
\begin{equation}
	s(E) \, \bigl[f(E) + f(-E)\bigr] \bigl[n_\text{B}(E) + \tfrac{1}{2}\bigr],
\end{equation}
which precisely cancels the contribution \labelcref{eqn:pole contribution} from the poles.

Finally, $\tfjm_4$ is proportional to $\gamma_1^2 \gamma_2^2$. In many cases, this vanishes exactly. For instance, if (at least) one of the particles running in the loop is a stable massless Goldstone boson with vanishing decay width $\Gamma = \gamma^2/m = 0$. Even if $\gamma_1^2 \neq 0 \neq \gamma_2^2$ are not zero for all $p$, $\tfjm_4$ will only receive contributions from those $p$ for which both $\gamma_1^2(p^2) \neq 0$ \textit{and} $\gamma_2^2(p^2) \neq 0$. Since $G(p)$ is without discontinuities if $p^2$ is positive, such $p$ are few. Thus the value of $\tfjm_4$ is expected to be small even in cases where it is non-zero. It will therefore be neglected. In summary, we have approximately $\tfjm_{11} \approx \tfjm_1$ and the result of the Matsubara summation in $\tfjm_{11}$ is simply \labelcref{eqn:p0 p0+q0 poles combined}.

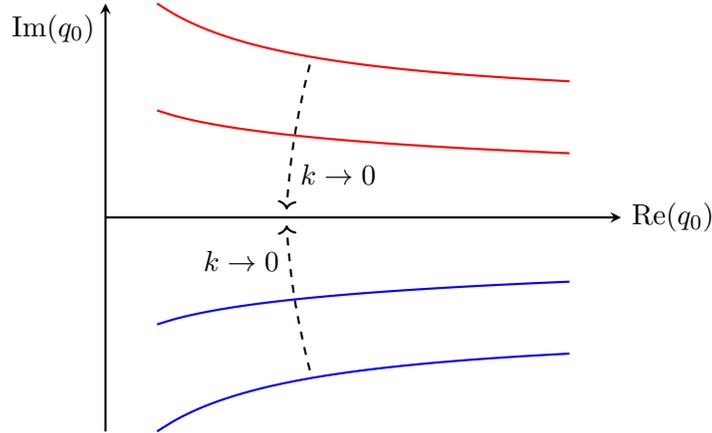
\begin{figure}[htb!]
	\centering
	\begin{tikzpicture}
        \begin{axis}[
            xmin = 0,xmax = 10,
            xlabel = $\Re(q_0)$,
            ylabel = $\Im(q_0)$,
            smooth,axis lines=middle,thick,
            every tick/.style = {thick},
            xlabel style = right,
            ylabel style = below left,
            ticks=none]
                        
            \addplot[color=red,domain = 1:9]{1/(sqrt(x)+2)};
            \addplot[color=red,domain = 1:9]{1/(x+2) + 1/3};
                        
            \addplot[color=blue,domain = 1:9]{-1/(sqrt(x)+2)};
            \addplot[color=blue,domain = 1:9]{-1/(x+2) - 1/3};
        
            \draw[dashed,shorten >=3,shorten <=3] (axis cs:4,0.5) edge [->,bend right=5] node[near end,right] {$k \to 0$} (axis cs:3.5,0);
            \draw[dashed,shorten >=3,shorten <=3] (axis cs:4,-0.5) edge [->,bend left=5] node[near end,left] {$k \to 0$} (axis cs:3.5,0);
            
        \end{axis}
    \end{tikzpicture}
	\caption{Branch cut structure of $\tfjm_1$ in the $q_0$-plane}
	\label{fig:cuts}
\end{figure}

Looking more closely at the expression \labelcref{eqn:p0 p0+q0 poles combined} reveals that for $k > 0$, $\tfjm_1$ contains not only a single discontinuity on the real frequency axis but several jumps along lines that are approximately parallel to the real $q_0$-axis as shown in \cref{fig:cuts}. The cuts are located at
\begin{equation}
	q_0
	= \csqrt{\vec{p}^2 + \alpha_1^{i\vphantom{j}r} k^2} + \csqrt{\vec{p}^2 + \alpha_2^{js} k^2}
	\qquad\text{and}\qquad
	q_0
	= \Bigl|\csqrt{\vec{p}^2 + \alpha_1^{i\vphantom{j}r} k^2} - \csqrt{\vec{p}^2 + \alpha_2^{js} k^2}\Bigr|.
\end{equation}
(Note that we were dealing with branch cuts in the $p_0$-plane when we decomposed $\tfjm_{11}$ into $\tfjm_1 + \tfjm_2 + \tfjm_3 + \tfjm_4$ and said that $\tfjm_1$ is branch cut-free. Now, we are looking at $\tfjm_1$'s branch cut structure in the $q_0$-plane.)

The cuts approach the real $q_0$-axis with decreasing $k$ where they merge for $k \to 0$. Instead of taking into account the complete analytic structure of $\tfjm_1$, we approximate the regularized propagator as having only a single cut on the real $q_0$-axis. To simplify the calculation, we sum the contributions from the different discontinuities at real $q_0$ even if they are shifted away from the real axis for $k \neq 0$. For $k \to 0$ this simplification converges on the correct result and for $k > 0$ it yields a reasonable approximation.

We project the cuts to the real $q_0$-axis by replacing in \labelcref{eqn:p0 p0+q0 poles combined}
\begin{equation}
	\alpha_1^{ir}
	\to \Re \alpha_1^{ir},
	\qquad
	\alpha_2^{js}
	\to \Re \alpha_2^{js}.
\end{equation}
This gives
\begin{mleqn}\label{eqn:disc j}
	\disc_{q_0} \tfjm_1
	= \frac{1}{4} \sum_{i,j,r,s} \int_{\vec p} \frac{\beta_1^{ir}}{2 \csqrt{\vec{p}^2 + \alpha_1^{i\vphantom{j}r} k^2}} \frac{\beta_2^{js}}{2 \csqrt{\vec{p}^2 + \alpha_2^{js} k^2}} \biggl\{\Bigl[n_\text{B}\Bigl(\!\csqrt{\vec{p}^2 + \alpha_1^{i\vphantom{j}r} k^2}\Bigr) + n_\text{B}\Bigl(\!\csqrt{\vec{p}^2 + \alpha_2^{js} k^2}\Bigr)\Bigr]\\
	\times \sign\bigl(\Re\alpha_1^{ir} - \Re\smash{\alpha_2^{js}}\bigr) \pi \delta\Bigl(q_0 - \Bigl|\csqrt{\vec{p}^2 + \Re\alpha_1^{ir} k^2} - \csqrt{\vec{p}^2 + \Re\smash{\alpha_2^{js}} k^2}\Bigr|\Bigr)\\
	+ \Bigl[n_\text{B}\Bigl(\!\csqrt{\vec{p}^2 + \alpha_1^{i\vphantom{j}r} k^2}\Bigr) + n_\text{B}\Bigl(\!\csqrt{\vec{p}^2 + \alpha_2^{js} k^2}\Bigr) + 1\Bigr] \pi \delta\Bigl(q_0 - \csqrt{\vec{p}^2 + \Re\alpha_1^{ir} k^2} - \csqrt{\vec{p}^2 + \Re\smash{\alpha_2^{js}} k^2}\Bigr)\mathrlap{\biggr\}.}
\end{mleqn}

\section{Momentum Integration}
\label{sec:momentum integration}

\subsection{Effective Potential}

Since the integrand \labelcref{eqn:delta i0 result} of $\Delta \tfim_0$ depends only on the magnitude of $\vec{p}$, the $d$-dimensional momentum integration is best carried out in spherical coordinates with trivial angular integration \cite{floerchinger2017unpublished},
\begin{equation}
	\int_{\vec p}
	= \int_{\reals^d} \frac{\dif^d \vec p}{(2 \pi)^d}
	= S_d \int_0^\infty \frac{\dif p}{(2 \pi)^d} \, p^{d-1},
	\qquad
	S_d
	= \frac{2 \pi^{d/2}}{\Gamma\bigl(\frac{d}{2}\bigr)},
\end{equation}
where $S_d$ denotes the surface area of the $d$-dimensional unit sphere and $p = |\vec{p}|$ is no longer the 4-momentum but the spatial momentum magnitude. The temperature-dependent part of the $\vec{p}$-integral in \labelcref{eqn:delta i0 result} will be evaluated numerically. Considering only the $T$-independent part, we can proceed analytically.
\begin{equation}\label{eqn:mom int}
	\int \dif p \, p^{d-1} \, \frac{1}{2} \csqrt{p^2 + \mu_j}
	= \begin{cases}
		\frac{1}{16} \bigl[p \, \csqrt{p^2 + \mu_j} \, (2 p^2 + \mu_j) - \mu_j^2 \ln\bigl(p + \csqrt{p^2 + \mu_j}\bigr)\bigr] & d = 3,\\
		\frac{1}{6} \bigl[p^2 + \mu_j\bigr]^{3/2} & d = 2,\\
		\frac{1}{4} \bigl[p \, \csqrt{p^2 + \mu_j} + \mu_j \ln\bigl(p + \csqrt{p^2 + \mu_j}\bigr)\bigr] & d = 1.
	\end{cases}
\end{equation}
Approximating \labelcref{eqn:mom int} for $d = 3$ at the upper integration boundary where $p^2 \gg \mu_j$ gives
\begin{mleqn}\label{eqn:ub3}
	\frac{1}{16} \bigl[p& \, \csqrt{p^2 + \mu_j} \, (2 p^2 + \mu_j) - \mu_j^2 \ln\bigl(p + \csqrt{p^2 + \mu_j}\bigr)\bigr]\\
	&= \frac{1}{16} \biggl[2 p^4 \sqrt{1 + \frac{\mu_j}{p^2}} \biggl(1 + \frac{\mu_j}{2 p^2}\biggr) - \mu_j^2 \ln\biggl(p + p \underbrace{\sqrt{1 + \frac{\mu_j}{p^2}}}_{\approx 1}\biggr)\biggr]\\
	&\approx \frac{p^4}{8} \biggl(1 + \frac{\mu_j}{2 p^2}\biggr) \biggl(1 + \frac{\mu_j}{2 p^2}\biggr) - \frac{\mu_j^2}{16} \ln(2 p)
	\approx \frac{p^4}{8} + \frac{\mu_j p^2}{8} + \frac{\mu_j^2}{32} - \frac{\mu_j^2}{16} \ln(2 p),
\end{mleqn}
where we used $(1 + x)^n \approx 1 + n x$ for $x \ll 1$. In $d = 2$, the same boundary contributes
\begin{equation}\label{eqn:ub2}
	\frac{1}{6} (p^2 + \mu_j)^{3/2}
	= \frac{p^3}{6} \biggl(1 + \frac{\mu_j}{p^2}\biggr)^{3/2}
	\approx \frac{p^3}{6} + \frac{\mu_j p}{4},
\end{equation}
and in $d = 1$
\begin{equation}\label{eqn:ub1}
	\frac{1}{4} \bigl[p \, \csqrt{p^2 + \mu_j} + \mu_j \ln\bigl(p + \csqrt{p^2 + \mu_j}\bigr)\bigr]
	\approx \frac{p^2}{4} + \frac{\mu_j}{8} + \frac{\mu_j}{4} \ln(2 p).
\end{equation}
Since $w_j = -w_{j+5}$ all the leading terms in \labelcref{eqn:ub3,eqn:ub2,eqn:ub1} without a factor of $\mu_j$ cancel under the sum over $j$. Terms containing $\mu_j$ cancel as well up to a shift of the effective potential. To see why, consider
\begin{equation}
	\alpha^{++} k^2 \, \alpha^{+-} k^2 + \alpha^{-+} k^2 \, \alpha^{--} k^2
	= \frac{2 m^2}{z} + \frac{2 k^2}{c},
\end{equation}
and similarly for the $(k \to \Lambda)$-term. Using the definition \labelcref{eqn:alpha}, we can calculate
\begin{mleqn}
	&(\alpha^{++})^2 + (\alpha^{+-})^2 + (\alpha^{-+})^2 + (\alpha^{--})^2\\
	&=\biggl[\frac{1}{2} \biggl(\frac{1}{c} + \frac{\tilde m^2}{z} - i \, \frac{\tilde\gamma^2}{z}\biggr) + \bigl(A + i \, B\bigr)\biggr]^2
	+ \biggl[\frac{1}{2} \biggl(\frac{1}{c} + \frac{\tilde m^2}{z} + i \, \frac{\tilde\gamma^2}{z}\biggr) + \bigl(A - i \, B\bigr)\biggr]^2\\
	&\hphantom{{}={}}+ \biggl[\frac{1}{2} \biggl(\frac{1}{c} + \frac{\tilde m^2}{z} - i \, \frac{\tilde\gamma^2}{z}\biggr) - \bigl(A + B\bigr)\biggr]^2
	+ \biggl[\frac{1}{2} \biggl(\frac{1}{c} + \frac{\tilde m^2}{z} + i \, \frac{\tilde\gamma^2}{z}\biggr) - \bigl(A - B\bigr)\biggr]^2\\
	&= \biggl(\frac{1}{c} + \frac{\tilde m^2}{z}\biggr)^2 - \frac{\tilde\gamma^4}{z^2} + 4 (A^2 - B^2)
	\crefrel{eqn:-A^2+B^2}{=} -\frac{4}{c z} + 2 \biggl(\frac{1}{c^2} + \frac{\tilde m^4}{z^2} - \frac{\tilde\gamma^4}{z^2}\biggr),
\end{mleqn}
where in the last step we used $4 (A^2 - B^2) = -\frac{4}{c \, z} + \Bigl(\frac{1}{c} - \frac{\tilde m^2}{z}\Bigr)^2 - \frac{\tilde\gamma^4}{z^2}$. In combination one finds
\begin{equation}\label{eqn:irrel shift}
	\sum_{j=1}^{10} w_j \, \mu_j^2 = \frac{4}{c \, z} (\Lambda^4 - k^4).
\end{equation}
This can be neglected since it is independent of $\tilde m^2$ and $\tilde\gamma^2$ and therefore only amounts to a temperature-independent shift of the effective potential, immaterial for most practical purposes. Thus, for $d \in \{1,2,3\}$, the contribution from the upper boundary of the momentum integration vanishes. The lower boundary, on the other hand, is non-zero and contributes with a minus sign,
\begin{mleqn}
	&-\frac{1}{16} \bigl[p \, \csqrt{p^2 + \mu_j} \, (2 p^2 + \mu_j) - \mu_j^2 \ln\bigl(p + \csqrt{p^2 + \mu_j}\bigr)\bigr]
	&&\xrightarrow{p \to 0} \frac{1}{32} \mu_j^2 \ln(\mu_j),
	&&d = 3,\\
	&-\frac{1}{6} \bigl[p^2 + \mu_j\bigr]^{3/2}
	&&\xrightarrow{p \to 0} -\frac{1}{6} \mu_j^{3/2},
	&&d = 2,\\
	&-\frac{1}{4} \bigl[p \, \csqrt{p^2 + \mu_j} + \mu_j \ln\bigl(p + \csqrt{p^2 + \mu_j}\bigr)\bigr]
	&&\xrightarrow{p \to 0} -\frac{1}{8} \mu_j \ln(\mu_j),
	\qquad &&d = 1,
\end{mleqn}
and in combination for $\Delta \tfim_0$,
\begin{equation}
	\Delta \tfim_0
	= \frac{S_d}{(2 \pi)^d} \sum_{j=1}^{10} w_j
	\begin{cases}
		\frac{1}{32} \mu_j^2 \ln(\mu_j) & d = 3,\\
		-\frac{1}{6} \mu_j^{3/2} & d = 2,\\
		-\frac{1}{8} \mu_j \ln(\mu_j) & d = 1.
	\end{cases}
\end{equation}
The prefactor is
\begin{equation}
	\frac{S_d}{(2 \pi)^d}
	= \frac{1}{(2 \pi)^d} \frac{2 \pi^{d/2}}{\Gamma\bigl(\frac{d}{2}\bigr)} = 
	\begin{cases}
		\frac{1}{2 \pi^2} & d = 3,\\
		\frac{1}{2 \pi} & d = 2,\\
		\frac{1}{\pi} & d = 1.
	\end{cases}
\end{equation}
We thus get the following explicit expressions for the dimensionless threshold functions $\tfimdl_j = k^{2j - d - 1} \tfim_j$ for different values of $d$ \cite{floerchinger2017unpublished}.

\paragraph{$\boldsymbol{d + 1 = 4}$} Dropping again the shift of the effective potential \labelcref{eqn:irrel shift}, $\Delta \tfimdl_0$ takes the form
\begin{align}\label{eqn:d=3 delta i0}
	&\Delta \tfimdl_0
	= k^{-4} \, \Delta \tfim_0
	= k^{-4} \biggl[\frac{m^4 - \gamma^4}{32 \pi^2 z^2} \, \ln(k^2/\Lambda^2) + \mathcal{K}_4^k \, k^4 - \mathcal{K}_4^\Lambda \, \Lambda^4\biggl],\\
	&\tfimdl_0
	= k^{-4} \, \tilde\partial_t \, \Delta \tfim_0
	= \frac{\tilde m^4 - \tilde\gamma^4}{16 \pi^2 z^2} + \Bigl(4 - 2 \tilde m^2 \, \partial_{\tilde m^2} - 2 \tilde\gamma^2 \, \partial_{\tilde\gamma^2}\Bigr) \mathcal{K}_4,
\end{align}
Recall that $\tilde\partial_t = k \, \tilde\partial_k$ only targets explicit $k$-dependence of $R_k$ in $\Delta I_0$. The kernel $\mathcal{K}_3^k$ is defined as
\begin{mleqn}
	\mathcal{K}_4
	&= \frac{1}{64 \pi^2} \sum_{j=1}^4 w_j \mu_j^2 \ln(\mu_j)\\
	&= \frac{1}{64 \pi^2} \Bigl[(\alpha^{++})^2 \ln(\alpha^{++}) + (\alpha^{+-})^2 \ln(\alpha^{+-}) + (\alpha^{-+})^2 \ln(\alpha^{-+}) + (\alpha^{--})^2 \ln(\alpha^{--})\Bigr].
\end{mleqn}
Higher orders can be generated by taking derivatives with respect to to $\tilde{m}^2 = m^2/k^2$.
\begin{align}
	\tfimdl_1
	&= \partial_{\tilde{m}^2} \tfimdl_0
	= \frac{\tilde{m}^2}{8 \pi^2 z^2} + \Bigl(2 \partial_{\tilde{m}^2} - 2 \tilde{m}^2 \, \partial_{\tilde{m}^2}^2 - 2 \tilde\gamma^2 \, \partial_{\tilde{m}^2} \, \partial_{\tilde\gamma^2}\Bigr) \mathcal{K}_4,\\
	\tfimdl_2
	&= -\partial_{\tilde{m}^2} \tfimdl_1
	= -\frac{1}{8 \pi^2 z^2} + \Bigl(2 \tilde{m}^2 \, \partial_{\tilde{m}^2}^3 + 2 \tilde\gamma^2 \, \partial_{\tilde{m}^2}^2 \, \partial_{\tilde\gamma^2}\Bigr) \mathcal{K}_4,\\
	\tfimdl_3
	&= -\frac{1}{2} \, \partial_{\tilde{m}^2} \tfimdl_2
	= -\frac{1}{2} \Bigl(2 \partial_{\tilde{m}^2}^3 + 2 \tilde{m}^2 \, \partial_{\tilde{m}^2}^4 + 2 \tilde\gamma^2 \, \partial_{\tilde{m}^2}^3 \, \partial_{\tilde\gamma^2}\Bigr) \mathcal{K}_4,\\
	\tfimdl_4
	&= -\frac{1}{3} \partial_{\tilde{m}^2} \tfimdl_3
	= \frac{1}{6} \Bigl(4 \partial_{\tilde{m}^2}^4 + 2 \tilde{m}^2 \, \partial_{\tilde{m}^2}^5 + 2 \tilde\gamma^2 \, \partial_{\tilde{m}^2}^4 \, \partial_{\tilde\gamma^2}\Bigr) \mathcal{K}_4.
\end{align}

\paragraph{$\boldsymbol{d + 1 = 3}$} In two spatial dimensions,
\begin{equation}
	\mathcal{K}_3
	= -\frac{1}{12 \pi} \sum_{j=1}^4 w_j \mu_j^{3/2}
	= -\frac{1}{12 \pi} \Bigl[(\alpha^{++})^{3/2} + (\alpha^{+-})^{3/2} + (\alpha^{-+})^{3/2} + (\alpha^{--})^{3/2}\Bigr]
\end{equation}
in terms of which the threshold functions read
\begin{align}
	\Delta \tfimdl_0
	&= k^{-3} \Bigl(\mathcal{K}_3^k \, k^3 - \mathcal{K}_3^\Lambda \, \Lambda^3\Bigr),\\
	\tfimdl_0
	&= k^{-3} \, \tilde\partial_t \, \Delta \tfim_0
	= \Bigl(3 - 2 \tilde{m}^2 \partial_{\tilde{m}^2} - 2 \tilde\gamma^2 \, \partial_{\tilde\gamma^2}\Bigr) \mathcal{K}_3,\\
	\tfimdl_1
	&= \partial_{\tilde{m}^2} \tfimdl_0
	= \Bigl(\partial_{\tilde{m}^2} - 2 \tilde{m}^2 \, \partial_{\tilde{m}^2}^2 - 2 \tilde\gamma^2 \, \partial_{\tilde{m}^2} \, \partial_{\tilde\gamma^2}\Bigr) \mathcal{K}_3,\\
	\tfimdl_2
	&= -\partial_{\tilde{m}^2} \tfimdl_1
	= \Bigl(\partial_{\tilde{m}^2}^2 + 2 \tilde{m}^2 \, \partial_{\tilde{m}^2}^3 + 2 \tilde\gamma^2 \, \partial_{\tilde{m}^2}^2 \, \partial_{\tilde\gamma^2}\Bigr) \mathcal{K}_3,\\
	\tfimdl_3
	&= -\frac{1}{2} \, \partial_{\tilde{m}^2} \tfimdl_2
	= -\frac{1}{2} \Bigl(3 \, \partial_{\tilde{m}^2}^3 + 2 \tilde{m}^2 \, \partial_{\tilde{m}^2}^4 + 2 \tilde\gamma^2 \, \partial_{\tilde{m}^2}^3 \, \partial_{\tilde\gamma^2}\Bigr) \mathcal{K}_3,\\
	\tfimdl_4
	&= -\frac{1}{3} \, \partial_{\tilde{m}^2} \tfimdl_3
	= \frac{1}{6} \Bigl(5 \, \partial_{\tilde{m}^2}^4 + 2 \tilde{m}^2 \, \partial_{\tilde{m}^2}^5 + 2 \tilde\gamma^2 \, \partial_{\tilde{m}^2}^4 \, \partial_{\tilde\gamma^2}\Bigr) \mathcal{K}_3.
\end{align}

\paragraph{$\boldsymbol{d + 1 = 2}$} For a single dimension of space, the threshold functions take the form
\begin{align}
	\Delta \tfimdl_0
	&= k^{-2} \biggl[-\frac{m^2}{4 \pi z} \ln(k^2/\Lambda^2) + \mathcal{K}_2^k \, k^2 - \mathcal{K}_2^\Lambda \, \Lambda^2\biggr],\\
	\tfimdl_0
	&= k^{-2} \, \tilde\partial_t \, \Delta \tfim_0
	= -\frac{m^2}{2 z k^2} + \Bigl(2 - 2 \tilde{m}^2 \, \partial_{\tilde{m}^2} - 2 \tilde\gamma^2 \, \partial_{\tilde\gamma^2}\Bigr) \mathcal{K}_2,\\
	\tfimdl_1
	&= -\partial_{\tilde{m}^2} \tfimdl_0
	= \frac{1}{2 z} + \Bigl(2 \tilde{m}^2 \, \partial_{\tilde{m}^2}^2 + 2 \tilde\gamma^2 \, \partial_{\tilde{m}^2} \, \partial_{\tilde\gamma^2}\Bigr) \mathcal{K}_2,\\
	\tfimdl_2
	&= -\partial_{\tilde{m}^2} \tfimdl_1
	= -\Bigl(2 \partial_{\tilde{m}^2}^2 + 2 \tilde{m}^2 \, \partial_{\tilde{m}^2}^3 + 2 \tilde\gamma^2 \, \partial_{\tilde{m}^2}^2 \, \partial_{\tilde\gamma^2}\Bigr) \mathcal{K}_2,\\
	\tfimdl_3
	&= -\frac{1}{2} \, \partial_{\tilde{m}^2} \tfimdl_2
	= \frac{1}{2} \Bigl(4 \, \partial_{\tilde{m}^2}^3 + 2 \tilde{m}^2 \, \partial_{\tilde{m}^2}^4 + 2 \tilde\gamma^2 \, \partial_{\tilde{m}^2}^3 \, \partial_{\tilde\gamma^2}\Bigr) \mathcal{K}_2,\\
	\tfimdl_4
	&= -\frac{1}{3} \, \partial_{\tilde{m}^2} \tfimdl_3
	= -\frac{1}{6} \Bigl(6 \partial_{\tilde{m}^2}^4 + 2 \tilde{m}^2 \, \partial_{\tilde{m}^2}^5 + 2 \tilde\gamma^2 \, \partial_{\tilde{m}^2}^4 \, \partial_{\tilde\gamma^2}\Bigr) \mathcal{K}_2,
\end{align}
where
\begin{mleqn}
	\mathcal{K}_2
	&= -\frac{1}{8 \pi} \sum_{j=1}^4 w_j \mu_j \ln(\mu_j)\\
	&= -\frac{1}{8 \pi} \Bigl[\alpha^{++} \ln(\alpha^{++}) + \alpha^{+-} \ln(\alpha^{+-}) + \alpha^{-+} \ln(\alpha^{-+}) + \alpha^{--} \ln(\alpha^{--})\Bigr].
\end{mleqn}

\paragraph{$\boldsymbol{d + 1 = 1}$} Even the case of a time dimension all by itself with zero dimensions of space has experimental relevance (for instance in the context of quantum dots coupled to reservoirs \cite{floerchinger2012analytic}). The threshold functions in this case become
\begin{align}
	\Delta \tfimdl_0
	&= k^{-1} \bigl[\mathcal{K}_1^k \, k - \mathcal{K}_1^\Lambda \, \Lambda\bigr],\\
	\tfimdl_0
	&= k^{-1} \, \tilde\partial_t \, \Delta \tfim_0
	= \bigl(1 - 2 \tilde{m}^2 \, \partial_{\tilde{m}^2} - 2 \tilde\gamma^2 \, \partial_{\tilde\gamma^2}\bigr) \mathcal{K}_1,\\
	\tfimdl_1
	&= -\partial_{\tilde{m}^2} \tfimdl_0
	= \bigl(\partial_{\tilde{m}^2} + 2 \tilde{m}^2 \, \partial_{\tilde{m}^2}^2 + 2 \tilde\gamma^2 \, \partial_{\tilde{m}^2} \, \partial_{\tilde\gamma^2}\bigr) \mathcal{K}_1,\\
	\tfimdl_2
	&= -\partial_{\tilde{m}^2} \tfimdl_1
	= -\bigl(3 \, \partial_{\tilde{m}^2}^2 + 2 \tilde{m}^2 \, \partial_{\tilde{m}^2}^3 + 2 \tilde\gamma^2 \, \partial_{\tilde{m}^2}^2 \, \partial_{\tilde\gamma^2}\bigr) \mathcal{K}_1,\\
	\tfimdl_3
	&= -\frac{1}{2} \, \partial_{\tilde{m}^2} \tfimdl_2
	= \frac{1}{2} \bigl(5 \, \partial_{\tilde{m}^2}^3 + 2 \tilde{m}^2 \, \partial_{\tilde{m}^2}^4 + 2 \tilde\gamma^2 \, \partial_{\tilde{m}^2}^3 \, \partial_{\tilde\gamma^2}\bigr) \mathcal{K}_1,\\
	\tfimdl_4
	&= -\frac{1}{3} \, \partial_{\tilde{m}^2} \tfimdl_3
	= -\frac{1}{6} \bigl(7 \, \partial_{\tilde{m}^2}^4 + 2 \tilde{m}^2 \, \partial_{\tilde{m}^2}^5 + 2 \tilde\gamma^2 \, \partial_{\tilde{m}^2}^4 \, \partial_{\tilde\gamma^2}\bigr) \mathcal{K}_1,
\end{align}
with
\begin{equation}
	\mathcal{K}_1
	= \frac{1}{2} \sum_{j=1}^4 w_j \csqrt{\mu_j}
	= \frac{1}{2} \Bigl[\csqrt{\alpha^{++}} + \csqrt{\alpha^{+-}} + \csqrt{\alpha^{-+}} + \csqrt{\alpha^{--}}\Bigr].
\end{equation}

\subsection{Propagator}

Like $\tfim_j$, $\tfjm$ only depends on the magnitude of $\vec{p}$, so we again use spherical coordinates in the form
\begin{equation}
	\int_{\vec p}
	= \frac{2 \pi^\frac{d}{2}}{\Gamma\bigl(\frac{d}{2}\bigr)} \frac{1}{2} \int_0^\infty \frac{\dif p^2}{(2 \pi)^d} (\vec{p}^2)^{\frac{d-2}{2}},
\end{equation}
with $\dif p^2 = 2 p \, \dif p$. Integrating the first Dirac delta in \labelcref{eqn:disc j} gives \cite{floerchinger2017unpublished}
\begin{equation}\label{eqn:root}
	q_0
	= \csqrt{\vec{p}^2 + \Re\alpha_1^{ir} k^2} + \csqrt{\vec{p}^2 + \Re\alpha_2^{js} k^2}
\end{equation}
which we can solve for $\vec{p}^2$ to get
\begin{equation}\label{eqn:p sol}
	\vec{p}^2
	= \frac{1}{4 q_0^2} \Bigl[q_0^4 -2 q_0^2 \bigl(\Re\alpha_1^{ir} + \Re\alpha_2^{js}\bigr) k^2 + \bigl(\Re\alpha_1^{ir} - \Re\alpha_2^{js}\bigr)^2 k^4\Bigr].
\end{equation}
Since \labelcref{eqn:root} is a simple root of $f(\vec{p}^2) = q_0 - \csqrt{\vec{p}^2 + \Re\alpha_1^{ir} k^2} - \csqrt{\vec{p}^2 + \Re\alpha_2^{js} k^2}$, the prefactor arising from the Dirac delta $\delta\bigl(f(\vec{p}^2)\bigr) = \frac{1}{|f^\prime(\vec{p}^2)|} \delta(\vec{p}^2 - q_0^2)$ is
\begin{mleqn}
	\frac{1}{|f^\prime(\vec{p}^2)|}
	&= \biggl(\frac{1}{2 \csqrt{\vec{p}^2 + \Re\alpha_1^{ir} k^2}} + \frac{1}{2 \csqrt{\vec{p}^2 + \Re\alpha_2^{js} k^2}}\biggr)^{-1}\\
	&= \frac{2}{q_0} \, \csqrt{\vec{p}^2 + \Re\alpha_1^{ir} k^2} \csqrt{\vec{p}^2 + \smash{\Re\alpha_2^{js}} k^2},
\end{mleqn}
and only contributes for $q_0 > \csqrt{\vec{p}^2 + \Re\alpha_1^{ir} k^2} + \csqrt{\vec{p}^2 + \Re\alpha_2^{js} k^2}$, i.e. for
\begin{equation}
	\theta\Bigl(q_0 - \csqrt{\vec{p}^2 + \Re\alpha_1^{ir} k^2} - \csqrt{\vec{p}^2 + \smash{\Re\alpha_2^{js}} k^2}\Bigr)
	= 1,
\end{equation}
where $\theta$ denotes the Heaviside step function. The second Dirac delta gives for e.g. $\Re\alpha_1^{ir} > \Re\alpha_2^{js}$
\begin{equation}
	q_0
	= \csqrt{\vec{p}^2 + \Re\alpha_1^{ir} k^2} - \csqrt{\vec{p}^2 + \smash{\Re\alpha_2^{js}} k^2}.
\end{equation}
It contributes only for $0 = q_0 = \bigl|\csqrt{\vec{p}^2 + \Re\alpha_1^{ir} k^2} - \csqrt{\vec{p}^2 + \Re\alpha_2^{js} k^2}\bigr|$.

Interestingly, the corresponding solution for $\vec{p}^2$ is the same as in \labelcref{eqn:p sol} and so is the prefactor $|f^\prime(\vec{p}^2)|^{-1}$. (The ranges are such that they are distinct except when one of the $\alpha$s vanishes.)

In summary, we find after momentum integration
\small\begin{align}\label{eqn:disc j integrated}
	\disc \tfjm
	&= -\frac{1}{4} \sum_{i,j,r,s} \frac{\pi^\frac{d}{2}}{\Gamma\bigl(\frac{d}{2}\bigr)} \frac{(\vec{p}^2)^{\frac{d-2}{2}}}{(2 \pi)^d} \frac{\csqrt{\vec{p}^2 + \Re\alpha_1^{ir} k^2}}{\csqrt{\vec{p}^2 + \alpha_1^{i\vphantom{j}r} k^2}} \frac{\csqrt{\vec{p}^2 + \Re\smash{\alpha_2^{js}} k^2}}{\csqrt{\vec{p}^2 + \alpha_2^{js} k^2}} \frac{\pi \, \beta_1^{ir} \beta_2^{js}}{2 q_0} \\
	&\hphantom{=}\times \Biggl[\Bigl[1 + n_\text{B}\Bigl(\!\csqrt{\vec{p}^2 + \alpha_1^{i\vphantom{j}r} k^2}\Bigr) + n_\text{B}\Bigl(\!\csqrt{\vec{p}^2 + \alpha_2^{js} k^2}\Bigr)\Bigr] \theta\Bigl(q_0 - \csqrt{\vec{p}^2 + \Re\alpha_1^{ir} k^2} - \csqrt{\vec{p}^2 + \Re\smash{\alpha_2^{js}} k^2}\Bigr)\notag\\
	&\hphantom{=\times \Biggl[}+ \Bigl[n_\text{B}\Bigl(\!\csqrt{\vec{p}^2 + \alpha_1^{i\vphantom{j}r} k^2}\Bigr) - n_\text{B}\Bigl(\!\csqrt{\vec{p}^2 + \alpha_2^{js} k^2}\Bigr)\Bigr] 
	\theta\Bigl(\csqrt{\vec{p}^2 + \Re\smash{\alpha_2^{js}} k^2} - \csqrt{\vec{p}^2 + \Re\alpha_1^{ir} k^2} - q_0\Bigr)\notag\\
	&\hphantom{=\times \Biggl[}+ \Bigl[n_\text{B}\Bigl(\!\csqrt{\vec{p}^2 + \alpha_2^{js} k^2}\Bigr) - n_\text{B}\Bigl(\!\csqrt{\vec{p}^2 + \alpha_1^{i\vphantom{j}r} k^2}\Bigr)\Bigr]
	\theta\Bigl(\csqrt{\vec{p}^2 + \Re\alpha_1^{ir} k^2} - \csqrt{\vec{p}^2 + \Re\alpha_2^{js} k^2} - q_0\Bigr)\Biggr],\notag
\end{align}
\normalsize
which is to be evaluated at $\vec{p}^2$ as given in \labelcref{eqn:p sol}. If $\alpha_1^{ir} \approx \Re\alpha_1^{ir}$ the square root fractions in front are approximately one. Also, at $T = 0$ all terms $\propto n_\text{B} \propto e^{-p_0/T}$ drop out and \labelcref{eqn:disc j integrated} simplifies to
\begin{mleqn}\label{eqn:zero temp disc j}
	\disc \tfjm\bigr|_{T=0}
	= -\frac{1}{4} \sum_{i,j,r,s} \frac{\pi^\frac{d}{2}}{\Gamma\bigl(\frac{d}{2}\bigr)} \frac{\pi \, \beta_1^{ir} \beta_2^{js}}{2 (2 \pi)^d \, q_0} \, \theta\Bigl(q_0 - \csqrt{\vec{p}^2 + \Re\alpha_1^{ir} k^2} - \csqrt{\vec{p}^2 + \Re\smash{\alpha_2^{js}} k^2}\Bigr)\\
	\times \biggl(\frac{q_0^4 -2 q_0^2 \bigl(\Re\alpha_1^{ir} \Re\alpha_2^{js}\bigr) + \bigl(\Re\alpha_1^{ir} - \Re\alpha_2^{js}\bigr) k^4}{4 q_0^2}\biggr)^{\frac{d-2}{2}}.
\end{mleqn}

\section{Numerical Results}
\label{sec:numerical results}

\begin{figure}[htb!]
	\centering
	\begin{subfigure}{0.45\linewidth}
		\includegraphics[width=\linewidth]{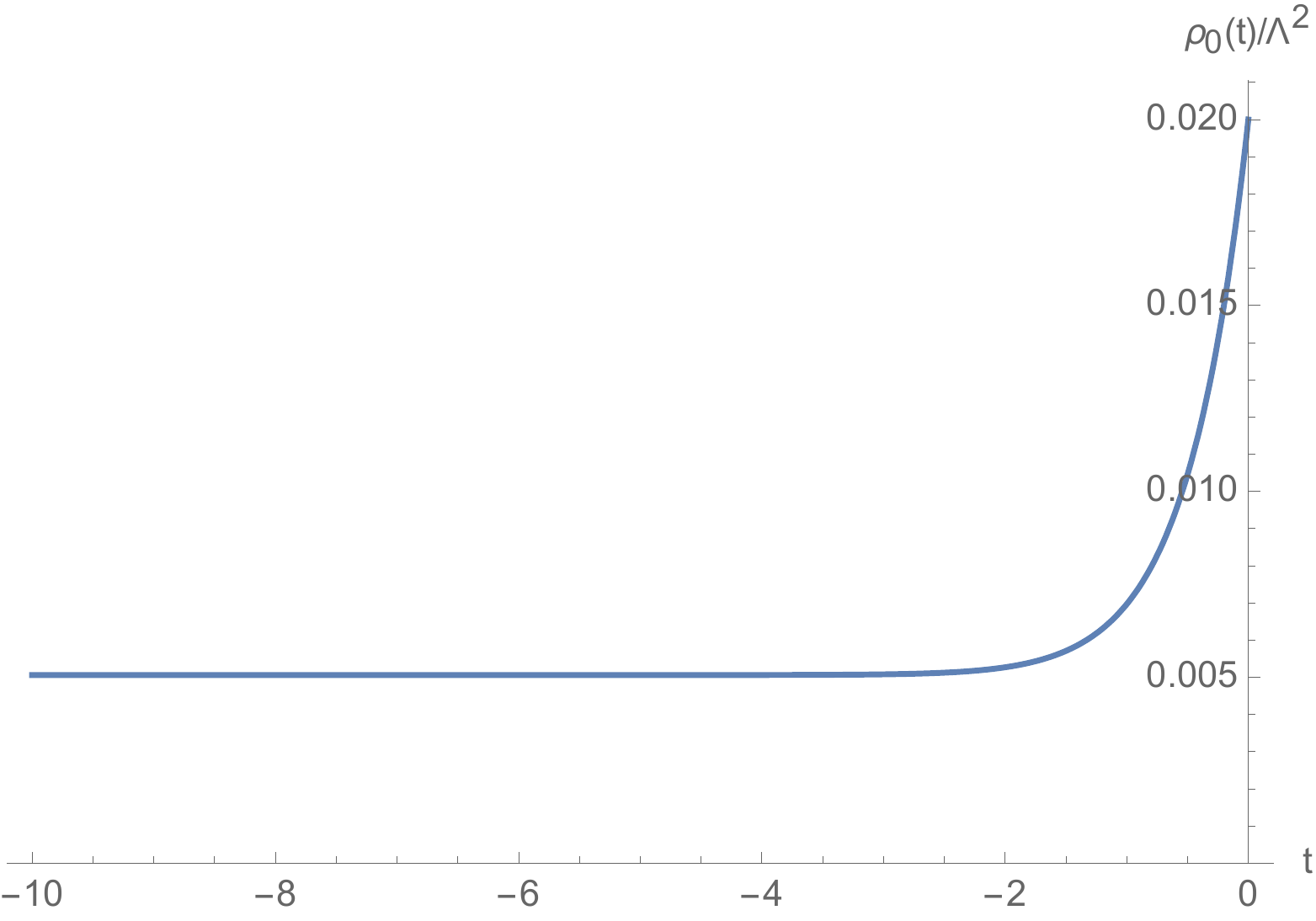}
		\caption{$\rho_0$}
		\label{fig:rho}
	\end{subfigure}
	\hfill
	\begin{subfigure}{0.45\linewidth}
		\includegraphics[width=\linewidth]{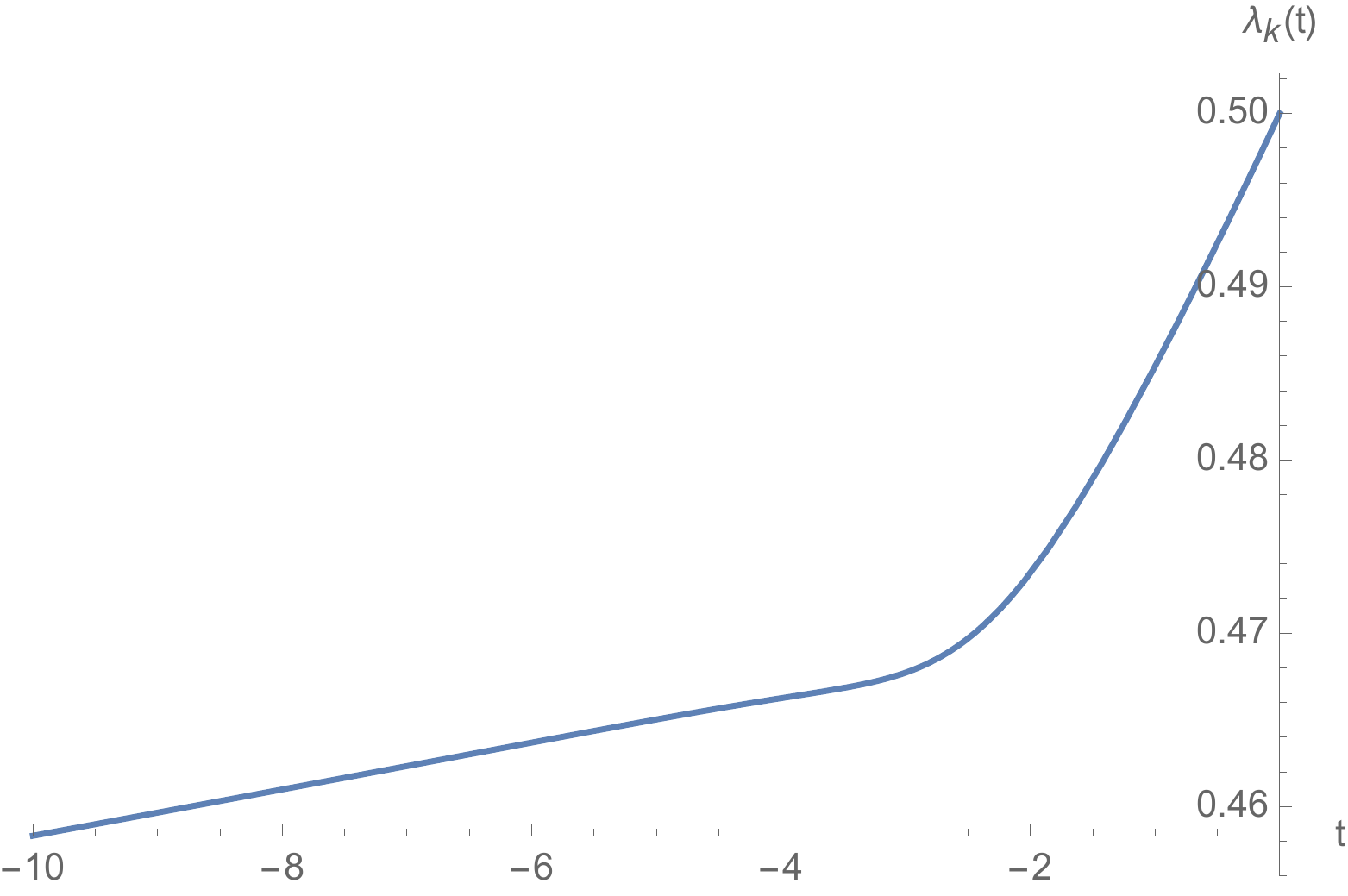}
		\caption{$\lambda$}
		\label{fig:lambda}
	\end{subfigure}
	\begin{subfigure}{0.45\linewidth}
		\includegraphics[width=\linewidth]{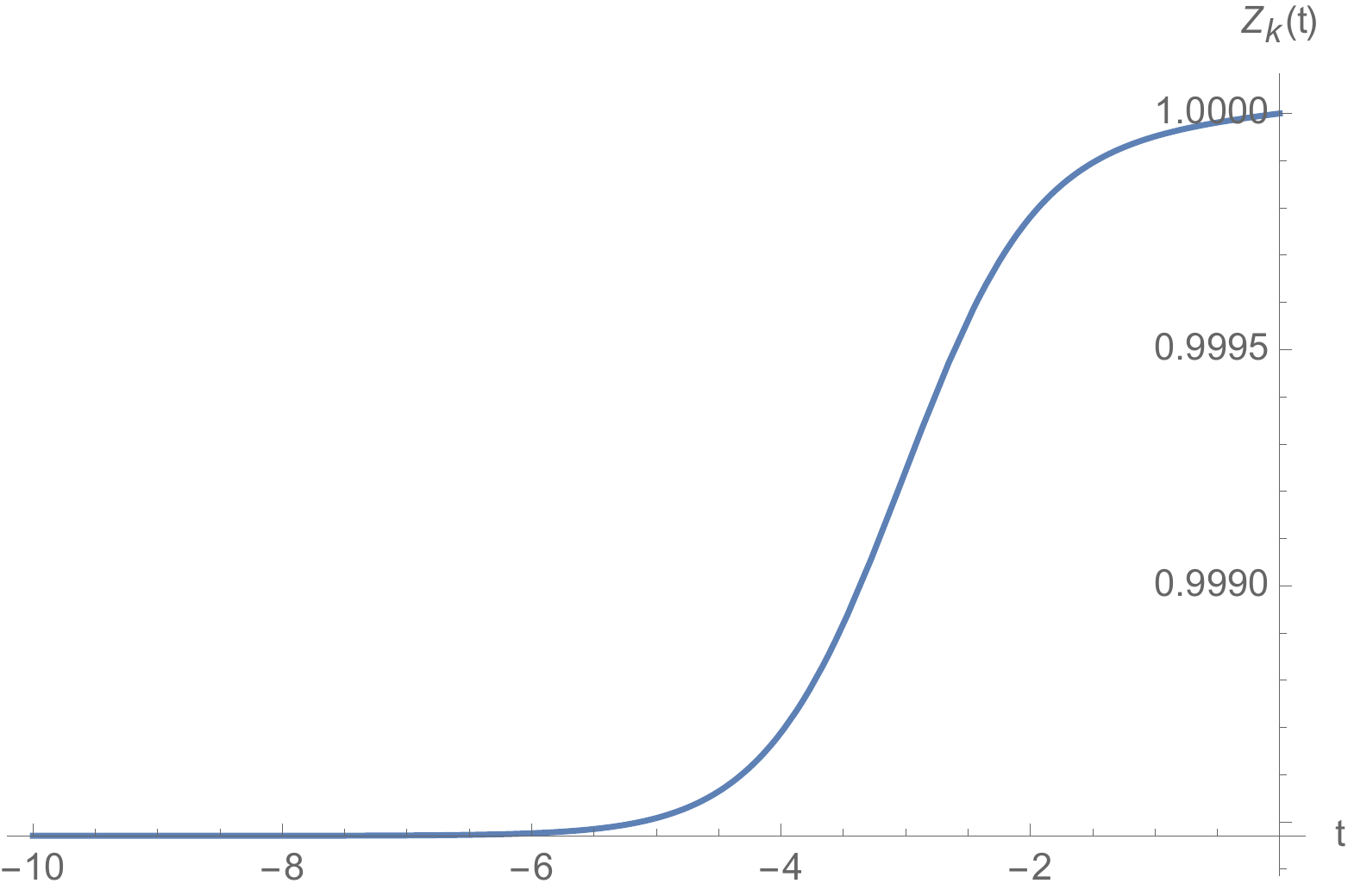}
		\caption{$Z_k$}
		\label{fig:zk}
	\end{subfigure}
	\hfill
	\begin{subfigure}{0.45\linewidth}
		\includegraphics[width=\linewidth]{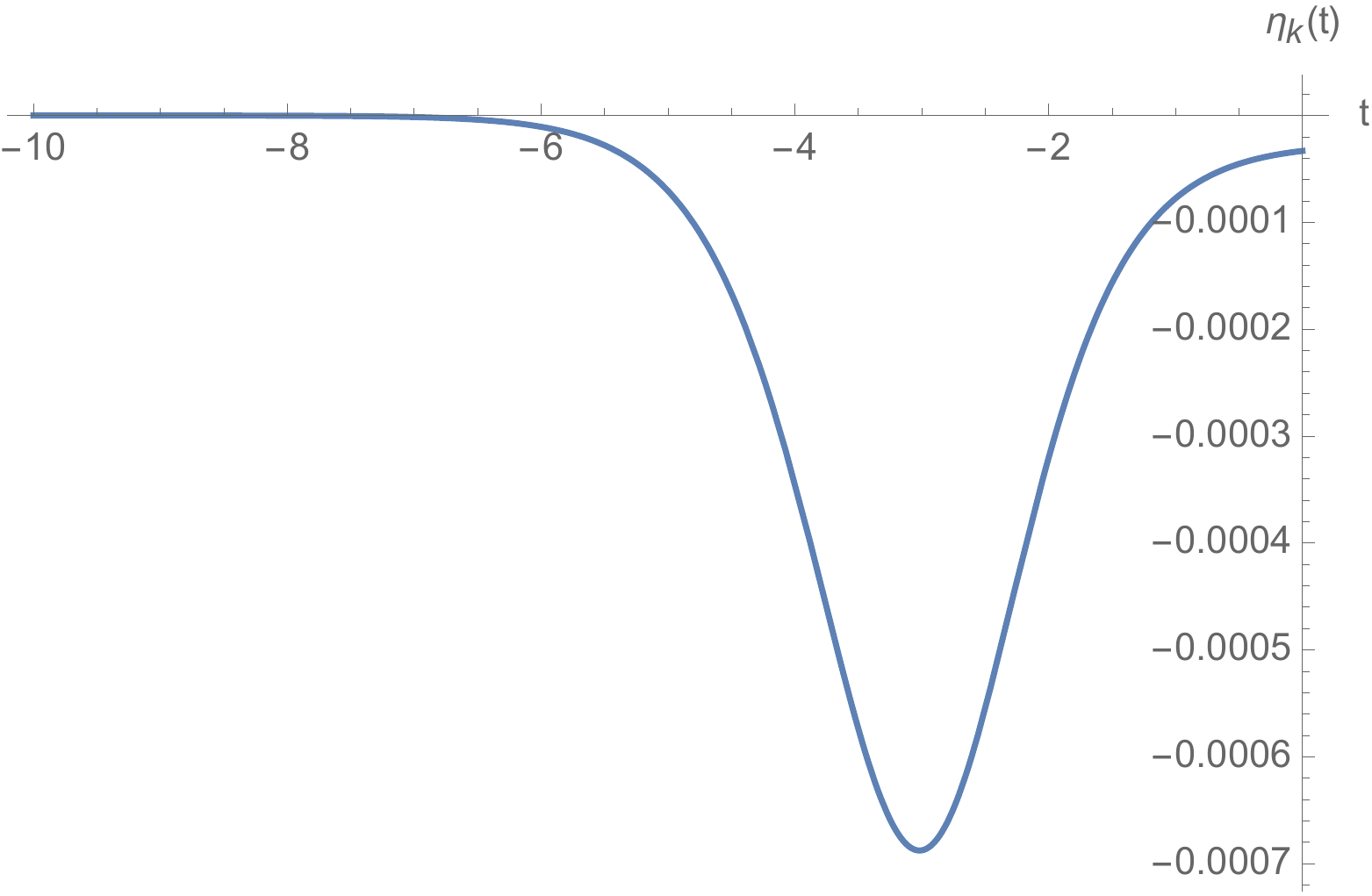}
		\caption{$\eta$}
		\label{fig:eta}
	\end{subfigure}
	\begin{subfigure}{0.45\linewidth}
		\includegraphics[width=\linewidth]{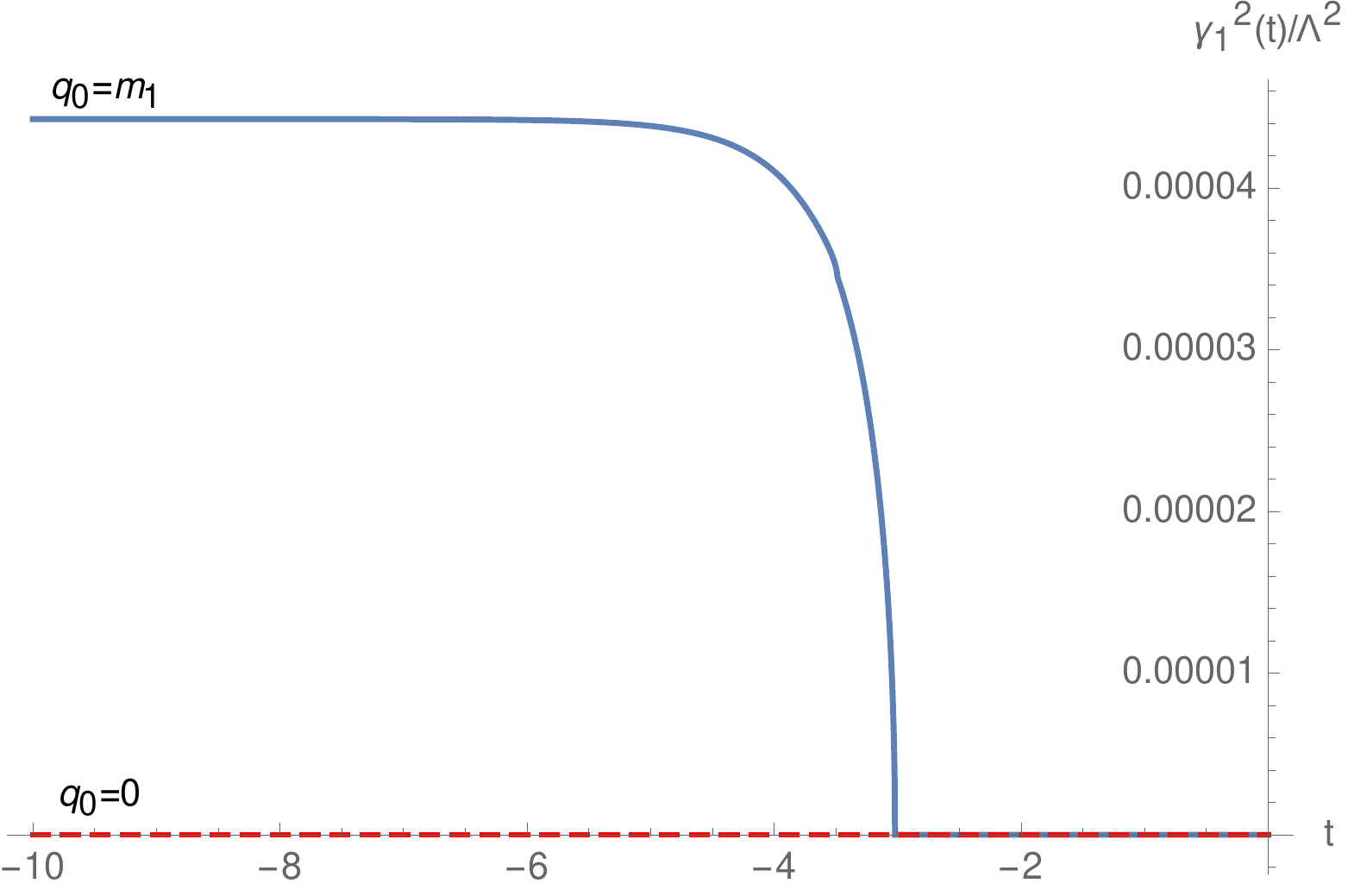}
		\caption{$\gamma_1^2$}
		\label{fig:gamma}
	\end{subfigure}
	\hfill
	\begin{subfigure}{0.45\linewidth}
		\includegraphics[width=\linewidth]{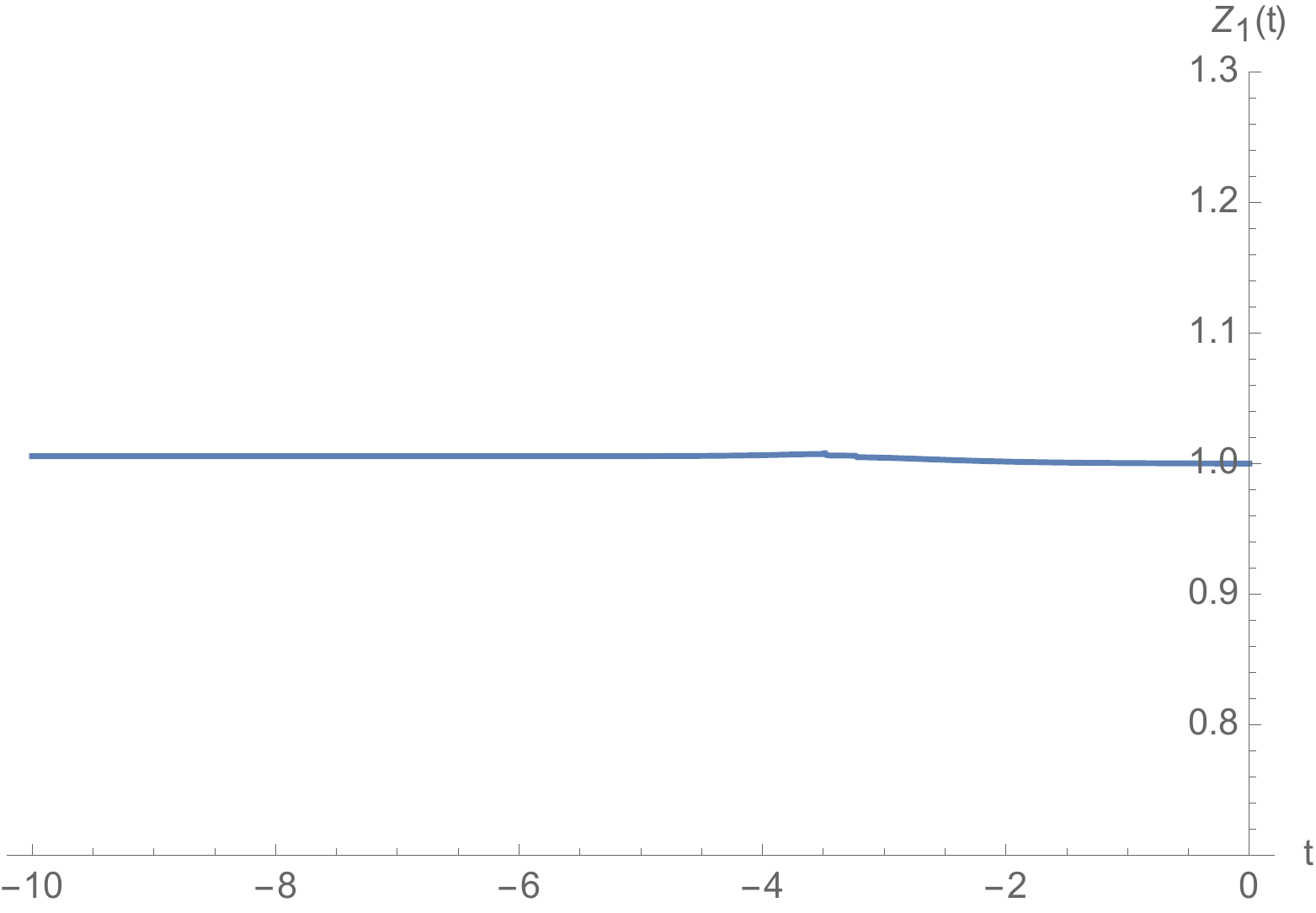}
		\caption{$Z_1$}
		\label{fig:z1}
	\end{subfigure}
	\caption{Flow of the $O(2)$ model at $T = 0$ in $3+1$ spacetime dimensions}
	\label{fig:o(n) flow}
\end{figure}

\Cref{eqn:mk mink flow,eqn:rhok mink flow,eqn:lambdak mink flow,eqn:z1 flow at rho0,eqn:g1 flow at rho0,eqn:zk flow at rho0} constitute a closed set of integro-differential equations. After inserting our results of \cref{sec:matsubara summation,sec:momentum integration} for the Matsubara summation and momentum integration of the threshold functions, the system reduces to a coupled set of non-linear ordinary differential equations which can be solved numerically once we specify initial values for the parameters of our truncation at the ultraviolet cutoff scale $k = \Lambda$. For our calculations we choose
\begin{equation}\label{eqn:ini conds}
	\tilde\rho_0(\Lambda)
	= \frac{1}{50},
	\qquad
	\lambda(\Lambda)
	= \frac{1}{2},
	\qquad
	\tilde\gamma_1^2(\Lambda)
	= 0,
	\qquad
	Z_1(\Lambda)
	= 1,
	\qquad
	N = 2,
\end{equation}
where $\tilde\rho_0(\Lambda) = \rho_0(\Lambda)/\Lambda^2$ and $\tilde\gamma_1^2(\Lambda) = \gamma_1^2(\Lambda)/\Lambda^2$. The resulting real-time flow of the propagator and the effective potential in the truncation \labelcref{eqn:ren gammak2} of the scalar $O(N)$-model at zero temperature in $3+1$-dimensional spacetime is shown in \cref{fig:o(n) flow}.

\Cref{fig:rho} displays the scale dependence of the effective potential's minimum location $\rho_0(k)$. Near the ultraviolet cutoff $\Lambda$ it exhibits a sharp fall-off to about a fourth of its initial value and then becomes scale independent for $k \lesssim \Lambda/e^2$.

\Cref{fig:lambda} reveals a weak logarithmic flow of the quartic coupling $\lambda$ from large values at microscopic distances to smaller values at $k \ll \Lambda$. The logarithmic running implies $\lambda(k) \to 0$ for $k \to 0$ which indicates that a field theory involving only scalars must be free in four spacetime dimensions. The underlying effect of charge screening forces the quartic coupling to zero at $k = 0$, a feature known as ``triviality''.

We also point out that the flow behavior of $\lambda$ separates into two regimes. Above the transition region $k^2 \approx m_1^2$, the logarithmic running is significantly stronger since fluctuations of both the radial and Goldstone modes contribute to the scale dependence. Below $k^2 \approx m_1^2$ contributions from the radial mode diminish since its fluctuations are suppressed by the non-zero mass $m_1^2 = 2 \lambda \rho_0/Z_1$.

\Cref{fig:zk,fig:eta} show the flow of the overall wave function renormalization $Z_k$ and its derivative, the anomalous dimension $\eta = - 1/Z_k \, \partial_t Z_k$. The latter exhibits a drop to small negative values with a minimum at $k \approx \Lambda/e^3$ after which it smoothly returns to zero for larger scales.

Of particular importance to our investigation is the flow of the discontinuity $\gamma_1^2$ shown in \cref{fig:gamma}, both for external energies that correspond tp on-shell ($q_0 = m_1$) and virtual ($q_0 = 0$) radial excitations. As we would expect, the on-shell discontinuity adheres to its initial value of zero until $k < q_0 = m_1 = \sqrt{2 \lambda \rho_0/Z_1}$ at which point it abruptly rises to non-zero values before again becoming scale independent for $k \lesssim \Lambda/e^5$. The physical origin of the discontinuity $\gamma_1^2$ in the radial mode's propagator is its decay channel into two massless Goldstone bosons via the non-zero $\smash{\Gamma_{k,1aa}^{(3)}}$, $a \in \{2,\dots,N\}$. The flow \labelcref{fig:gamma} signals that on-shell radial excitations are unstable at energies below the mass $m_1$ but become stable once $k > m_1$. On the other hand, for virtual $q_0 = 0$, the discontinuity $\gamma_1^2$ is zero on all scales, suggesting that zero-momentum radial fluctuations do not decay. We emphasize that our computation took this non-zero real-time decay width into account in a self-consistent manner.

\Cref{fig:z1} shows the scale dependence of the radial wave function renormalization $Z_1$. The fact that $Z_1$ never strays far from unity shows that at zero temperature on-shell radial excitations renormalize just like the Goldstone bosons.

All results in \cref{fig:o(n) flow} agree with those obtained in \cite{floerchinger2012analytic}.

\section{Conclusions and Outlook}
\label{sec:conclusions}

We discussed a method to analytically continue functional renormalization group equations from discrete imaginary Matsubara frequencies to the continuous real frequency axis. This method has been developed in \cite{floerchinger2012analytic}. The present work contains further details on the derivation of the flow equations. We showcased how this method works in practice for the example of a theory of $N$ relativistic scalar fields with $O(N)$ symmetry, putting particular emphasis on the propagator $G_r(p)$ of the massive radial mode $\varphi_1$ in the regime where spontaneous symmetry breaking reduces the $O(N)$ to $O(N-1)$. An important characteristic of such excitations -- one that has received little attention up to this point -- is the imaginary discontinuity $\gamma_1^2$ of the inverse propagator at on-shell external momenta $q^2 = m_1^2/Z_1 = 2 \rho_0 \, \lambda/Z_1$. It is closely related to a non-zero particle decay width $\Gamma = \gamma_1^2/m_1$ describing the fission of a massive radial field into two massless Goldstone bosons.

To obtain a truncation able to account for this new type of singularity in the analytic structure of the propagator $G_k(p)$, we employed a Minkowski-space derivative expansion of the average action $\Gamma_k$ around singular points of the propagator. This expansion scheme is very close to the actual dynamics \cite{floerchinger2012analytic} in the sense that loop integrals on the r.h.s. of flow equations are strongly dominated by the on-shell physical excitations corresponding to these poles and branch cuts. Such an expansion may therefore exhibit improved convergence behavior compared to a Euclidean space derivative expansion around vanishing frequency.

We exploited this convenient circumstance by regulating our flow equations with a class of algebraic regulators due to Flörchinger \cite{floerchinger2012analytic} that would otherwise have been inadequate. These regulators exhibit a much milder decay in the ultraviolet and consequently inferior separation of momentum modes compared to typical Euclidean-space (exponential or Litim-type) regulators. Besides being fully compatible with Lorentz invariance, choosing the simplest representative of this class of regulators also allowed us to resort to contour integration methods to perform the summation over Matsubara frequencies both in flow equations for parameters of the effective potential and the radial propagator analytically. The resulting flow equations proved to be both infrared and ultraviolet finite without the need for further regularization, supporting our claim of improved convergence of the derivative expansion in Minkowski space.

An interesting prospect for future applications of this method is the first-principles calculation of transport properties. Since our formalism is based on a linear response framework, quantities such as viscosities, conductivities, permittivities, relaxation times, etc. now lie within the scope of functional renormalization.

We were mainly concerned with the investigation of conceptual issues and therefore restricted our treatment to the simple case of relativistic scalar fields. However, the method can also be applied to more complicated theories mixing both bosonic and fermionic degrees of freedom of different spin. With increased effort, it allows the treatment of such systems at arbitrary temperature and chemical potential.

In summary, the analytic continuation of functional renormalization group equations brings the renormalization group closer to the physical dynamics in Minkowski space. It enables the computation of new observables previously inaccessible to the FRG. Thanks to the enhanced performance of the derivative expansion in Minkowski space we believe it will lead to more accurate results despite little computational effort.

\appendix
\crefalias{appendix}{section}

\section{Propagator}

\subsection{Analytic Structure}
\label{app:prop analytic structure}

The Källen-Lehmann spectral representation can be written
\begin{equation}\label{eqn:kl rep}
	G(p)
	= \int_0^\infty \dif \mu^2 \, \frac{\rho(\mu^2)}{p^2 + \mu^2},
\end{equation}
where $p^2 = -p_0^2 + \vec p^2$ decomposes into real and imaginary parts,
\begin{mleqn}
	&\Re(p^2)
	= -\Re(p_0^2) + \vec{p}^2
	= -\Re(p_0)^2 + \Im(p_0)^2 + \vec{p}^2,\\
	&\Im(p^2)
	= -\Im(p_0^2)
	= -2 \Re(p_0) \Im(p_0).
\end{mleqn}
For $\Im(p_0) \approx 0$ and non-vanishing $\Re(p_0)$ and/or $\vec p^2$, we have $|\Re(p^2)| \gg |\Im(p^2)|$. Thus close to the real $p_0$-axis, we recognize a form to which we can apply the Sokhotski–Plemelj theorem for the real line ($\mathcal{P}$ is the Cauchy principal value)
\begin{equation}
	\lim_{\epsilon \to 0} \int_a^b \dif x \, \frac{f(x)}{x + y \pm i \epsilon}
	= \mathcal{P} \int_a^b \frac{f(x)}{x + y} \, \dif x \, \mp \, i \pi \, f(-y),
	\mathrlap{\qquad (a < y < b)}
\end{equation}
by identifying
\begin{equation}
	x = \mu^2,
	\quad
	a = 0,
	\quad
	b = \infty,
	\quad
	f(x) = \rho(\mu^2),
	\quad
	y = \Re(p^2),
	\quad
	i \epsilon = i \Im(p^2).
\end{equation}
Thus the propagator can also be written
\begin{equation}\label{eqn:sp thm applied}
	G(p)
	= \mathcal{P} \int_0^\infty \dif \mu^2 \, \frac{\rho(\mu^2)}{p^2 + \mu^2}
	+ i \pi \, s(p_0) \, \rho(-p^2),
\end{equation}
where $s(p_0) = \sign(\Re p_0 \Im p_0)$ and we approximated $\Re(p^2) \approx p^2$. \labelcref{eqn:sp thm applied} reveals that the propagator has a branch cut along the real axis for all values of $-\Im(p_0)^2 - \vec{p}^2$ for which $\rho(-\Im(p_0)^2 - \vec{p}^2) \neq 0$. (Of course, $s(p_0)$ also switches sign when moving across the real axis. However, we cannot infer from this that the propagator also has a branch cut along the imaginary axis. \labelcref{eqn:sp thm applied} is valid only close to the real axis and does not allow any insight into the analytic structure of $G(p)$ for $\Im(p_0) \not\approx 0$. In fact, we already know there can't be a cut on the imaginary axis. The Källen-Lehmann decomposition \labelcref{eqn:kl rep} clearly shows that the Minkowski-space propagator is completely regular throughout the complex plane except for $p_0 \in \reals$.)

The propagator's analytic structure also exhibits poles. To see this, we perform a partial fraction decomposition,
\begin{mleqn}
	\frac{1}{p^2 + \mu^2}
	&\overset{!}{=} \frac{a}{-p_0 + \csqrt{\vec{p}^2 + \mu^2}} + \frac{b}{-p_0 - \csqrt{\vec{p}^2 + \mu^2}}\\[1ex]
	&= \frac{a (-p_0 - \sqrt{\vec{p}^2 + \mu^2}) + b (-p_0 + \sqrt{\vec{p}^2 + \mu^2})}{p_0^2 - \vec{p}^2 - \mu^2},
\end{mleqn}
i.e.
\begin{equation}
	1
	\overset{!}{=} a (p_0 + \sqrt{\vec{p}^2 + \mu^2}) + b (p_0 - \sqrt{\vec{p}^2 + \mu^2}).
\end{equation}
Inserting $p_0 = \pm\csqrt{\vec{p}^2 + \mu^2}$, we find the coefficients
\begin{equation}
	a
	= \frac{1}{2 \csqrt{\vec{p}^2 + \mu^2}}
	= -b.
\end{equation}
Plugging this back into \labelcref{eqn:kl rep}, we can make the pole structure explicit,
\begin{equation}\label{eqn:explicit pole structure}
	G(p)
	= \int_0^\infty \dif \mu^2 \, \frac{\rho(\mu^2)}{2 \csqrt{\vec{p}^2 + \mu^2}} \Biggl[\frac{1}{-p_0 + \csqrt{\vec{p}^2 + \mu^2} \pm i \epsilon} - \frac{1}{-p_0 - \csqrt{\vec{p}^2 + \mu^2} \pm i \epsilon}\Biggr].
\end{equation}
We added the infinitesimal $i \epsilon$-terms in \labelcref{eqn:explicit pole structure} by hand to move the singularities slightly away from the real $p_0$-axis. Different combinations of signs for these terms correspond to differently time-ordered propagators. $(+,+)$ gives the advanced, $(-,-)$ the retarded, $(-,+)$ the time-ordered (Feynman) and $(+,-)$ the anti-time-ordered propagator.

To complete our discussion, we give a quick proof of the Sokhotski–Plemelj theorem on the real line where $a < y < b$. Consider
\begin{equation}
	\lim_{\epsilon \to 0} \int_a^b \frac{f(x)}{x + y \pm i \epsilon} \, \dif x
	= \mp i \pi \lim_{\epsilon \to 0} \int_a^b \frac{\epsilon \, f(x) \, \dif x}{\pi [(x + y)^2 + \epsilon^2]} + \lim_{\epsilon \to 0} \int_a^b \frac{(x + y)^2}{(x + y)^2 + \epsilon^2} \, \frac{f(x)}{x + y} \, \dif x.
\end{equation}
In the first term, $\lim_{\epsilon \to 0} \epsilon/\{\pi [(x + y)^2 + \epsilon^2]\} = \delta(x + y)$ is a nascent delta function, giving simply $\mp i \pi \, f(-y)$ in the limit $\lim_{\epsilon \to 0}$. In the second term, $\frac{(x + y)^2}{(x + y)^2 + \epsilon^2}$ approaches $1$ for $|x + y| \gg \epsilon$, $0$ for $|x + y| \ll \epsilon$ and is symmetric about 0. For $\lim_{\epsilon \to 0}$, it thus gives the Cauchy principal value.

\subsection{Decomposition}
\label{app:propagator decomposition}

The sum of $P_k$ and $R_k$ defined as in \labelcref{eqn:inverse prop expanded,eqn:choice of rk},
\begin{equation}
	P_k
	= Z_k \Bigl[z \, p^2 + m^2 - i s(p_0) \, \gamma^2\Bigr],
	\qquad
	R_k(p)
	= \frac{Z_k \, k^2}{1 + c \, \frac{p^2}{k^2}},
\end{equation}
gives \cite{floerchinger2012analytic,floerchinger2017unpublished}
\begin{equation}\label{eqn:pkprk}
	P_k + R_k
	= Z_k \, \frac{z \, p^4 + p^2 \bigl[m^2 - i \, s(p_0) \, \gamma^2 + z \, \frac{k^2}{c}\bigr] + \frac{k^2}{c} \bigl[m^2 - i \, s(p_0) \gamma^2\bigr] + \frac{k^4}{c}}{p^2 + \frac{k^2}{c}} .
\end{equation}
To obtain an expression that closely resembles a sum of free propagators, the idea is now to decompose the regularized propagator into
\begin{equation}\label{eqn:prop decomposition}
	\frac{1}{P_k + R_k}
	= \frac{1}{Z_k}\biggl(\frac{\beta^+}{p^2 + \alpha^+ k^2} + \frac{\beta^-}{p^2 + \alpha^- k^2}\biggr),
\end{equation}
\labelcref{eqn:prop decomposition} implies
\begin{equation}\label{eqn:pkprk ito ab}
	P_k + R_k
	= Z_k \, \frac{p^4 + p^2 \, (\alpha^+ + \alpha^-) \, k^2 + \alpha^+ \, \alpha^- \, k^4}{(\beta^+ + \beta^-) \, p^2 + (\alpha^- \, \beta^+ + \alpha^+ \, \beta^-) \, k^2}.
\end{equation}
Comparing \labelcref{eqn:pkprk,eqn:pkprk ito ab}, we can read off the relations
\begin{align}
	&\alpha^+ + \alpha^-
	= \frac{\tilde m^2}{z} - i \, s(p_0) \, \frac{\tilde\gamma^2}{z} + \frac{1}{c},
	&&\beta^+ + \beta^-
	= \frac{1}{z},\\
	\label{eqn:ABCD relations}
	&\alpha^+ \, \alpha^-
	= \frac{1}{c \, z} \Bigl(\tilde m^2 - i \, s(p_0) \, \tilde\gamma^2 + 1\Bigr),
	&&\alpha^+ \, \beta^- + \alpha^- \, \beta^+
	= \frac{1}{c \, z}.
\end{align}
We have four equations and four unknowns. Solving for $\alpha^\pm$, $\beta^\pm$ yields
\begin{align}
	\label{eqn:alpha}
	&\alpha^\pm
	= \frac{1}{2} \biggl(\frac{1}{c} + \frac{\tilde m^2}{z} - i \, s(p_0) \, \frac{\tilde\gamma^2}{z}\biggr) \pm \bigl(A + i \, s(p_0) \, B\bigr),\\
	\label{eqn:beta}
	&\beta^\pm
	= \frac{1}{2 \, z} \pm \bigl(C + i \, s(p_0) \, D\bigr),
\end{align}
where $A$, $B$, $C$, $D$ are independent of $p$ and in particular of $s(p_0)$. Furthermore, $B = D = 0$ for $\gamma^2 = 0$, i.e. for $p_0 \notin \reals_+$, as we will see below. $A$ and $B$ are obtained by inserting $\alpha^\pm$ from \labelcref{eqn:alpha} into the left equality in \labelcref{eqn:ABCD relations}, yielding
\begin{mleqn}
	\alpha^+ \, \alpha^-
	&= \frac{1}{4} \biggl(\frac{1}{c} + \frac{\tilde m^2}{z} - i \, s(p_0) \, \frac{\tilde\gamma^2}{z}\biggr)^2 - \bigl(A + i \, s(p_0) \, B\bigr)^2,\\
	&= \frac{1}{4} \biggl(\frac{1}{c} + \frac{\tilde m^2}{z}\biggr)^2 - \frac{\tilde\gamma^4}{4 z^2} - A^2 + B^2 - i \, s(p_0) \biggl[\frac{1}{2} \biggl(\frac{1}{c} + \frac{\tilde m^2}{z}\biggr) \frac{\tilde\gamma^2}{z} + 2 A \, B\biggr]^2,\\
	&\overset{!}{=} \frac{1}{c \, z} + \frac{m^2}{c \, z \, k^2} - i \, s(p_0) \, \frac{\gamma^2}{c \, z \, k^2},
\end{mleqn}
Equating the real and imaginary parts on both sides of $\overset{!}{=}$ results in
\begin{align}\label{eqn:-A^2+B^2}
	-A^2 + B^2
	&= \frac{1}{c \, z} + \frac{m^2}{c \, z \, k^2} - \frac{1}{4} \biggl(\frac{1}{c} + \frac{\tilde m^2}{z}\biggr)^2 + \frac{\tilde\gamma^4}{4 z^2}\notag\\
	&= \frac{1}{c \, z} - \frac{1}{4} \biggl(\frac{1}{c} - \frac{\tilde m^2}{z}\biggr)^2 + \frac{\tilde\gamma^4}{4 z^2},\\
	2 A \, B
	&= \frac{\gamma^2}{c \, z \, k^2} - \frac{1}{2} \biggl(\frac{1}{c} + \frac{\tilde m^2}{z}\biggr) \frac{\tilde\gamma^2}{z}\notag\\
	&= \frac{1}{2} \biggl(\frac{1}{c} - \frac{\tilde m^2}{z}\biggr) \frac{\tilde\gamma^2}{z}.
\end{align}
With these relations, we can express $(i \, A + B)^2$ and $(i \, A - B)^2$ as
\begin{align}
	&\begin{aligned}
		(i \, A + B)^2
		&= -A^2 + B^2 + 2 i \, A \, B
		= \frac{1}{c \, z} - \frac{1}{4} \biggl(\frac{1}{c} - \frac{\tilde m^2}{z}\biggr)^2 + \frac{\tilde\gamma^4}{4 z^2} + \frac{i}{2} \biggl(\frac{1}{c} - \frac{\tilde m^2}{z}\biggr) \frac{\tilde\gamma^2}{z}\\
		&= \frac{1}{c \, z} - \frac{1}{4} \biggl(\frac{1}{c} - \frac{\tilde m^2}{z} - i \, \frac{\tilde\gamma^2}{z}\biggr)^2,
	\end{aligned}\\
	&(i \, A - B)^2
	= -A^2 + B^2 - 2 i \, A \, B
	= \frac{1}{c \, z} - \frac{1}{4} \biggl(\frac{1}{c} - \frac{\tilde m^2}{z} + i \, \frac{\tilde\gamma^2}{z}\biggr)^2.
\end{align}
Taking the square root gives
\begin{align}
	&i \, A + B
	= \pm \sqrt{\frac{1}{c \, z} - \frac{1}{4} \biggl(\frac{1}{c} - \frac{\tilde m^2}{z} - i \, \frac{\tilde\gamma^2}{z}\biggr)^2},\\
	&i \, A - B
	= \pm \sqrt{\frac{1}{c \, z} - \frac{1}{4} \biggl(\frac{1}{c} - \frac{\tilde m^2}{z} + i \, \frac{\tilde\gamma^2}{z}\biggr)^2},
\end{align}
and so
\begin{align}\label{eqn:a}
	A
	&= \frac{r}{2i} \Biggl[\sqrt{\frac{1}{c \, z} - \frac{1}{4} \biggl(\frac{1}{c} - \frac{\tilde m^2}{z} - \frac{i \tilde\gamma^2}{z}\biggr)^2} + s \, \sqrt{\frac{1}{c \, z} - \frac{1}{4} \biggl(\frac{1}{c} - \frac{\tilde m^2}{z} + \frac{i \tilde\gamma^2}{z}\biggr)^2}\Biggr],\\
	\label{eqn:b}
	B
	&= \frac{r}{2} \Biggl[\sqrt{\frac{1}{c \, z} - \frac{1}{4}\biggl(\frac{1}{c} - \frac{\tilde m^2}{z} - \frac{i \tilde\gamma^2}{z} \biggr)^2} - s \, \sqrt{\frac{1}{c \, z} - \frac{1}{4} \biggl(\frac{1}{c} - \frac{\tilde m^2}{z} + \frac{i \tilde\gamma^2}{z} \biggr)^2}\Biggr],
\end{align}
where the signs $r, s \in \{\pm 1\}$ can be chosen for convenience.

We choose the branch cut of the complex square root to lie on the negative real axis $\reals_-$. In this case, the two square roots in $A$ and $B$ are equal in the limit $\gamma^2 \to 0$ if
\begin{equation}
	\frac{1}{c \, z} - \frac{1}{4} \biggl(\frac{1}{c} - \frac{\tilde m^2}{z}\biggr)^2 \geq 0.
\end{equation}
Otherwise, they differ by a factor of $-1$. Enforcing $B \to 0$ for $\gamma^2 \to 0$ is therefore equivalent to the choice
\begin{equation}\label{eqn:s2 def}
	s = \sign\biggl[\frac{1}{c \, z} - \frac{1}{4} \biggl(\frac{1}{c} - \frac{\tilde m^2}{z}\biggr)^2\biggr].
\end{equation}
Note that with the definition \labelcref{eqn:s2 def} $A$ and $B$ are real for $s = -1$ and imaginary for $s = 1$.

The choice for $r$ is irrelevant since sending $r \to -r$ simply switches $\alpha^+ \!\!\leftrightarrow\! \alpha^-$. For definiteness, we choose $r = 1$ and $s$ as in \labelcref{eqn:s2 def}.

Conditions for $C$ and $D$ derive from $\alpha^+ \, \beta^- + \alpha^- \, \beta^+ = \frac{1}{c \, z}$ (see \labelcref{eqn:ABCD relations}) by inserting \labelcref{eqn:alpha,eqn:beta},
\begin{mleqn}
	&\biggl[\frac{1}{2} \biggl(\frac{1}{c} + \frac{\tilde m^2}{z} - i \, s(p_0) \, \frac{\tilde\gamma^2}{z}\biggr) + \bigl(A + i \, s(p_0) \, B\bigr)\biggr] \biggl[\frac{1}{2 \, z} - \bigl(C + i \, s(p_0) \, D\bigr)\biggr]\\
	&\hphantom{=}+ \biggl[\frac{1}{2} \biggl(\frac{1}{c} + \frac{\tilde m^2}{z} - i \, s(p_0) \, \frac{\tilde\gamma^2}{z}\biggr) - \bigl(A + i \, s(p_0) \, B\bigr)\biggr] \biggl[\frac{1}{2 \, z} + \bigl(C + i \, s(p_0) \, D\bigr)\biggr]\\
	&= \frac{1}{2 \, z} \biggl(\frac{1}{c} + \frac{\tilde m^2}{z} - i \, s(p_0) \, \frac{\tilde\gamma^2}{z}\biggr) - 2 A \, C + 2 B \, D - i \, s(p_0) \bigl[2 B \, C + 2 A \, D\bigr]
	\overset{!}{=} \frac{1}{c \, z}.
\end{mleqn}
Equating again real and imaginary parts on both sides of $\overset{!}{=}$ we find
\begin{align}
	\frac{1}{2 \, z} \biggl(\frac{1}{c} + \frac{\tilde m^2}{z}\biggr) - 2 A \, C + 2 B \, D &= \frac{1}{c \, z},\\
	\frac{\gamma^2}{2 z^2 \, k^2} + 2 B \, C + 2 A \, D &= 0.
\end{align}
or in matrix form,
\begin{equation}
	\begin{pmatrix}
		A & -B \\ B & A
	\end{pmatrix}
	\begin{pmatrix}
		C \\ D
	\end{pmatrix}
	= -\frac{1}{4 \, z} \begin{pmatrix}
		\frac{1}{c} - \frac{\tilde m^2}{z}\\
		\frac{\tilde\gamma^2}{z}
	\end{pmatrix}.
\end{equation}
Inverting $\Bigl(\begin{smallmatrix} A & -B \\ B & A \end{smallmatrix}\Bigr)^{-1} = \frac{1}{A^2+B^2} \bigl(\begin{smallmatrix} A & B \\ -B & A \end{smallmatrix}\bigr)$ yields
\begin{align}\label{eqn:cd}
	&C
	= -\frac{A \Bigl(\frac{1}{c}-\frac{\tilde m^2}{z}\Bigr) + B \,\frac{\tilde\gamma^2}{z}}{4 \, z\, (A^2 + B^2)},
	&D
	= \frac{B \Bigl(\frac{1}{c} - \frac{\tilde m^2}{z}\Bigr) - A \, \frac{\tilde\gamma^2}{z}}{4 \, z\, (A^2 + B^2)}.
\end{align}
For $\gamma^2 = 0$, we have $B = 0$ and thus also $D = 0$.

\paragraph{Explicit expressions} Plugging \labelcref{eqn:a,eqn:b,eqn:cd} into \labelcref{eqn:alpha}, we get the following expressions for $\alpha^\pm$,
\begin{mleqn}
	\alpha^\pm
	= \tfrac{1}{2} \Bigl(\tfrac{1}{c} + \tfrac{\tilde m^2}{z} - i \, s(p_0) \, \tfrac{\tilde\gamma^2}{z}\Bigr)
	\pm \tfrac{i}{2} \bigl[-1 + s(p_0)\bigr] \sqrt{\tfrac{1}{c \, z} - \tfrac{1}{4} \Bigl(\tfrac{1}{c} - \tfrac{\tilde m^2}{z} - \tfrac{i \tilde\gamma^2}{z}\Bigr)^2}\\
	\mp \tfrac{i}{2} \bigl[1 + s(p_0)\bigr] \sign\Bigl[\tfrac{1}{c \, z} - \tfrac{1}{4} \Bigl(\tfrac{1}{c} - \tfrac{\tilde m^2}{z}\Bigr)^2\Bigr] \sqrt{\tfrac{1}{c \, z} - \tfrac{1}{4} \Bigl(\tfrac{1}{c} - \tfrac{\tilde m^2}{z} + \tfrac{i \tilde\gamma^2}{z}\Bigr)^2}.
\end{mleqn}
If instead we choose $s = -1$ we get
\begin{mleqn}
	\alpha^\pm
	= \tfrac{1}{2} \Bigl(\tfrac{1}{c} + \tfrac{\tilde m^2}{z} - i \, s(p_0) \, \tfrac{\tilde\gamma^2}{z}\Bigr)
	\pm \tfrac{i}{2} \bigl[- 1 + s(p_0)\bigr] &\sqrt{\tfrac{1}{c \, z} - \tfrac{1}{4} \Bigl(\tfrac{1}{c} - \tfrac{\tilde m^2}{z} - \tfrac{i \tilde\gamma^2}{z}\Bigr)^2}\\
	\pm \tfrac{i}{2} \bigl[1 + s(p_0)\bigr] &\sqrt{\tfrac{1}{c \, z} - \tfrac{1}{4} \Bigl(\tfrac{1}{c} - \tfrac{\tilde m^2}{z} + \tfrac{i \tilde\gamma^2}{z}\Bigr)^2}.
\end{mleqn}
which gives rise to the convenient relation
\begin{equation}
	\alpha^\pm\bigl(s(p_0) = 1\bigr)
	= \alpha^{\pm\ast}\bigl(s(p_0) = -1\bigr),
\end{equation}
or $\alpha^{\pm+} = \bigl(\alpha^{\pm-}\bigr)^\ast$ using the notation introduced in \labelcref{eqn:integral decomposition}. Similarly, we can insert \labelcref{eqn:cd} into \labelcref{eqn:beta} to get,
\begin{align}
	\beta^\pm
	= \frac{1}{2 \, z} \pm \frac{\bigl(-A + i s(p_0) \, B\bigr) \bigl(\frac{1}{c} - \frac{\tilde m^2}{z} + i s(p_0) \, \frac{\tilde\gamma^2}{z}\bigr)}{4 z (A^2 + B^2)}
\end{align}
which using \labelcref{eqn:a,eqn:b} yields the following explicit expressions for $\beta^\pm$,
\begin{mleqn}
	\beta^\pm
	= \frac{1}{2 z} \pm \frac{1}{4 z} \Bigl(\tfrac{1}{c} + \tfrac{\tilde m^2}{z} - i \, s(p_0) \, \tfrac{\tilde\gamma^2}{z}\Bigr)
	\biggl[\tfrac{i}{2} \bigl[1 + s(p_0)\bigr] \sqrt{\tfrac{1}{c \, z} - \tfrac{1}{4} \Bigl(\tfrac{1}{c} - \tfrac{\tilde m^2}{z} + \tfrac{i \tilde\gamma^2}{z}\Bigr)^2}^{-1}\\
	+ \tfrac{i}{2} \bigl[1 - s(p_0)\bigr] \sign\Bigl[\tfrac{1}{c \, z} - \tfrac{1}{4} \Bigl(\tfrac{1}{c} - \tfrac{\tilde m^2}{z}\Bigr)^2\Bigr] \sqrt{\tfrac{1}{c \, z} - \tfrac{1}{4} \Bigl(\tfrac{1}{c} - \tfrac{\tilde m^2}{z} - \tfrac{i \tilde\gamma^2}{z}\Bigr)^2}^{-1}\biggr],
\end{mleqn}
whereas for $s = -1$ we have
\begin{mleqn}
	\beta^\pm
	= \frac{1}{2 z} \pm \frac{1}{4 z} \Bigl(\tfrac{1}{c} + \tfrac{\tilde m^2}{z} - i \, s(p_0) \, \tfrac{\tilde\gamma^2}{z}\Bigr)
	\biggl[&\tfrac{i}{2} \bigl[1 + s(p_0)\bigr] \sqrt{\tfrac{1}{c \, z} - \tfrac{1}{4} \Bigl(\tfrac{1}{c} - \tfrac{\tilde m^2}{z} + \tfrac{i \tilde\gamma^2}{z}\Bigr)^2}^{-1}\\
	&\pm \tfrac{i}{2} \bigl[1 - s(p_0)\bigr] \sqrt{\tfrac{1}{c \, z} - \tfrac{1}{4} \Bigl(\tfrac{1}{c} - \tfrac{\tilde m^2}{z} - \tfrac{i \tilde\gamma^2}{z}\Bigr)^2}^{-1}\biggr].
\end{mleqn}

\section{Numerical Implementation}

All numerical operations were carried out in \textit{Mathematica}. To implement the closed set of flow equations \labelcref{eqn:mk mink flow,eqn:rhok mink flow,eqn:lambdak mink flow,eqn:z1 flow at rho0,eqn:g1 flow at rho0,eqn:zk flow at rho0} numerically, we defined several auxiliary functions.

\subsection{Auxiliary functions}

A complex square root with branch cut along the negative real axis.
\begin{mmaCell}[moredefined={sqrt},morepattern={x_,y_}]{Code}
sqrt[(x_)?NumericQ,(y_)?NumericQ] = Piecewise[{{I*Sqrt[-x],Re[x] < 0
&& Re[y] >= 0},{(-I)*Sqrt[-x],Re[x] < 0 && Re[y] < 0}},Sqrt[x]]
\end{mmaCell}

\begin{mmaCell}{Output}
i \mmaSqrt{-x}	Re[x]<0 && Re[y]\(\geq\)0
-i \mmaSqrt{-x}	Re[x]<0 && Re[y]<0
\mmaSqrt{x}		 True
\end{mmaCell}
The second argument $y$ decides which branch to take (upper branch if $\Re(y) \geq 0$, lower if $\Re(y) < 0$). To allow Mathematica to perform symbolic simplifications on \mmaInlineCell[moredefined={sqrt}]{Code}{sqrt}, we supply
\begin{mmaCell}[moredefined={sqrt},morepattern={x_,y_,x}]{Code}
Derivative[1,0][sqrt][x_,y_] = 1/(2*sqrt[x,y]);
Derivative[0,1][sqrt][x_,y_] = 0;
\end{mmaCell}
The numerator coefficients $\alpha^{\pm\pm}$ of the propagator decomposition \labelcref{eqn:prop decomposition} are defined as
\begin{mmaCell}[morepattern={m2_, g2_, z_, c_, c, m2, z, g2}]{Input}
\mmaUnd{\(\pmb{\alpha}\)k}[1,1][m2_,g2_,z_,c_]:=\mmaFrac{1}{2}(\mmaFrac{1}{c}+\mmaFrac{m2}{z}-\mmaFrac{\mmaDef{i} g2}{z})
	+\mmaDef{i} sqrt[\mmaFrac{1}{c z}-\mmaFrac{1}{4}\mmaSup{(\mmaFrac{1}{c}-\mmaFrac{m2}{z}+\mmaFrac{\mmaDef{i} g2}{z})}{2},-(\mmaFrac{1}{c}-\mmaFrac{m2}{z})];
\mmaUnd{\(\pmb{\alpha}\)k}[1,2][m2_,g2_,z_,c_]:=\mmaFrac{1}{2}(\mmaFrac{1}{c}+\mmaFrac{m2}{z}+\mmaFrac{\mmaDef{i} g2}{z})
	-\mmaDef{i} sqrt[\mmaFrac{1}{c z}-\mmaFrac{1}{4}\mmaSup{(\mmaFrac{1}{c}-\mmaFrac{m2}{z}-\mmaFrac{\mmaDef{i} g2}{z})}{2},\mmaFrac{1}{c}-\mmaFrac{m2}{z}];
\mmaUnd{\(\pmb{\alpha}\)k}[2,1][m2_,g2_,z_,c_]:=\mmaFrac{1}{2}(\mmaFrac{1}{c}+\mmaFrac{m2}{z}-\mmaFrac{\mmaDef{i} g2}{z})
	-\mmaDef{i} sqrt[\mmaFrac{1}{c z}-\mmaFrac{1}{4}\mmaSup{(\mmaFrac{1}{c}-\mmaFrac{m2}{z}+\mmaFrac{\mmaDef{i} g2}{z})}{2},-(\mmaFrac{1}{c}-\mmaFrac{m2}{z})];
\mmaUnd{\(\pmb{\alpha}\)k}[2,2][m2_,g2_,z_,c_]:=\mmaFrac{1}{2}(\mmaFrac{1}{c}+\mmaFrac{m2}{z}+\mmaFrac{\mmaDef{i} g2}{z})
	+\mmaDef{i} sqrt[\mmaFrac{1}{c z}-\mmaFrac{1}{4}\mmaSup{(\mmaFrac{1}{c}-\mmaFrac{m2}{z}-\mmaFrac{\mmaDef{i} g2}{z})}{2},\mmaFrac{1}{c}-\mmaFrac{m2}{z}];
\mmaUnd{\(\pmb{\alpha}\)k}[1,0][m2_,g2_,z_,c_]:=\mmaFrac{1}{2}(\mmaFrac{1}{c}+\mmaFrac{m2}{z})+\mmaFrac{1}{2}(\mmaDef{i} sqrt[\mmaFrac{1}{c z}-\mmaFrac{1}{4}\mmaSup{(\mmaFrac{1}{c}-\mmaFrac{m2}{z}+\mmaFrac{\mmaDef{i} g2}{z})}{2}
	,-(\mmaFrac{1}{c}-\mmaFrac{m2}{z})]-\mmaDef{i} sqrt[\mmaFrac{1}{c z}-\mmaFrac{1}{4}\mmaSup{(\mmaFrac{1}{c}-\mmaFrac{m2}{z}-\mmaFrac{\mmaDef{i} g2}{z})}{2},\mmaFrac{1}{c}-\mmaFrac{m2}{z}]);
\mmaUnd{\(\pmb{\alpha}\)k}[2,0][m2_,g2_,z_,c_]:=\mmaFrac{1}{2}(\mmaFrac{1}{c}+\mmaFrac{m2}{z})-\mmaFrac{1}{2}(\mmaDef{i} sqrt[\mmaFrac{1}{c z}-\mmaFrac{1}{4}\mmaSup{(\mmaFrac{1}{c}-\mmaFrac{m2}{z}+\mmaFrac{\mmaDef{i} g2}{z})}{2}
	,-(\mmaFrac{1}{c}-\mmaFrac{m2}{z})]-\mmaDef{i} sqrt[\mmaFrac{1}{c z}-\mmaFrac{1}{4}\mmaSup{(\mmaFrac{1}{c}-\mmaFrac{m2}{z}-\mmaFrac{\mmaDef{i} g2}{z})}{2},\mmaFrac{1}{c}-\mmaFrac{m2}{z}]);
\end{mmaCell}
where \mmaInlineCell[morepattern={m2}]{Code}{m2}${} = \tilde m^2 = m^2/k^2$, \mmaInlineCell[morepattern={g2}]{Code}{g2}${} = \tilde\gamma^2 = \gamma^2/k^2$. Similarly, the numerator $\beta^{\pm\pm}$ coefficients read
\begin{mmaCell}[morepattern={m2_, g2_, z_, c_, z, c, m2, g2}]{Input}
\mmaUnd{\(\pmb{\beta}\)k}[1,1][m2_,g2_,z_,c_]:=\mmaFrac{1}{2 z}+\mmaFrac{\mmaDef{i} (\mmaFrac{1}{c}-\mmaFrac{m2}{z}+\mmaFrac{\mmaDef{i} g2}{z})}{(4 z) sqrt[\mmaFrac{1}{c z}-\mmaFrac{1}{4} \mmaSup{(\mmaFrac{1}{c}-\mmaFrac{m2}{z}+\mmaFrac{\mmaDef{i} g2}{z})}{2},-(\mmaFrac{1}{c}-\mmaFrac{m2}{z})]};
\mmaUnd{\(\pmb{\beta}\)k}[1,2][m2_,g2_,z_,c_]:=\mmaFrac{1}{2 z}-\mmaFrac{\mmaDef{i} (\mmaFrac{1}{c}-\mmaFrac{m2}{z}-\mmaFrac{\mmaDef{i} g2}{z})}{(4 z) sqrt[\mmaFrac{1}{c z}-\mmaFrac{1}{4} \mmaSup{(\mmaFrac{1}{c}-\mmaFrac{m2}{z}-\mmaFrac{\mmaDef{i} g2}{z})}{2},\mmaFrac{1}{c}-\mmaFrac{m2}{z}]};
\mmaUnd{\(\pmb{\beta}\)k}[2,1][m2_,g2_,z_,c_]:=\mmaFrac{1}{2 z}-\mmaFrac{\mmaDef{i} (\mmaFrac{1}{c}-\mmaFrac{m2}{z}+\mmaFrac{\mmaDef{i} g2}{z})}{(4 z) sqrt[\mmaFrac{1}{c z}-\mmaFrac{1}{4} \mmaSup{(\mmaFrac{1}{c}-\mmaFrac{m2}{z}+\mmaFrac{\mmaDef{i} g2}{z})}{2},-(\mmaFrac{1}{c}-\mmaFrac{m2}{z})]};
\mmaUnd{\(\pmb{\beta}\)k}[2,2][m2_,g2_,z_,c_]:=\mmaFrac{1}{2 z}+\mmaFrac{\mmaDef{i} (\mmaFrac{1}{c}-\mmaFrac{m2}{z}-\mmaFrac{\mmaDef{i} g2}{z})}{(4 z) sqrt[\mmaFrac{1}{c z}-\mmaFrac{1}{4} \mmaSup{(\mmaFrac{1}{c}-\mmaFrac{m2}{z}-\mmaFrac{\mmaDef{i} g2}{z})}{2},\mmaFrac{1}{c}-\mmaFrac{m2}{z}]};
\mmaUnd{\(\pmb{\beta}\)k}[1,0][m2_,g2_,z_,c_]:=\mmaFrac{1}{2 z}+\mmaFrac{\mmaDef{i} (\mmaFrac{1}{c}-\mmaFrac{m2}{z}+\mmaFrac{\mmaDef{i} g2}{z})}{(8 z) sqrt[\mmaFrac{1}{c z}-\mmaFrac{1}{4} \mmaSup{(\mmaFrac{1}{c}-\mmaFrac{m2}{z}+\mmaFrac{\mmaDef{i} g2}{z})}{2},-(\mmaFrac{1}{c}-\mmaFrac{m2}{z})]}
	-\mmaFrac{\mmaDef{i} (\mmaFrac{1}{c}-\mmaFrac{m2}{z}-\mmaFrac{\mmaDef{i} g2}{z})}{(8 z) sqrt[\mmaFrac{1}{c z}-\mmaFrac{1}{4} \mmaSup{(\mmaFrac{1}{c}-\mmaFrac{m2}{z}-\mmaFrac{\mmaDef{i} g2}{z})}{2},\mmaFrac{1}{c}-\mmaFrac{m2}{z}]};
\mmaUnd{\(\pmb{\beta}\)k}[2,0][m2_,g2_,z_,c_]:=\mmaFrac{1}{2 z}-\mmaFrac{\mmaDef{i} (\mmaFrac{1}{c}-\mmaFrac{m2}{z}+\mmaFrac{\mmaDef{i} g2}{z})}{(8 z) sqrt[\mmaFrac{1}{c z}-\mmaFrac{1}{4} \mmaSup{(\mmaFrac{1}{c}-\mmaFrac{m2}{z}+\mmaFrac{\mmaDef{i} g2}{z})}{2},-(\mmaFrac{1}{c}-\mmaFrac{m2}{z})]}
	+\mmaFrac{\mmaDef{i} (\mmaFrac{1}{c}-\mmaFrac{m2}{z}-\mmaFrac{\mmaDef{i} g2}{z})}{(8 z) sqrt[\mmaFrac{1}{c z}-\mmaFrac{1}{4} \mmaSup{(\mmaFrac{1}{c}-\mmaFrac{m2}{z}-\mmaFrac{\mmaDef{i} g2}{z})}{2},\mmaFrac{1}{c}-\mmaFrac{m2}{z}]};
\end{mmaCell}

\subsection{Threshold Functions}

The $T = 0$-part of the threshold functions $\tfim_j$ in $d + 1 = 4$ spacetime dimensions can be defined as
\begin{mmaCell}[morepattern={m2_, g2_, z_, c_, \#1}]{Input}
tfI[0][m2_,g2_,z_,0,c_,4]=\mmaFrac{\mmaSup{m2}{2}-\mmaSup{g2}{2}}{16\mmaSup{\mmaDef{\(\pmb{\pi}\)}}{2}\mmaSup{z}{2}}+(4 #1-2 m2 \mmaSub{\(\pmb{\partial}\)}{m2}#1-2 g2 \mmaSub{\(\pmb{\partial}\)}{g2}#1&)[ker[4]];
tfI[1][m2_,g2_,z_,0,c_,4]=-\mmaFrac{m2}{8\mmaSup{\mmaDef{\(\pmb{\pi}\)}}{2}\mmaSup{z}{2}}-(2 \mmaSub{\(\pmb{\partial}\)}{m2}#1-2 m2 \mmaSub{\(\pmb{\partial}\)}{\{m2,2\}}#1-2 g2 \mmaSub{\(\pmb{\partial}\)}{m2,g2}#1&)[ker[4]];
tfI[2][m2_,g2_,z_,0,c_,4]=\mmaFrac{1}{8\mmaSup{\mmaDef{\(\pmb{\pi}\)}}{2}\mmaSup{z}{2}}-(2 m2 \mmaSub{\(\pmb{\partial}\)}{\{m2,3\}}#1+2 g2 \mmaSub{\(\pmb{\partial}\)}{\{m2,2\},g2}#1&)[ker[4]];\medskip
tfI[3][m2_,g2_,z_,0,c_,4]=(2 \mmaSub{\(\pmb{\partial}\)}{\{m2,3\}}#1+2 m2 \mmaSub{\(\pmb{\partial}\)}{\{m2,4\}}#1+2 g2 \mmaSub{\(\pmb{\partial}\)}{\{m2,3\},g2}#1&)[ker[4]];\bigskip
tfI[4][m2_,g2_,z_,0,c_,4]=-(4 \mmaSub{\(\pmb{\partial}\)}{\{m2,4\}}#1+2 m2 \mmaSub{\(\pmb{\partial}\)}{\{m2,5\}}#1+2 g2 \mmaSub{\(\pmb{\partial}\)}{\{m2,4\},g2}#1&)[ker[4]];
\end{mmaCell}
where the derivatives act on the kernel
\begin{mmaCell}[morepattern={m2_, g2_, z_, c_, \#1}]{Input}
ker[4]=\mmaFrac{1}{64\mmaSup{\mmaDef{\(\pmb{\pi}\)}}{2}}\mmaUnderOver{\(\pmb{\sum}\)}{i,j}{2}\mmaSup{\mmaUnd{\(\pmb{\alpha}\)k}[i,j][m2,g2,z,c]}{2} Log[\mmaUnd{\(\pmb{\alpha}\)k}[i,j][m2,g2,z,c]];
\end{mmaCell}

The temperature-independent part of $\partial_{q^2} \tfjm$ in $d + 1 = 4$ we implemented as
\begin{mmaCell}[morepattern={q0_, m1_, m2_, g1_, g2_, z1_, z2_, c_, \#1},morelocal={a1, a2, b1, b2}]{Input}
dqTfJ[q0_,m1_,m2_,g1_,g2_,z1_,z2_,0,c_,4]=\mmaFrac{1}{2(2\mmaSup{\mmaDef{\(\pmb{\pi}\)}}{2})}Re[(-2 #1-q0 \mmaSub{\(\pmb{\partial}\)}{q0}#1-2m1 \mmaSub{\(\pmb{\partial}\)}{m1}#1-2m2 \mmaSub{\(\pmb{\partial}\)}{m2}#1-2g1 \mmaSub{\(\pmb{\partial}\)}{g1}#1-2g2 \mmaSub{\(\pmb{\partial}\)}{g2}#1&)[Module[\{a1,a2,b1,b2\},\mmaFrac{1}{4}\mmaUnderOver{\(\pmb{\sum}\)}{i,j}{2}\mmaUnderOver{\(\pmb{\sum}\)}{sig1}{2}\mmaUnderOver{\(\pmb{\sum}\)}{sig2}{2}(a1=\mmaUnd{\(\pmb{\alpha}\)k}[i,sig1][m1,g1,z1,c];a2=\mmaUnd{\(\pmb{\alpha}\)k}[j,sig2][m2,g2,z2,c];b1=\mmaUnd{\(\pmb{\beta}\)k}[i,sig1][m1,g1,z1,c];b2=\mmaUnd{\(\pmb{\beta}\)k}[j,sig2][m2,g2,z2,c];b1 b2(Piecewise[\(\pmb{\{}\)\mmaFrac{\mmaSup{(a1-a2)}{2}-(a1+a2)\mmaSup{q0}{2}}{4\mmaSup{q0}{4}\mmaSqrt{-\mmaSup{(a1-a2)}{2}+2(a1+a2)\mmaSup{q0}{2}-\mmaSup{q0}{4}}}(ArcTan[\mmaFrac{\mmaSup{q0}{2}+a1-a2}{\mmaSqrt{-\mmaSup{q0}{4}-\mmaSup{(a1-a2)}{2}+2\mmaSup{q0}{2}(a1+a2)}}]
+ArcTan[\mmaFrac{\mmaSup{q0}{2}-a1+a2}{\mmaSqrt{-\mmaSup{q0}{4}-\mmaSup{(a1-a2)}{2}+2\mmaSup{q0}{2}(a1+a2)}}])+\mmaFrac{1}{4\mmaSup{q0}{2}}-\mmaFrac{(a1-a2)(Log[a1]-Log[a2])}{8\mmaSup{q0}{4}},
a1-a2\(\pmb{\neq}\)0\(\pmb{\}}\),\(\pmb{\{}\)\mmaFrac{1}{4\mmaSup{q0}{2}}-\mmaFrac{a1 ArcTan[\mmaFrac{\mmaSup{q0}{2}}{\mmaSqrt{4a1\mmaSup{q0}{2}-\mmaSup{q0}{4}}}]}{\mmaSup{q0}{2} \mmaSqrt{4 a1 \mmaSup{q0}{2}-\mmaSup{q0}{4}}},a1-a2==0\(\pmb{\}}\)]))]]];
\end{mmaCell}
and for $q_0 = 0$ as appropriate for Goldstone bosons and the flow equation of $Z_k$ as
\begin{mmaCell}[morepattern={m1_, m2_, g1_, g2_, z1_, z2_, c_, \#1},morelocal={a1, a2, b1, b2}]{Input}
dqTfJ[0,m1_,m2_,g1_,g2_,z1_,z2_,0,c_,4]=\mmaFrac{1}{2(2\mmaSup{\mmaDef{\(\pmb{\pi}\)}}{2})}Re[(-2 #1-2m1 \mmaSub{\(\pmb{\partial}\)}{m1}#1-2m2 \mmaSub{\(\pmb{\partial}\)}{m2}#1-2g1 \mmaSub{\(\pmb{\partial}\)}{g1}#1-2g2 \mmaSub{\(\pmb{\partial}\)}{g2}#1&)[Module[\{a1,a2,b1,b2\},\mmaFrac{1}{4}\mmaUnderOver{\(\pmb{\sum}\)}{i,j}{2}\mmaUnderOver{\(\pmb{\sum}\)}{sig1}{2}\mmaUnderOver{\(\pmb{\sum}\)}{sig2}{2}(a1=\mmaUnd{\(\pmb{\alpha}\)k}[i,sig1][m1,g1,z1,c];a2=\mmaUnd{\(\pmb{\alpha}\)k}[j,sig2][m2,g2,z2,c];b1=\mmaUnd{\(\pmb{\beta}\)k}[i,sig1][m1,g1,z1,c];b2=\mmaUnd{\(\pmb{\beta}\)k}[j,sig2][m2,g2,z2,c];b1 b2 Piecewise[\(\pmb{\{}\)\mmaFrac{-\mmaSup{a1}{2}+\mmaSup{a2}{2}+2 a1 a2(Log[a1]-Log[a2])}{8\mmaSup{(a1-a2)}{3}},a1-a2\(\pmb{\neq}\)0\(\pmb{\}}\),\(\pmb{\{}\)-\mmaFrac{1}{24a1},a1-a2==0\(\pmb{\}}\)])]]];
\end{mmaCell}
The discontinuous part $\disc_{q_0} J$ is given by
\begin{mmaCell}[morepattern={q0_, m1_, m2_, g1_, g2_, z1_, z2_, c_, \#1},morelocal={a1, a2, b1, b2}]{Input}
discTfJ[q0_,m1_,m2_,g1_,g2_,z1_,z2_,0,c_,4]=(q0 \mmaSub{\(\pmb{\partial}\)}{q0}#1+2m1 \mmaSub{\(\pmb{\partial}\)}{m1}#1+2m2 \mmaSub{\(\pmb{\partial}\)}{m2}#1+2g1 \mmaSub{\(\pmb{\partial}\)}{g1}#1+2g2 \mmaSub{\(\pmb{\partial}\)}{g2}#1&)[Module[\{a1,a2,b1,b2\},\mmaUnderOver{\(\pmb{\sum}\)}{i,j}{2}(a1=\mmaUnd{\(\pmb{\alpha}\)k}[i,0][m1,g1,z1,c];a2=\mmaUnd{\(\pmb{\alpha}\)k}[j,0][m2,g2,z2,c];b1=\mmaUnd{\(\pmb{\beta}\)k}[i,0][m1,g1,z1,c];b2=\mmaUnd{\(\pmb{\beta}\)k}[j,0][m2,g2,z2,c];\mmaFrac{1}{16\mmaDef{\(\pmb{\pi}\)}q0}(b1 b2)\mmaSqrt{(\mmaSup{q0}{4}-2\mmaSup{q0}{2}(a1+a2)+\mmaSup{(a1-a2)}{2})/\mmaSup{q0}{2}} UnitStep[Re[q0]-Re[\mmaSqrt{a1}+\mmaSqrt{a2}]])]];
\end{mmaCell}

\subsection{Flow equations}

With the above definitions, the flow equations for $\rho_0$, $\lambda$, $Z_k$, $\gamma_1^2$ and $Z_1$ read
\begin{mmaCell}[morelocalconflict={g2_, Z1_, c_, d_, N_, g2, Z1, c, d},morepattern={q0_, q0, Zk_, Zk}]{Input}
\mmaUnd{\(\pmb{\rho}\)0Flow}[\mmaPat{\(\pmb{\rho}\)0_},\mmaPat{\(\pmb{\lambda}\)k_},\mmaPat{\(\pmb{\eta}\)k_},\mmaPat{g2_},\mmaPat{Z1_},\mmaPat{c_},\mmaPat{d_},\mmaPat{N_}]:=-(2+\mmaPat{\(\pmb{\eta}\)k})\mmaPat{\(\pmb{\rho}\)0}
	+(\mmaFrac{3}{2}tfI[1][2\mmaPat{\(\pmb{\rho}\)0} \mmaPat{\(\pmb{\lambda}\)k},\mmaPat{g2},\mmaPat{Z1},0,\mmaPat{c},\mmaPat{d}]+\mmaFrac{1}{2}(N-1)tfI[1][0,0,1,0,\mmaPat{c},\mmaPat{d}])\medskip
\mmaUnd{\(\pmb{\lambda}\)kFlow}[\mmaLCn{\(\pmb{\rho}\)0_},\mmaLCn{\(\pmb{\lambda}\)k_},\mmaLCn{\(\pmb{\eta}\)k_},g2_,Z1_,c_,d_,N_]:=2\mmaLCn{\(\pmb{\eta}\)k} \mmaLCn{\(\pmb{\lambda}\)k}
	+\mmaSup{\mmaLCn{\(\pmb{\lambda}\)k}}{2}(\mmaFrac{9}{2}tfI[2][2\mmaLCn{\(\pmb{\rho}\)0} \mmaLCn{\(\pmb{\lambda}\)k},g2,Z1,0,c,d]+\mmaFrac{1}{2}(N-1)tfI[2][0,0,1,0,c,d])\medskip
g2Flow[q0_,\mmaLCn{\(\pmb{\rho}\)0_},\mmaLCn{\(\pmb{\lambda}\)k_},\mmaLCn{\(\pmb{\eta}\)k_},g2_,Z1_,c_,d_,N_]:=(\mmaLCn{\(\pmb{\eta}\)k}-2)g2
	-2\mmaLCn{\(\pmb{\rho}\)0} \mmaSup{\mmaLCn{\(\pmb{\lambda}\)k}}{2}(9 discTfJ[q0,2\mmaLCn{\(\pmb{\rho}\)0} \mmaLCn{\(\pmb{\lambda}\)k},2\mmaLCn{\(\pmb{\rho}\)0} \mmaLCn{\(\pmb{\lambda}\)k},g2,g2,Z1,Z1,0,c,d]
		+(N-1)discTfJ[q0,0,0,0,0,1,1,0,c,d])\medskip
Z1Flow[\mmaLCn{q0_},\mmaLCn{\(\pmb{\rho}\)0_},\mmaLCn{\(\pmb{\lambda}\)k_},\mmaLCn{\(\pmb{\eta}\)k_},g2_,Z1_,c_,d_,N_]:=\mmaLCn{\(\pmb{\eta}\)k} Z1
	-2\mmaLCn{\(\pmb{\rho}\)0} \mmaSup{\mmaLCn{\(\pmb{\lambda}\)k}}{2}(9dqTfJ[\mmaLCn{q0},2\mmaLCn{\(\pmb{\rho}\)0} \mmaLCn{\(\pmb{\lambda}\)k},2\mmaLCn{\(\pmb{\rho}\)0} \mmaLCn{\(\pmb{\lambda}\)k},g2,g2,Z1,Z1,0,c,d]
		-(N-1)dqTfJ[\mmaLCn{q0},0,0,0,0,1,1,0,c,d])\medskip
ZkFlow[Zk_,\mmaLCn{\(\pmb{\rho}\)0_},\mmaLCn{\(\pmb{\lambda}\)k_},g2_,Z1_,c_,d_]:=2Zk \mmaLCn{\(\pmb{\rho}\)0} \mmaSup{\mmaLCn{\(\pmb{\lambda}\)k}}{2} dqTfJ[0,2\mmaLCn{\(\pmb{\rho}\)0} \mmaLCn{\(\pmb{\lambda}\)k},0,g2,0,Z1,1,0,c,d]
\end{mmaCell}
and can be solved with
\begin{mmaCell}[morelocal={runner, counter, domain, sol, c, d}]{Input}
AbsoluteTiming@Module[\{runner=0,counter=0,domain=\{t,-10,0\},sol=run[5]\},
With[\{\mmaLoc{\(\pmb{\eta}\)k}=-\mmaSup{Zk}{\mmaUnd{\(\pmb{\prime}\)}}[t]/Zk[t],c=1,d=4,N=2\},
run[5]=NDSolve[\{\mmaSup{\mmaUnd{\(\pmb{\rho}\)0}}{\mmaUnd{\(\pmb{\prime}\)}}[t]==\mmaUnd{\(\pmb{\rho}\)0Flow}[\mmaUnd{\(\pmb{\rho}\)0}[t],\mmaUnd{\(\pmb{\lambda}\)k}[t],\mmaLoc{\(\pmb{\eta}\)k},g2[t],Z1[t],c,d,N],
\mmaSup{\mmaUnd{\(\pmb{\lambda}\)k}}{\mmaUnd{\(\pmb{\prime}\)}}[t]==\mmaUnd{\(\pmb{\lambda}\)kFlow}[\mmaUnd{\(\pmb{\rho}\)0}[t],\mmaUnd{\(\pmb{\lambda}\)k}[t],\mmaLoc{\(\pmb{\eta}\)k},g2[t],Z1[t],c,d,N],
\mmaSup{Zk}{\mmaUnd{\(\pmb{\prime}\)}}[t]==ZkFlow[Zk[t],\mmaUnd{\(\pmb{\rho}\)0}[t],\mmaUnd{\(\pmb{\lambda}\)k}[t],g2[t],Z1[t],c,d],
\mmaSup{g2}{\mmaUnd{\(\pmb{\prime}\)}}[t]==g2Flow[\mmaSqrt{2 \mmaUnd{\(\pmb{\rho}\)0}[t] \mmaUnd{\(\pmb{\lambda}\)k}[t]/Z1[t]},\mmaUnd{\(\pmb{\rho}\)0}[t],\mmaUnd{\(\pmb{\lambda}\)k}[t],\mmaLoc{\(\pmb{\eta}\)k},g2[t],Z1[t],c,d,N],
\mmaSup{Z1}{\mmaUnd{\(\pmb{\prime}\)}}[t]==Z1Flow[\mmaSqrt{2 \mmaUnd{\(\pmb{\rho}\)0}[t] \mmaUnd{\(\pmb{\lambda}\)k}[t]/Z1[t]},\mmaUnd{\(\pmb{\rho}\)0}[t],\mmaUnd{\(\pmb{\lambda}\)k}[t],\mmaLoc{\(\pmb{\eta}\)k},g2[t],Z1[t],c,d,N],
WhenEvent[-Z1[t]+\mmaUnd{\(\pmb{\lambda}\)k}[t] \mmaUnd{\(\pmb{\rho}\)0}[t]>0,"CrossDiscontinuity"],
\mmaUnd{\(\pmb{\rho}\)0}[0]==0.02`,\mmaUnd{\(\pmb{\lambda}\)k}[0]==0.5`,Zk[0]==1,g2[0]==0,Z1[0]==1\},\{\mmaUnd{\(\pmb{\rho}\)0},\mmaUnd{\(\pmb{\lambda}\)k},Zk,g2,Z1\},domain,
StepMonitor\(\pmb{:\to}\)counter++ If[Abs[t]>runner,Print[Chop[\{counter,Round[t,1],-Z1[t]+\mmaUnd{\(\pmb{\lambda}\)k}[t] \mmaUnd{\(\pmb{\rho}\)0}[t],\mmaSup{\mmaDef{e}}{2 t} \mmaUnd{\(\pmb{\rho}\)0}[t],\mmaUnd{\(\pmb{\lambda}\)k}[t],Zk[t],\mmaSup{\mmaDef{e}}{2 t} g2[t],Z1[t]\}]];runner++]];
Plot[\mmaSup{\mmaDef{e}}{2 t} \mmaUnd{\(\pmb{\rho}\)0}[t]/.\(\pmb{\,}\)sol,domain,PlotRange\(\pmb{\to}\)\{0,All\},AxesLabel\(\pmb{\to}\)\{"t",\mmaSub{\(\pmb{\rho}\)}{0}(t)/\mmaSup{\(\pmb{\Lambda}\)}{2}\}] Plot[\mmaUnd{\(\pmb{\lambda}\)k}[t]/.\(\pmb{\,}\)sol,domain,AxesLabel\(\pmb{\to}\)\{"t",\mmaSub{\(\pmb{\lambda}\)}{k}(t)\}] 
Plot[-\mmaSup{Zk}{\mmaUnd{\(\pmb{\prime}\)}}[t]/Zk[t]/.\(\pmb{\,}\)sol,domain,AxesLabel\(\pmb{\to}\)\{"t",\mmaSub{\(\pmb{\eta}\)}{k}(t)\}]
Plot[\mmaSup{\mmaDef{e}}{2 t} g2[t]/.\(\pmb{\,}\)sol,domain,AxesLabel\(\pmb{\to}\)\{"t",\mmaSup{\mmaSub{\(\pmb{\gamma}\)}{1}}{2}(t)/\mmaSup{\(\pmb{\Lambda}\)}{2}\}]
Plot[Z1[t]/.\(\pmb{\,}\)sol,domain,AxesLabel\(\pmb{\to}\)\{"t",\mmaSub{Z}{1}(t)\}]]]
\end{mmaCell}
The plot commands contain \mmaInlineCell{Input}{\mmaSup{\mmaDef{e}}{2t} \mmaUnd{\(\pmb{\rho}\)0}[t]} and \mmaInlineCell{Input}{\mmaSup{\mmaDef{e}}{2t} g2[t]} because we performed the internal calculations using dimensionless variables $\tilde\rho_0(k) = \rho_0(k)/k^2 = \rho_0(k)/(\Lambda^2 e^{2 t})$, $\tilde\gamma_1^2 = \gamma_1^2/k^2 = \gamma_1^2/(\Lambda^2 e^{2 t})$. We use the event handler \mmaInlineCell{Input}{\mmaDef{WhenEvent}[Re[\mmaSqrt{-\mmaUnd{Z1}[t]+\mmaUnd{\(\pmb{\rho}\)0}[t]\mmaUnd{\(\pmb{\lambda}\)k}[t]}]==0,"CrossDiscontinuity"]} here to take care of a singularity of \mmaInlineCell{Input}{g2Flow} that is due to the term \mmaInlineCell{Input}{\mmaSqrt{-\mmaUnd{Z1}[t]+\mmaUnd{\(\pmb{\lambda}\)k}[t]\mmaUnd{\(\pmb{\rho}\)0}[t]}} in the denominator of \mmaInlineCell{Input}{discTfJ[\mmaSqrt{2\mmaUnd{\(\pmb{\rho}\)0}[t]\mmaUnd{\(\pmb{\lambda}\)k}[t]},0,0,0,0,1,1,0,1,4]} which incorporates fluctuations of the Goldstone bosons. \mmaInlineCell{Input}{\mmaSqrt{-\mmaUnd{Z1}[t]+\mmaUnd{\(\pmb{\lambda}\)k}[t]\mmaUnd{\(\pmb{\rho}\)0}[t]}} vanishes near $t=-3$. The singularity leads to a divergent derivative \mmaInlineCell{Input}{\mmaSup{g2}{\mmaUnd{\(\pmb{\prime}\)}}[t]} of the discontinuity $\gamma_1^2$ of the radial propagator which abruptly becomes zero (cf. \cref{fig:gamma}) above a certain renormalization scale $k^2 > m_1^2$ because the radial mode can exist as an on-shell stable particle above those energies.

\begin{multicols}{2}[{\printbibheading[title={References}]}]
	\printbibliography[heading=none]
\end{multicols}

\section*{Acknowledgments}

I would like to thank Stefan Flörchinger for providing me with this exciting research topic and for continued guidance during the preparation of this thesis. I also extend my gratitude to my other advisors Christof Wetterich and Michael Scherer for helpful feedback and thoughtful counsel. I thank Fabian Rennecke, Nicolas Wink, Manuel Scherzer, Sebastian Wetzel and Daniel Rosenblüh for many interesting discussions, patient explanations and enjoyable company. My sincerest appreciation goes to Thomas Mikhail and Lukas Barth for proofreading this thesis and providing me with valuable feedback, coming up with many critical questions and following through with enlightening discussions! For wonderful memories and many special moments during my studies in Heidelberg I thank my friends Friederike Erbe, Florian Kleinicke, Lars Hansen, Daniel Rosenblüh, Lukas Barth, Thomas Mikhail, Benjamin Freist, Adrian van Kan, Clara Miralles Vila, Clara Murgui Galvez and my sister Kaja Riebesell. Lastly, the biggest thank-you goes out to my parents Mona Botros and Ulf Riebesell, who supported me in every way possible and without whom I would not be who and where I am today!

\section*{Declaration of Authorship}

I hereby certify that this thesis has been composed by me and is based on my own work, except where stated otherwise.

\bigskip\noindent
Heidelberg, \rule{4cm}{0.1ex}
\hfill
\rule{6cm}{0.1ex}

\end{document}